\documentclass{aa}  

\usepackage{graphicx}
\usepackage{txfonts}

\begin{document}

\title{Unveiling dust, molecular gas, and high star-formation efficiency in extremely UV bright star-forming galaxies at $z\sim 2.1-3.6$}
\subtitle{}

\author{M. Dessauges-Zavadsky\inst{1},
            R. Marques-Chaves\inst{1},
            D. Schaerer\inst{1,2}, %\fnmsep\thanks{Just to show the usage of the elements in the author field}
            M.-Y. Xiao\inst{1},
            L. Colina\inst{3},
            J. Alvarez-Marquez\inst{3}, 
            \and
            I. P\'erez-Fournon\inst{4,5}
          }

\institute{D\'epartement d'Astronomie, Universit\'e de Gen\`eve, Chemin Pegasi 51, 1290 Versoix, Switzerland\\
\email{miroslava.dessauges@unige.ch}
\and
CNRS, IRAP, 14 Avenue E. Belin, 31400 Toulouse, France
\and 
Centro de Astrobiolog\'ia (CAB), CSIC-INTA, Ctra. de Ajalvir km 4, Torrej\'on de Ardoz, 28850, Madrid, Spain 
\and 
Instituto de Astrof\'isica de Canarias, C/V\'ia L\'actea, s/n, 38205 San Crist\'obal de La Laguna, Tenerife, Spain 
\and
Universidad de La Laguna, Dpto. Astrof\'isica, 38206 La Laguna, Tenerife, Spain
}

\date{Received; accepted}

\authorrunning{Dessauges-Zavadsky et~al.}
\titlerunning{Dust, molecular gas, and high star-formation efficiency in extremely UV bright galaxies}

% \abstract{}{}{}{}{}
% 5 {} token are mandatory
 
\abstract{We analysed the Atacama Large Millimetre/submillimetre Array (ALMA) far-infrared (FIR), 1.3~mm, dust continuum and CO emission of 12 starburst galaxies at $z\sim 2.1-3.6$ selected for their extreme brightness in the rest-frame UV, with absolute magnitudes of $-23.4$ to $-24.7$. We also analysed their Very Large Telescope (VLT) High Acuity Wide field $K$-band Imager (HAWK-I) $H$- and $K_{\rm s}$-band images. The targeted galaxies are characterised by negligible dust attenuations with blue UV spectral slopes ($-2.62$ to $-1.84$), very young stellar populations of $\sim 10$~Myr, and powerful starbursts with a high mean specific star-formation rate of $\rm 112~Gyr^{-1}$, placing them $\sim 1.5$~dex above the main sequence at similar redshifts and stellar masses ($M_{\rm stars} \sim (1.5-4.6)\times 10^9~M_{\odot}$). The FIR dust continuum emission revealed in nine galaxies gives IR luminosities of $(5.9-28.3)\times 10^{11}~L_{\odot}$, with six galaxies remaining dominated by unobscured UV star-formation rates, and high dust masses barely produced by supernovae within the 10~Myr timescale. The CO emission detected in eight galaxies leads to molecular gas masses higher than stellar masses, with the mean molecular gas mass fraction as high as $82\%$. The corresponding star-formation efficiencies reach $\gtrsim 40$\%, with amazingly short molecular gas depletion timescales between less than 13~Myr and 71~Myr. These unique properties never reported in previously studied galaxies highlight that these galaxies are likely caught at the very beginning of their stellar mass build-up and undergo a very efficient and fast conversion of gas into stars that can only result from the gas collapse within a very short free-fall time. We find that the feedback-free starburst model seems to be able to explain the formation of these galaxies. To reconcile the co-spatial FIR dust emission with the UV-bright unattenuated emission, we speculate about the presence of radiation-driven outflows that can temporarily remove dust at the location of the starburst and expel it at large distances in line with the measured high FIR effective radii (1.7~kpc to 5~kpc) in comparison to the very compact stellar radii that are a~few~hundreds~of parsecs).}
  


\maketitle

\nolinenumbers

%-----------------------------------------------------------------------

\section{Introduction}
\label{sect:introduction}

The census of star-forming galaxies at high redshifts ($z>2$) has been ongoing for decades, and different observational techniques have been developed to identify galaxies either as Lyman break galaxies (LBGs) or Lyman-$\alpha$ emitters (LAEs). Large volumes of space have been probed to investigate their space density as a function of their ultraviolet luminosity through luminosity functions \citep[e.g.][]{ReddySteidel09,Sobral+18a}. The bright end of the luminosity function (LF) of the UV and Ly$\alpha$ emission corresponds to massive star formation and high production of ionising photons. However, the identification of the most UV- or Ly$\alpha$-luminous star-forming galaxies remains challenging for three possible reasons: vigorous episodes of star formation are simply rare phenomena; galaxies with high star-formation rates (SFRs) consume their gas quickly, implying short timescales for their UV- or Ly$\alpha$-luminous phases; or galaxies at the bright end of the LF produce significant quantities of dust during their intense star formation so that these most vigorous and intrinsically luminous star-forming galaxies quickly have their emission at UV wavelengths heavily obscured by dust \citep[e.g.][]{Casey+14}. 

Probing the bright end of UV and Ly$\alpha$ LFs thus requires wide-area surveys. Several dedicated surveys at $z\sim 2-3$ covering a few square degrees ($1-4~\rm deg^2$) have searched for extremely UV-luminous galaxies but failed to discover galaxies more luminous than two times the typical luminosity ($L^{\star}$) of UV and Ly$\alpha$ of LBGs and LAEs, which corresponds to the unobscured absolute magnitude $M^{\star}_{\rm UV} = -20.7$ %-21.5 
and $\log(L^{\star}_{\rm Ly\alpha}/\rm erg~s^{-1}) = 42.8$, %43.3
respectively \citep[e.g.][]{Ouchi+08,Zheng+16,Sobral+18b}. The Baryon Oscillation Spectroscopic Survey Emission-Line Lens Survey for the GALaxy-Ly$\alpha$ EmitteR sYstems \citep[BELLS GALLERY;][]{Shu+16} led to the discovery of 
%187 galaxy-scale strong gravitational lens candidates with LAEs at $2<z<3$ as background sources. Five of them were found to have 
five LAEs at $2<z<3$ with intrinsic $M_{\rm UV}$ above $M^{\star}_{\rm UV}$, with two of them being brighter than $M_{\rm UV} < -23$ and having $L_{\rm Ly\alpha} > 2\times L^{\star}_{\rm Ly\alpha}$, and this was without the need to invoke an active galactic nucleus (AGN) component \citep{Marques+17,Marques+20a}. Additionally, five strongly lensed UV-luminous non-active LBGs with $-23.5 < M_{\rm UV} < -21.1$ were also reported in the literature but with, on average, strongly suppressed Ly$\alpha$ lines, higher metallicities, stronger interstellar medium (ISM) absorption lines, and redder slopes, suggesting higher dust attenuation than the extremely UV-bright LAEs \citep{Pettini+00,Quider+09,Quider+10,Dessauges+10,Dessauges+11,Patricio+16,
Marques+18}.

Remarkably, UV-luminous galaxies above $M^{\star}_{\rm UV}$ are also detected at $z>6$ \citep{Sobral+15,Matsuoka+18,Hashimoto+19,Endsley+21,Bouwens+22a}, and recent works now include the highest redshift sources known at the epoch of reionisation (EoR) discovered with the James Webb Space Telescope (JWST), leading to important and unexpected implications \citep[e.g.][]{Bouwens+23,Atek+23,Bunker+23,Casey+23,Castellano+24,Carniani+24}. The volume density inferred for these UV-luminous sources with $M_{\rm UV}$ reaching $-22.5$ is much higher by factors of $\sim 10-100$ than that predicted by models \citep[e.g.][]{Mason+18}, implying steeper UV LFs than predicted such that the bright end of the LFs does not significantly evolve between $8<z<16$ \citep[e.g.][]{Naidu+22,Finkelstein+24,Chemerynska+24}. The excess of these UV-bright sources at the EoR, which still needs to be confirmed with spectroscopic redshift assessments, currently puts strain on the standard $\Lambda$CDM cosmology unless new concepts in our understanding of star-formation processes and baryon physics are invoked \citep{Boylan23}. Different scenarios have thus been proposed to explain the extremely UV-luminous and massive galaxies detected at the EoR and their excess, including a feedback-free starburst yielding a very high star-formation efficiency, namely a very efficient conversion of accreted gas into stars within a very short free-fall timescale \citep{Dekel+23,Li+24}; a temporary removal of dust as a consequence of radiation-driven outflows that yields very low dust attenuation and makes the galaxies appear brighter \citep{Ferrara+23,Ziparo+23,Ferrara24}; a top-heavy IMF boosting the UV radiation and the luminosity to mass ratio \citep[e.g.][]{BekkiTsujimoto23,Trinca+24}; and a stochastic variability of the SFR \citep[e.g.][]{Mirocha+23,Gelli+24}. Which of these scenarios holds now needs to be determined. Finally, although AGN contamination could possibly also explain the excess of these UV-bright sources at the EoR, this scenario was discarded by \citet{FinkelsteinBagley22} and only the exceptionally UV-luminous galaxy GN-z11 at $z=10.6$ potentially hosts an AGN \citep{Maiolino+24}.

The recent search for extremely UV-luminous galaxies with $M_{\rm UV} < -23$ undertaken at $z\gtrsim 2$ within the $\sim 9000~\rm deg^2$-wide extended Baryon Oscillation Spectroscopic Survey \citep[eBOSS;][]{Abolfathi+18} of the Sloan Digital Sky Survey \citep[SDSS;][]{Eisenstein+11} delivered about 70 galaxies at the very bright end of the UV LF, and they are among the most UV-bright galaxies known at cosmic noon (Marques-Chaves et~al.\ in prep.). The first detailed studies of 13 of these galaxies are presented in \citet{Marques+20b}, \citet{Alvarez+21}, \citet{Marques+21}, \citet{Marques+22}, and \citet{Upadhyaya+24}. We selected 12 of the most UV-luminous galaxies with $M_{\rm UV}$ ranging from $-23.4$ to $-24.5$ 
%and $\log(L_{\rm Ly\alpha}/\rm erg~s^{-1}) = 42.9-44.1$
at $z=2.08-3.61$, accessible for observations with ALMA. They are characterised by unattenuated starlight with steep UV spectral slopes ranging from $-2.62$ to $-1.84$ and are dominated by young stellar populations with average ages of $\sim 10$~Myr, as testified by their rest-frame UV spectra revealing pure stellar features. Powered by powerful starbursts with very high specific SFRs ($\sim 100~\rm Gyr^{-1}$), the corresponding stellar masses are in the range of $(1.47-4.59)\times 10^9~M_{\odot}$, and they have sub-solar metallicities. Evidence of outflows, and even inflows, is found for some of these galaxies, as well as signatures of very massive stars (VMSs) with masses of $100~M_{\odot}-400~M_{\odot}$. Two galaxies are identified as very strong Lyman continuum (LyC) leakers. Considering these physical properties altogether, the galaxies are interpreted as being in an intense starburst phase with the bulk of their stellar mass being formed in a few millions of years.

Globally, most of the physical properties of these galaxies resemble the recently discovered galaxies at the EoR. Therefore, the scenarios proposed to explain the UV brightness and excess of galaxies out at $z>8$ could also explain the tremendous UV luminosities of these vigorous starburst galaxies at cosmic noon. The ALMA observations we acquired in the far-infrared (FIR) dust continuum at 1.3~mm (band~6) and in the CO(3--2) or CO(4--3) line emission (band~3) are key to inferring their dust and cold molecular gas masses -- two physical parameters that are essential to fully probe the star-formation process ongoing in the galaxies and bring answers to the questions as to how much dust was produced within the $\sim 10$~Myr burst timescale, where dust is located so that it is reconciled with the steep UV spectral slopes, how much obscured star formation contributes to the whole SFR budget in these highly star-forming galaxies, how much molecular gas mass is available to feed their star formation, what are the subsequent molecular gas mass depletion timescale and star-formation efficiency, with the latter expected to be higher with respect to that of nearby galaxies given the high stellar mass build-up achieved within 10~Myr only. Efficiencies higher than 20\% are indeed advocated by different studies to explain diverse observational findings at very high redshifts \citep[e.g.][]{Xiao+24,deGraaff+24,Weibel+24} and make them compatible with simulations \citep[e.g.][]{Kannan+23,Boylan23}. These answers are helpful to determine whether the molecular gas mass reservoir is consumed rapidly so that galaxies quench by starvation or, on the contrary, the molecular gas mass reservoir is massive enough to produce high amounts of dust so that galaxies turn into luminous dusty star-forming galaxies.

In Sect.~\ref{sect:targets} we describe the selection and the physical properties of the 12 UV-bright star-forming galaxies at $z = 2.08-3.61$ studied in this work. In Sect.~\ref{sect:observations} we present their ALMA and Very Large Telescope (VLT) HAWK-I observations and the corresponding data reduction and imaging. We then analyse the CO line, the FIR dust continuum, and the rest-frame UV or optical emission, and we measure CO luminosities, molecular gas masses, IR luminosities, dust masses, and FIR dust continuum as well as the rest-frame UV/optical sizes. Section~\ref{sect:results} discusses the derived measurements in the general context of main sequence (MS) star-forming galaxies and starburst galaxies at similar and higher redshifts. We focus on the dust-obscured star formation in Sect.~\ref{sect:obscured-SFR}, the molecular gas mass content and depletion timescale in Sect.~\ref{sect:Mgas-tdepl}, the star-formation efficiency in Sect.~\ref{sect:SFE}, the dust mass content in Sect.~\ref{sect:Mdust}, the FIR dust continuum and rest-frame UV/optical sizes in Sect.~\ref{sect:sizes}, and the rest-frame UV/optical morphology and spatial offsets in Sect.~\ref{sect:UVmorphology}. In Sect.~\ref{sect:discussion} we try to obtain a complete understanding of these extremely UV-luminous galaxies, considering all of their physical properties. Finally, in Sect.~\ref{sect:conclusions} we summarize our results.

Throughout the paper, we assume the $\Lambda$CDM cosmology with $\Omega_{\rm m} = 0.3$, $\Omega_{\Lambda} = 0.7$, and $H_0 = 70~\rm km~s^{-1}~Mpc^{-1}$. We adopt the \citet{Chabrier03} initial mass function (IMF).

%-----------------------------------------------------------------------
\begin{table*}
\caption{Rest-frame UV properties of the galaxy sample.}             
\label{tab:restUV-properties}      
\centering          
\begin{tabular}{c c c c c c c c} 
\hline\hline       
Target & $z_{\rm nebular}$\tablefootmark{a} & $M_{\rm UV}$ & $\beta_{\rm UV}$ & 
$L_{\rm UV}$ & $\rm SFR_{UV}$\tablefootmark{b} & $M_{\rm stars}$\tablefootmark{c} & 
sSFR\tablefootmark{d} \\
 & & & & $(10^{11}~L_{\odot})$ & $(\rm M_{\odot}~yr^{-1})$ & $(10^9~M_{\odot})$ & ($\rm Gyr^{-1}$) \\
\hline
J1322+0423  &   2.0800 & $-23.49$ & $-2.06\pm 0.12$ & $5.64\pm 0.12$ & $154\pm 7$ & $3.42\pm 0.40$ & $52\pm 7$ \\
J0146--0220  & 2.1595 & $-23.68$ & $-1.98\pm 0.12$ & $6.76\pm 0.06$ & $127\pm 6$ & $3.41\pm 0.43$ & $61\pm 23$ \\
J1415+2036 & 2.2435 & $-23.53$ & $-3.49\pm 0.11^{\dagger}$ & $5.87\pm 0.12$ & $188\pm 11$ & $1.88\pm 0.20$ & $118\pm 14$ \\
J1249+1550 & 2.2928 & $-23.41$ & $-1.84\pm 0.12$         & $5.24\pm 0.11$ & $118\pm 5$ & $4.25\pm 0.53$ & $48\pm 18$ \\
J0006+2452 & 2.3796 & $-24.17$ & $-2.30\pm 0.10$ & $10.6\pm 0.2$ & $232\pm 7$ & $3.16\pm 0.31$ & $163\pm 59$ \\
J0850+1549  & 2.4235 & $-23.76$ & $-2.62\pm 0.14$ & $7.21\pm 0.17$ & $189\pm 8$ & $1.89\pm 0.26$ & $131\pm 49$ \\
J1220--0051  & 2.4269 & $-23.50$ & $-2.43\pm 0.11$ & $5.68\pm 0.12$ & $144\pm 5$ & $1.47\pm 0.16$ & $163\pm 59$ \\
J0950+0523  & 2.4548 & $-23.69$ & $-2.41\pm 0.15$ & $6.80\pm 0.18$ & $179\pm 9$ & $1.91\pm 0.29$ & $191\pm 73$ \\ 
J1220+0842  & 2.4698 & $-24.36$ & $-2.36\pm 0.09$ & $12.6\pm 0.2$ & $302\pm 7$ & $3.60\pm 0.32$ & $100\pm 9$ \\
J1157+0113 & 2.5450 & $-23.40$ & $-2.15\pm 0.34$ & $5.19\pm 0.16$ & $104\pm 8$  & $1.94\pm 0.68$ & $126\pm 62$ \\
J0121+0025      & 3.2445 & $-24.11$ & $-2.19\pm 0.20$ & $10.0\pm 0.1$ & $269\pm 18$ & $4.59\pm 0.92$ & $74\pm 27$  \\ 
J1316+2614      & 3.6122 & $-24.65$ & $-2.59\pm 0.05$ & $16.4\pm 0.2$ & $415\pm 16$ & $4.15\pm 1.15$ & $118\pm 47$ \\
\hline                  
\end{tabular}
\tablefoot{
\tablefoottext{a}{Redshifts derived from optical nebular emission lines.}
\tablefoottext{b}{Unobscured SFRs (uncorrected for dust attenuation) obtained from the UV luminosities by applying the specific $L_{\rm UV}$--$\rm SFR_{UV}$ conversion factor $\kappa_{\rm UV} = 1.3\times 10^{-28}~M_{\odot}~\rm yr^{-1}/(erg~s^{-1}~Hz^{-1})$, derived using the BPASS models that assume the \citet{Chabrier03} IMF, a metallicity of $0.5~Z_{\odot}$, and a continuous star formation over 10~Myr (Sect.~\ref{sect:targets}).}
\tablefoottext{c}{Stellar masses of the dominant young stellar population derived by assuming a continuous star formation over 10~Myr, that is, $M_{\rm stars} = {\rm SFR_{UV}}\times 10~\rm Myr$, where $\rm SFR_{UV}$ are corrected for the dust attenuation using the observed $\beta_{\rm UV}$ and assuming the \citet{Calzetti+00} attenuation law (Sect.~\ref{sect:targets}).}
\tablefoottext{d}{Specific star-formation rates derived using the total $\rm SFR_{UV+IR}$, with $\rm SFR_{IR}$ given in Table~\ref{tab:COdust-properties}.}
\tablefoottext{$\dagger$}{The UV spectral slope of J1415+2036 is uncertain and likely unrealistic, so far measured using the low S/N SDSS spectrum. However, if confirmed, it would indicate a rest-frame UV spectrum of the galaxy dominated by stellar radiation (without nebular emission).}
}
\end{table*}
%-----------------------------------------------------------------------

%ds\section{Target selection}
\section{Target sample and their properties}
\label{sect:targets}

This work presents ALMA and VLT HAWK-I observations of 12 extremely UV-bright star-forming galaxies at $z=2.08-3.61$. These sources are part of a large sample of about 70 UV-luminous galaxies (Marques-Chaves et~al.\ in prep.) selected from the $\sim 9000~\rm deg^2$-wide eBOSS/SDSS survey \citep{Abolfathi+18} to have high UV absolute magnitudes and narrow Ly$\alpha$ emission profiles in their BOSS spectra (covering the spectral range $\lambda \simeq 3600-10000$~\AA\ with a resolving power $R\simeq 2000$). The targets studied here were selected for ALMA and HAWK-I observations with declinations $\leq 30^{\circ}$ and other properties that are representative of the parent sample. They show extremely bright $M_{\rm UV}$ ranging from $-23.4$ to $-24.6$, Ly$\alpha$ luminosities of $\log(L_{\rm Ly\alpha}/\rm erg~s^{-1}) = 42.9-44.1$, and Ly$\alpha$ rest-frame equivalent widths of $EW_{\rm Ly\alpha} = 12-55~\AA$. The rest-frame UV and optical properties of a fraction (7/12) of these targets were analysed in detail in previous works through deep spectroscopy with the 10.4~m Gran Telescopio Canarias \citep{Marques+20b,Marques+21,Marques+22,Marques+24,Alvarez+21,
Upadhyaya+24}, and for the remaining targets through shallow SDSS spectroscopy. We measured steep UV spectral slopes ($\beta_{\rm UV}$) from $-2.62$ to $-1.84$, suggesting residual/low dust attenuations $E(B-V) \leq 0.1$ when assuming the \citet{Calzetti+00} extinction law with an intrinsic $\beta_0 = -2.44$. 

In the references listed above it is also shown that the rest-frame UV spectra of the galaxies studied here reveal pure stellar features, such as wind lines and photospheric absorption, known to be produced only by the atmospheres of the most massive stars; as such, they are characterised by very young stellar populations with an average age of $\sim 10$~Myr for the whole sample (but some galaxies are even younger and some slightly older). They have metallicities in the range of $12+\log({\rm O/H}) = 8.13-8.49$, with a mean of 8.36, that is, $Z/Z_{\odot}\simeq 0.5$. 
%\citep{Marques+20b,Marques+21,Marques+22,Upadhyaya+24}. 
Using their UV luminosities ($L_{\rm UV}$), we measured high unobscured star-formation rates ${\rm SFR_{UV}} = 104-415~M_{\odot}~\rm yr^{-1}$ by applying the specific $L_{\rm UV}$--$\rm SFR_{UV}$ conversion factor $\kappa_{\rm UV} = 1.3\times 10^{-28}~M_{\odot}~\rm yr^{-1}/(erg~s^{-1}~Hz^{-1})$, derived using the Binary Population and Spectral Synthesis (BPASS) binary models \citep{Eldridge+17,StanwayEldridge18,Byrne+22}, and assuming a continuous star formation over 10~Myr, a metallicity of $0.5~Z_{\odot}$, and the \citet{Chabrier03} IMF. However, one needs to keep in mind that $\kappa_{\rm UV}$ is strongly dependent on the burst age for 10~Myr and below, such that $\rm SFR_{UV}$ gets more than twice higher for a two times shorter continuous star formation, as explained in \citet{Marques+24}. The stellar masses of the young stellar population were then derived following 
%assuming the continuous star formation over 10~Myr, i.e., 
$M_{\rm stars} = {\rm SFR_{UV}}\times 10~\rm Myr$, where $\rm SFR_{UV}$ were corrected for the dust attenuation using the observed $\beta_{\rm UV}$; 
%and assuming the \citet{Calzetti+00} attenuation law; 
we obtained $M_{\rm stars} = (1.47-4.59)\times 10^9~M_{\odot}$. Analysing the spectral energy distributions (SEDs) of the galaxies, \citet{Marques+20b,Marques+21,Marques+22} find that the SEDs are dominated by a young and intense burst of star formation, and could only infer loose upper limits on the mass of the old stellar population $\log (M_{\rm stars}^{\rm old}/M_{\odot}) < 9.8-10.2$ ($3\sigma$). The new Hubble Space Telescope (HST) photometry recently acquired for J1316+2614 provides a more stringent upper limit on $\log (M_{\rm stars}^{\rm old}/M_{\odot}) < 9.46$ ($3\sigma$), yielding a fraction of the starburst mass to the total (young+old) stellar mass $>62\%$ \citep[see Fig.~9][]{Marques+24}. The absence of relevant old stellar population in these galaxies 
%(with $M_{\rm old}/ M_{\rm young} \la 0.28-0.5$)
%ds (between a factor of less than 2 to 3.5) 
is interpreted as indicating that the galaxies are in an intense starburst phase with the bulk of their stellar mass being formed in a few millions of years. Table~\ref{tab:restUV-properties} summarizes all the rest-frame UV physical properties derived for our galaxies. 

The galaxies studied here also appear as very strong producers of ionising radiation given their extreme $M_{\rm UV}$, and two are confirmed to be strong LyC leakers \citep{Marques+21,Marques+22}, with J1316+2614 at $z=3.6122$ reaching a LyC escape fraction as high as $f_{\rm esc}({\rm LyC}) \simeq 90$\%. Moreover, they have complex gas kinematics, showing signatures of outflows from the weak ISM absorption lines having blueshifted centroids with respect to the systemic redshift \citep{Alvarez+21,Marques+21}, and even signatures of inflow for J1316+2614 from the blue-dominated Ly$\alpha$ emission with respect to the systemic redshift \citep{Marques+22}. 

%-----------------------------------------------------------------------
\begin{figure*}[!]
\hspace{0.4cm}
\includegraphics[width=0.22\textwidth,clip]{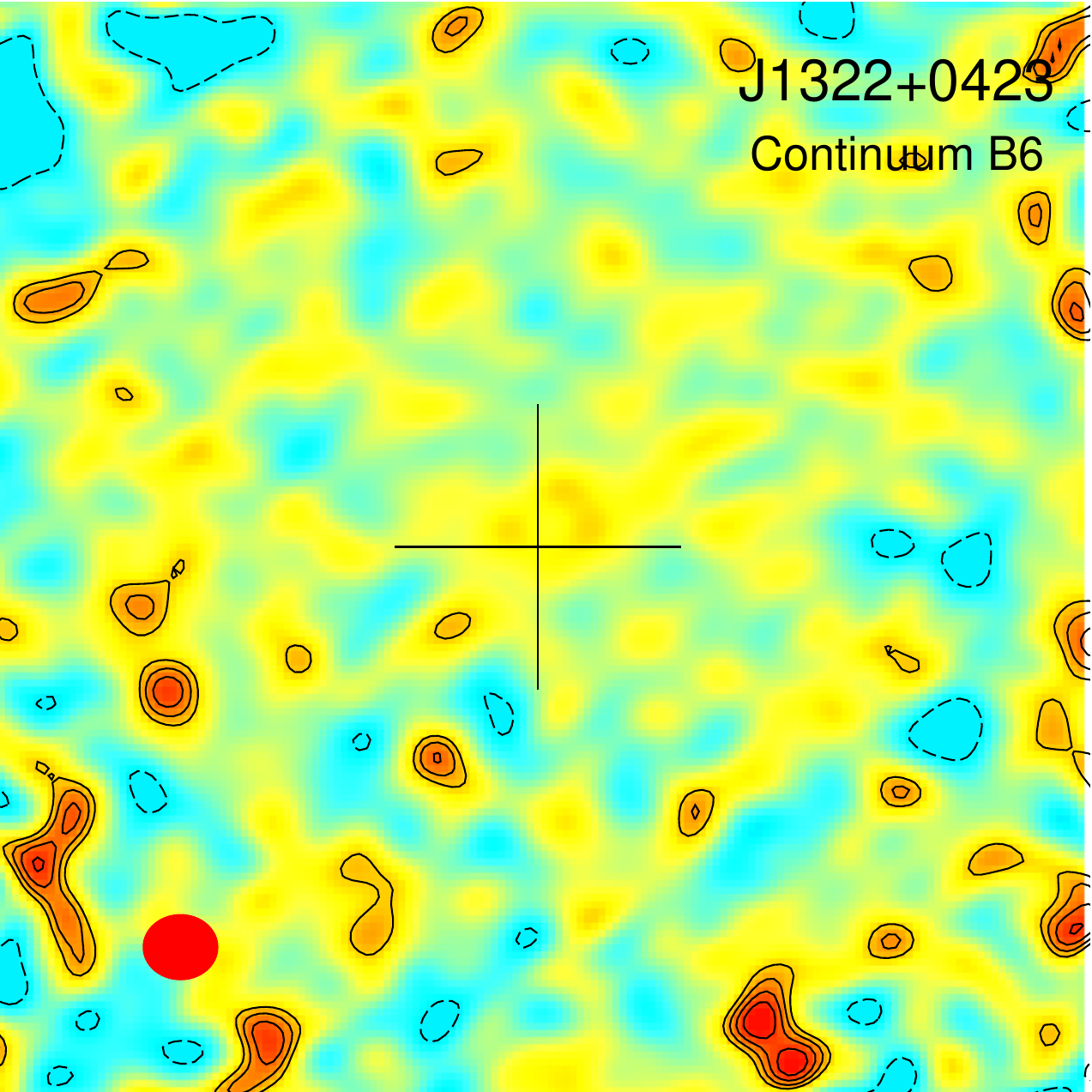}
\includegraphics[width=0.22\textwidth,clip]{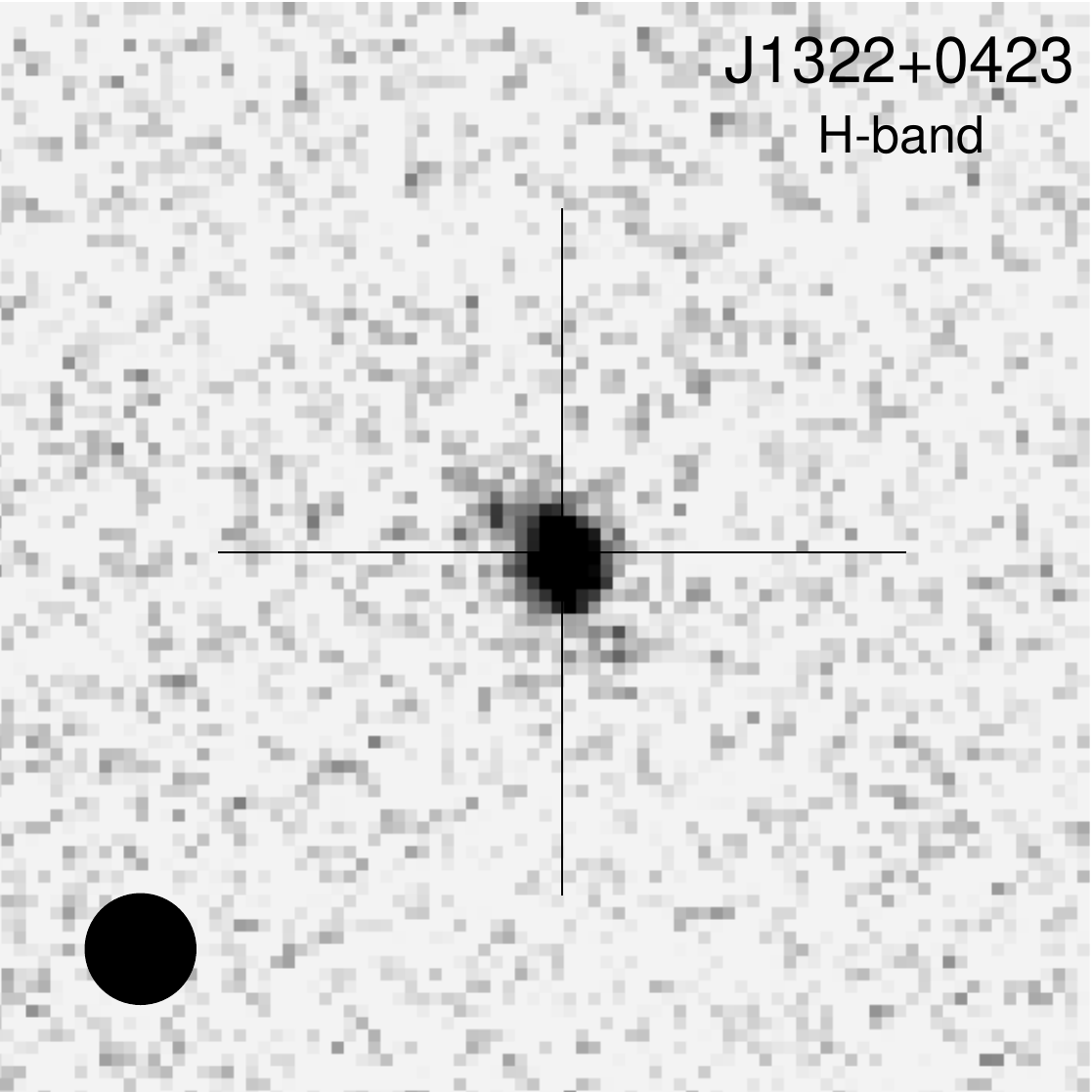}
\includegraphics[width=0.22\textwidth,clip]{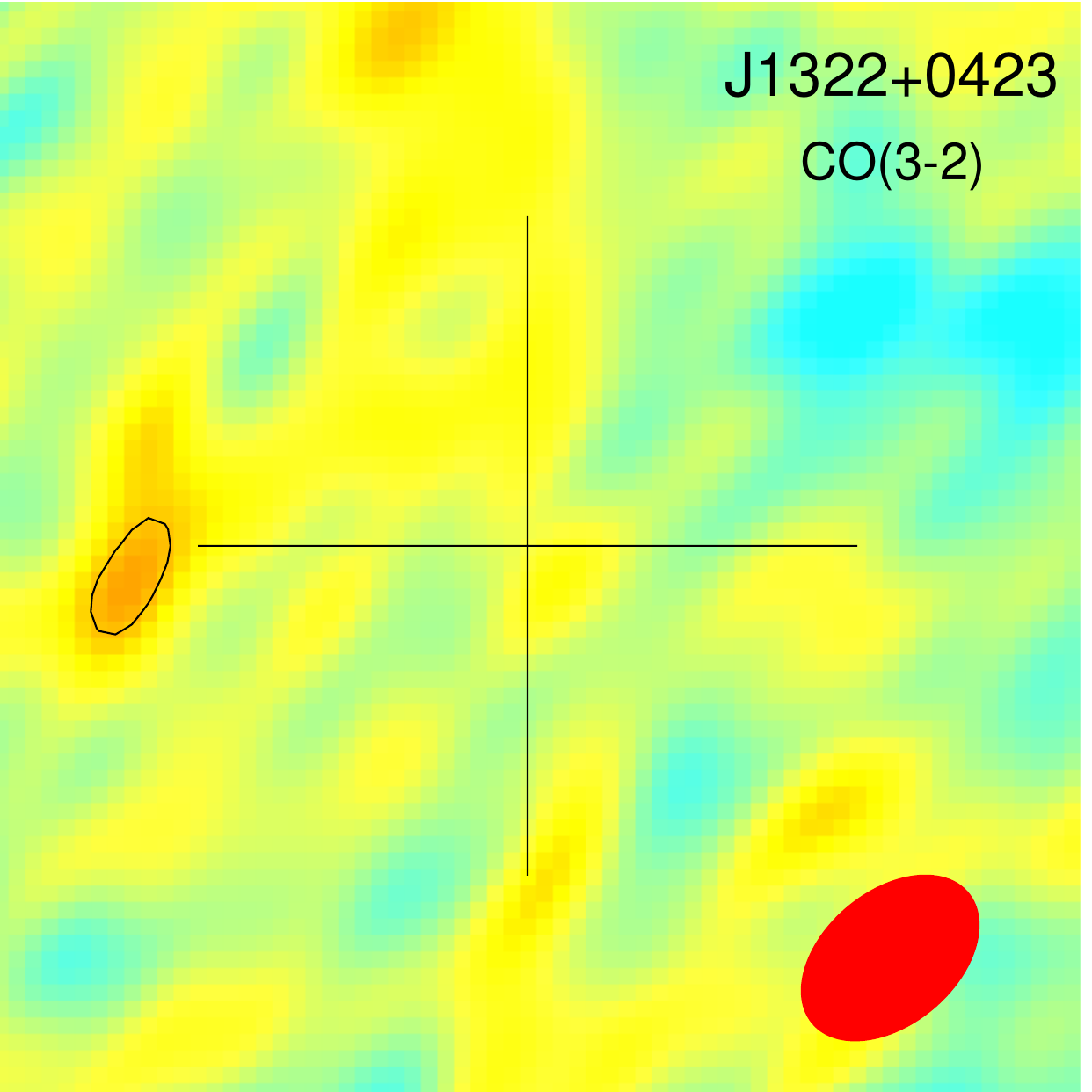}

\hspace{0.4cm}
\includegraphics[width=0.22\textwidth,clip]{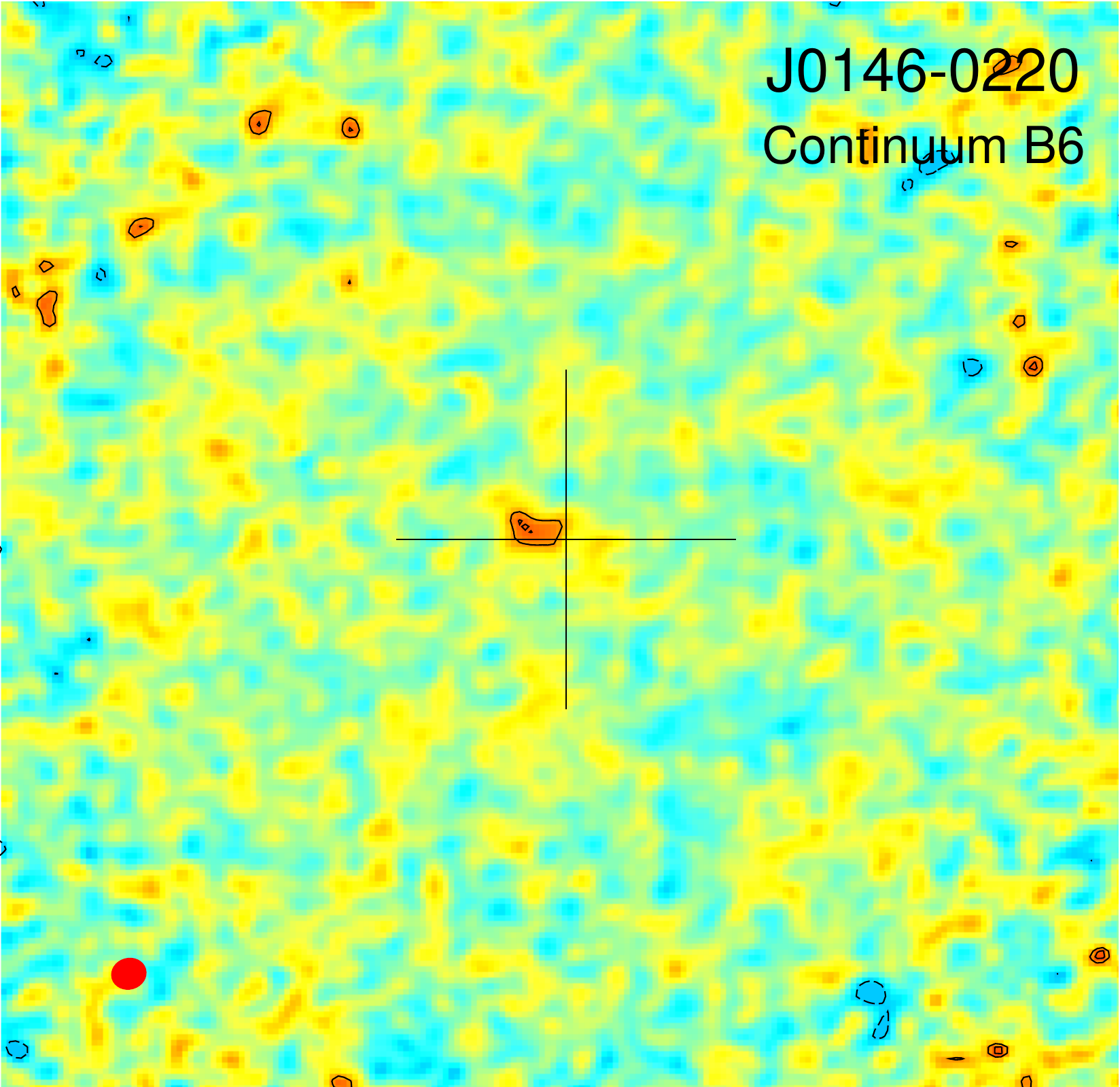}
\includegraphics[width=0.22\textwidth,clip]{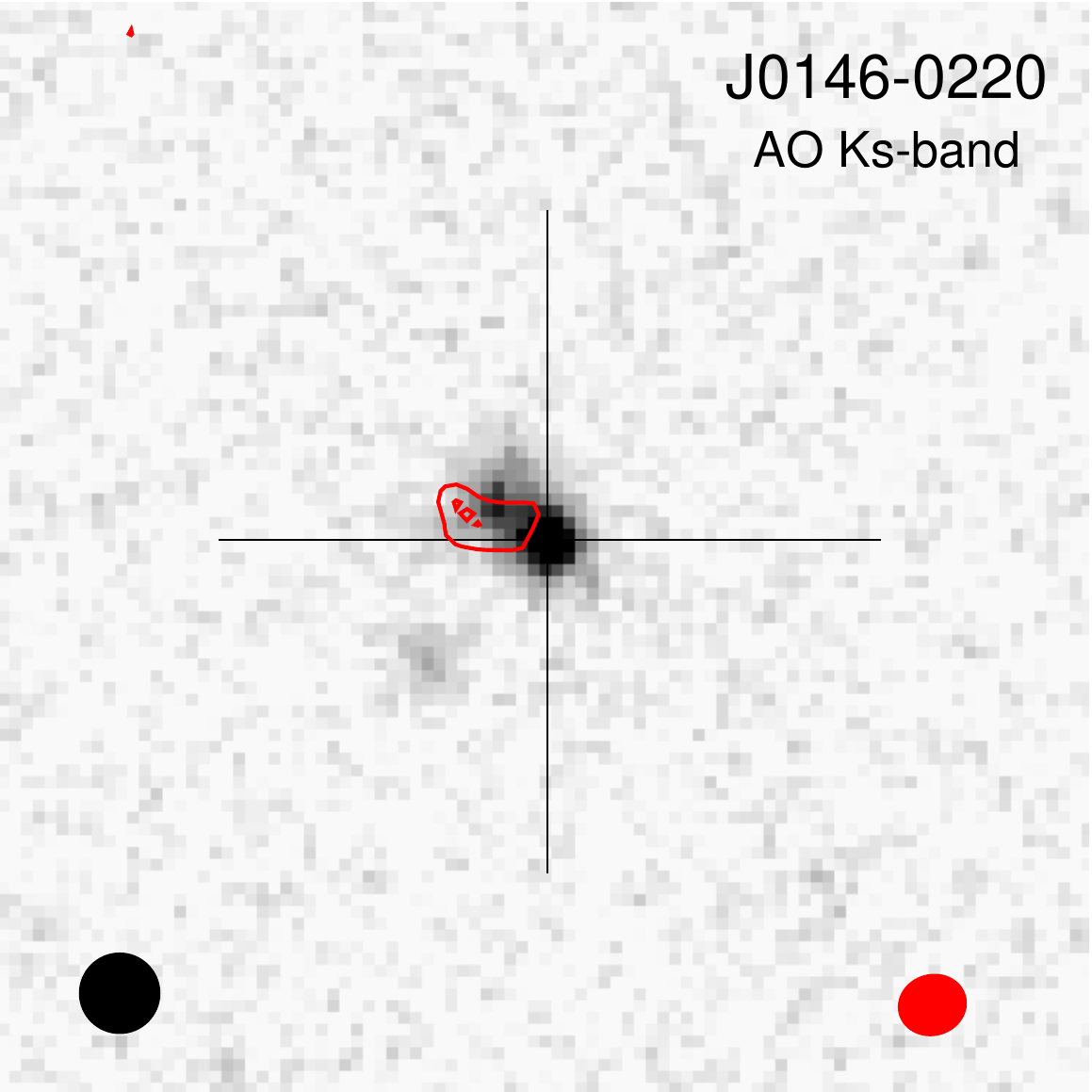}
\includegraphics[width=0.22\textwidth,clip]{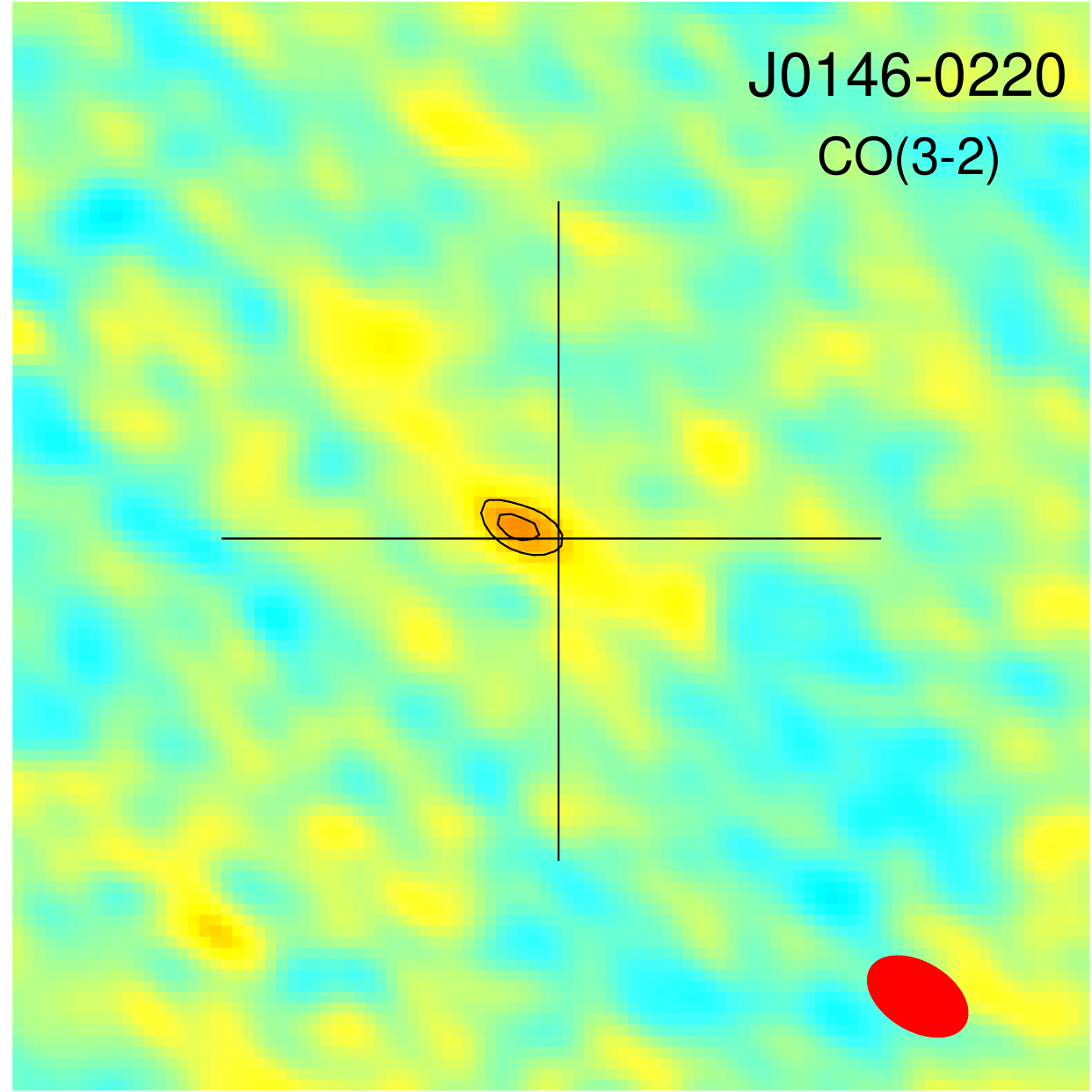}
\includegraphics[width=0.28\textwidth,clip]{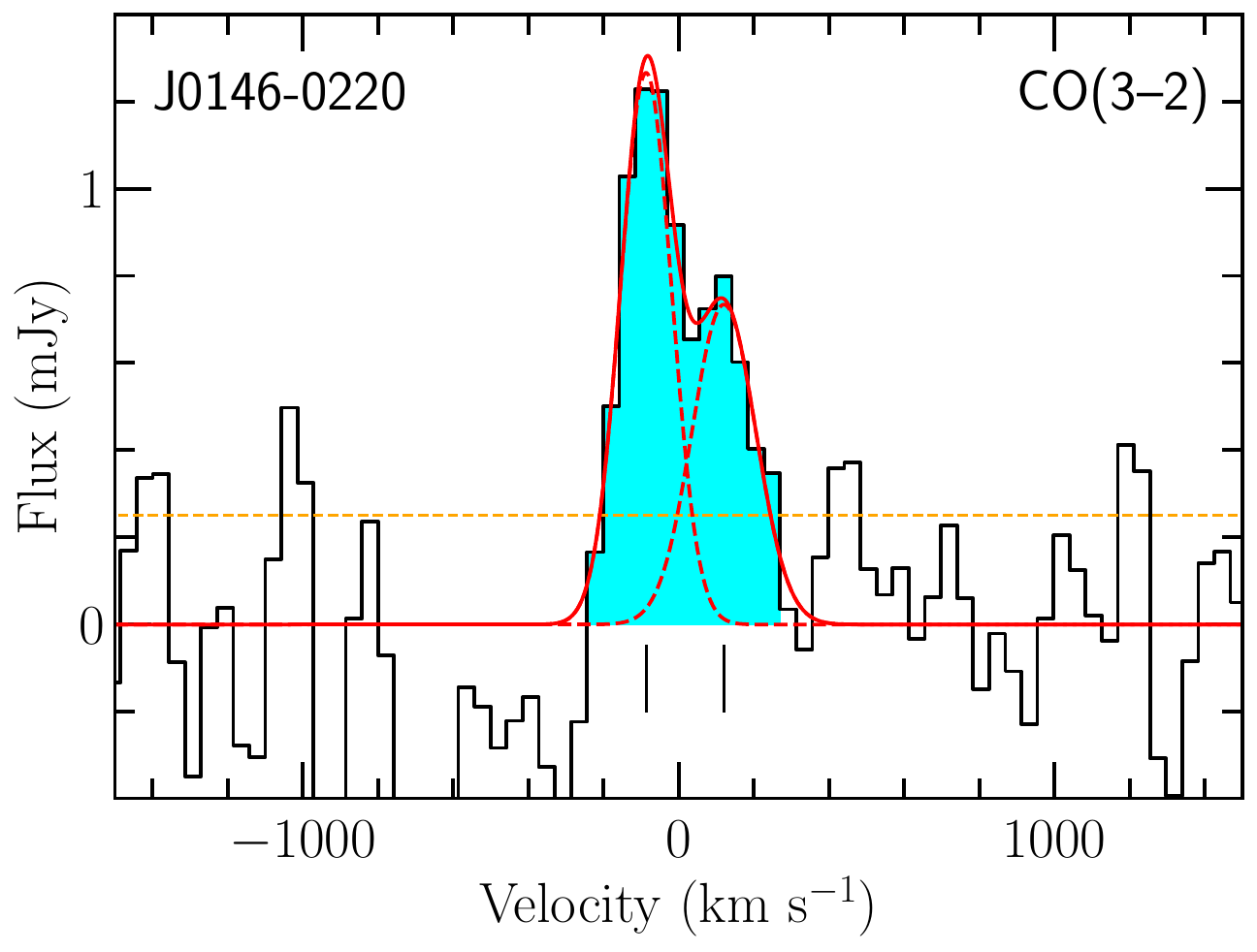 }

\hspace{0.4cm}
\includegraphics[width=0.22\textwidth,clip]{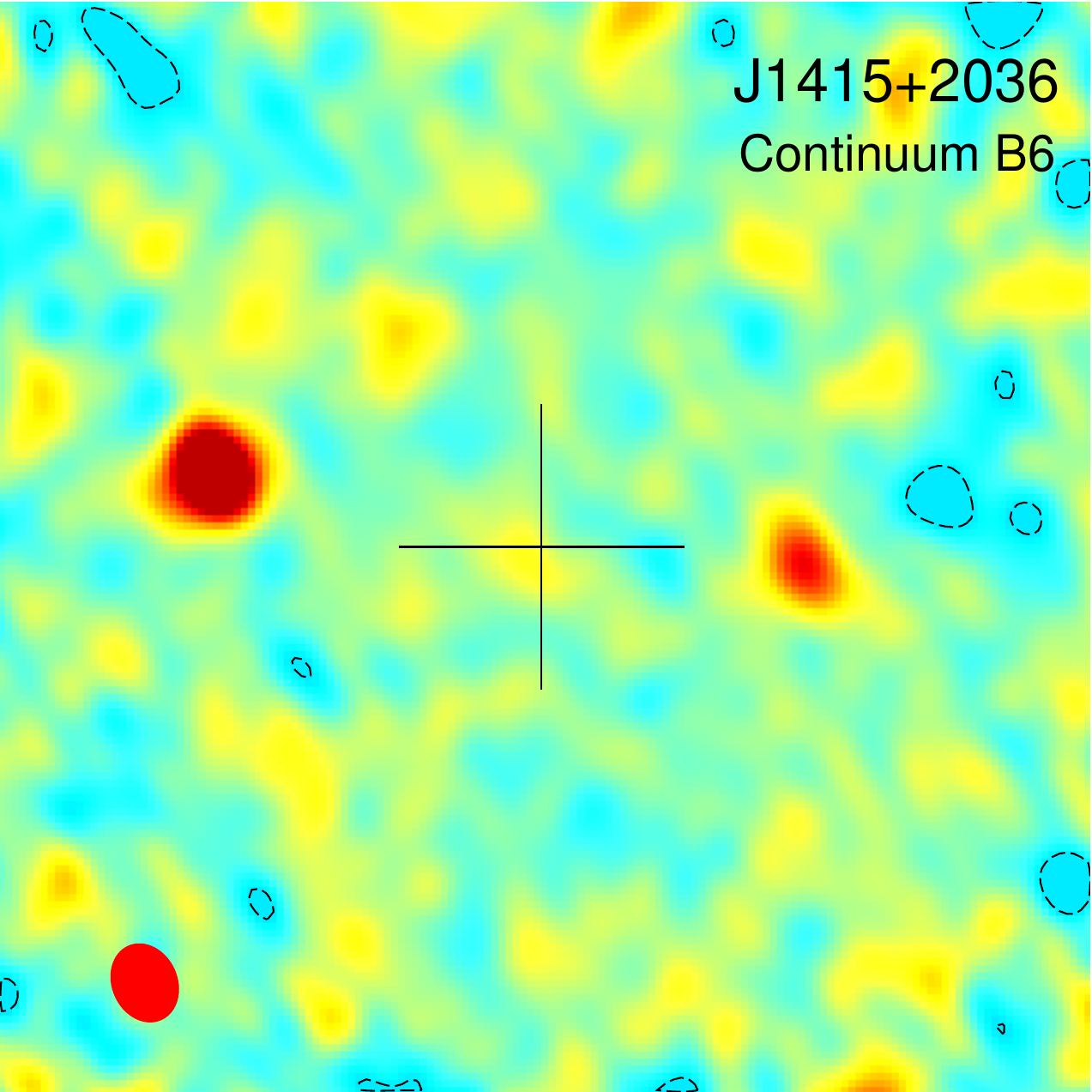}
\includegraphics[width=0.22\textwidth,clip]{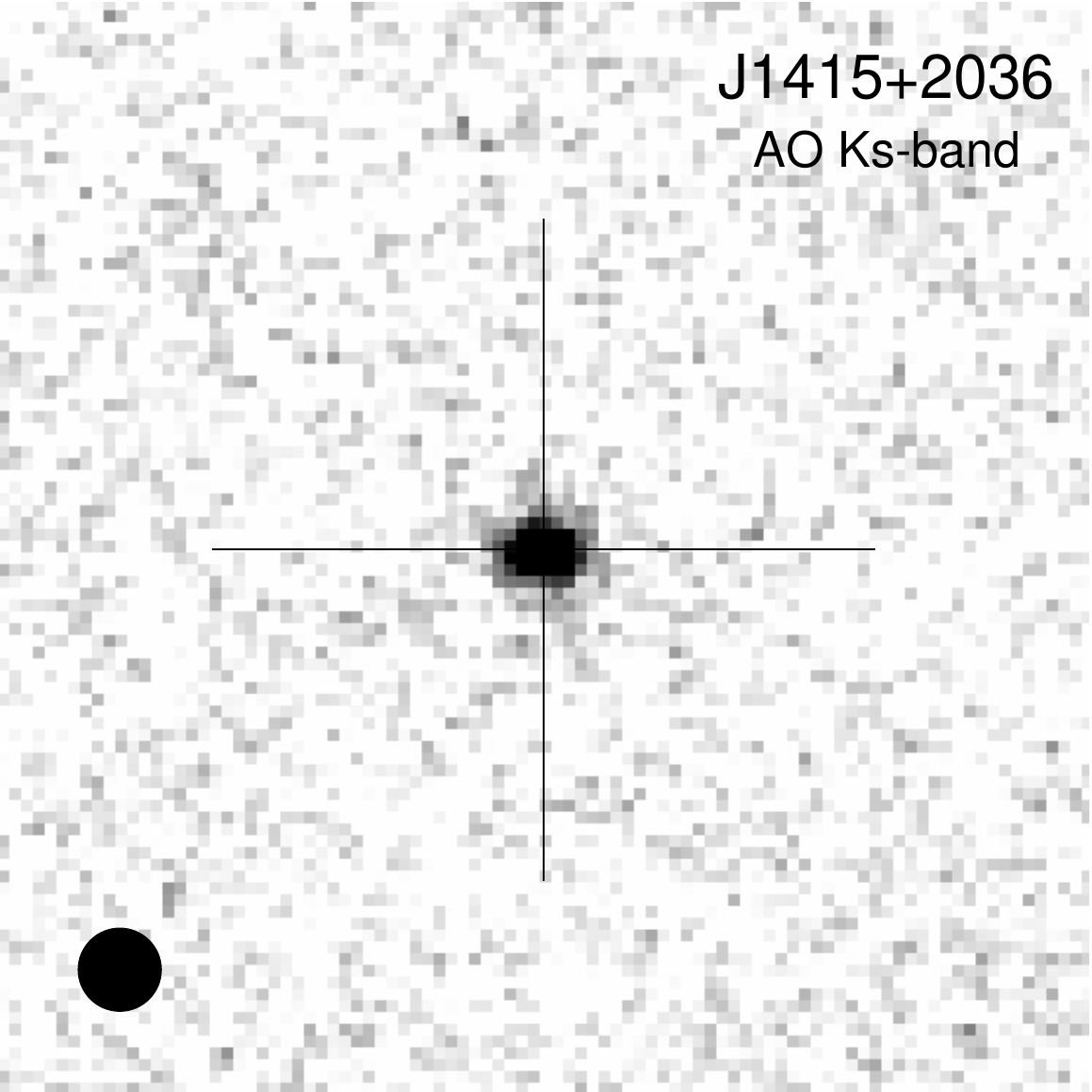}
\includegraphics[width=0.22\textwidth,clip]{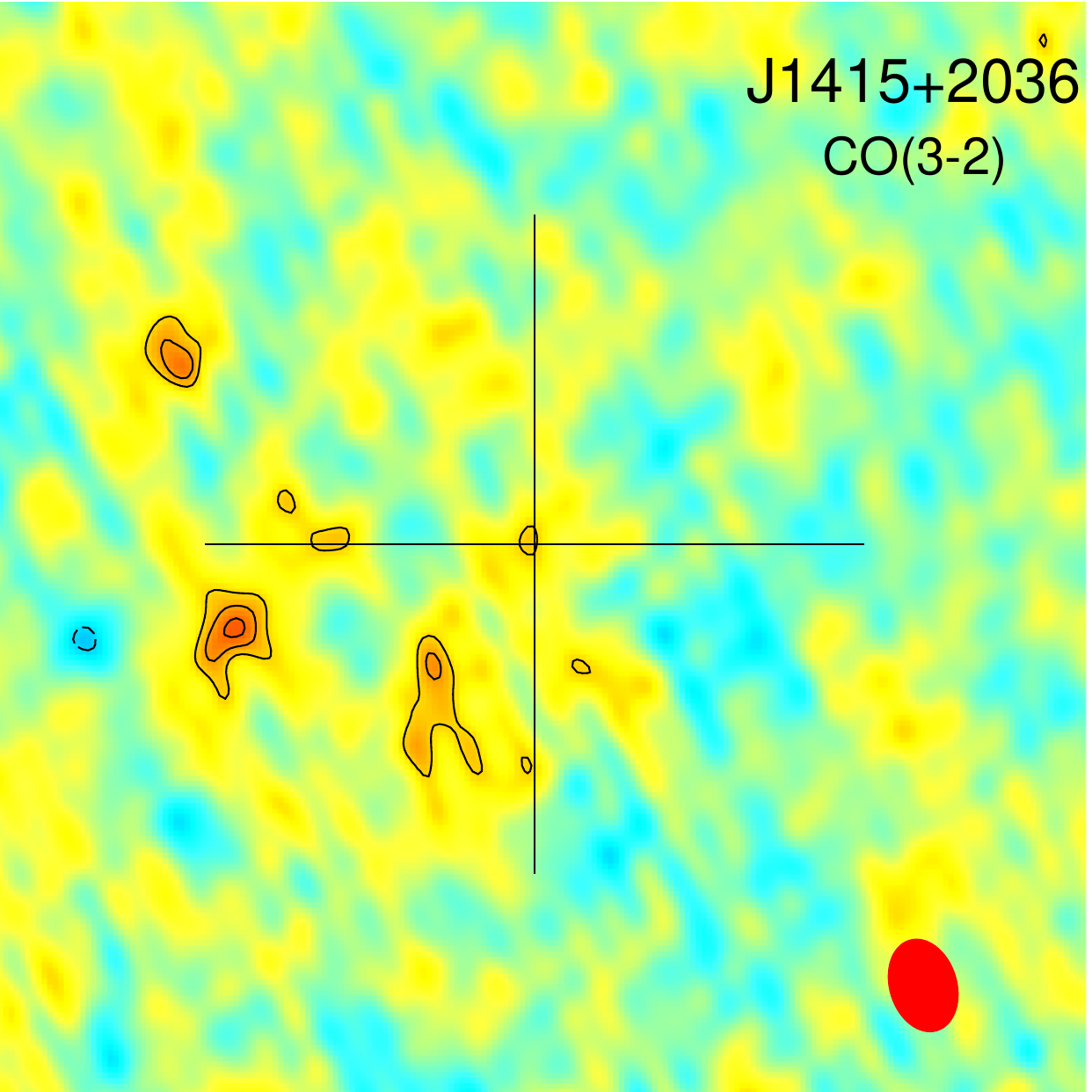}
\includegraphics[width=0.28\textwidth,clip]{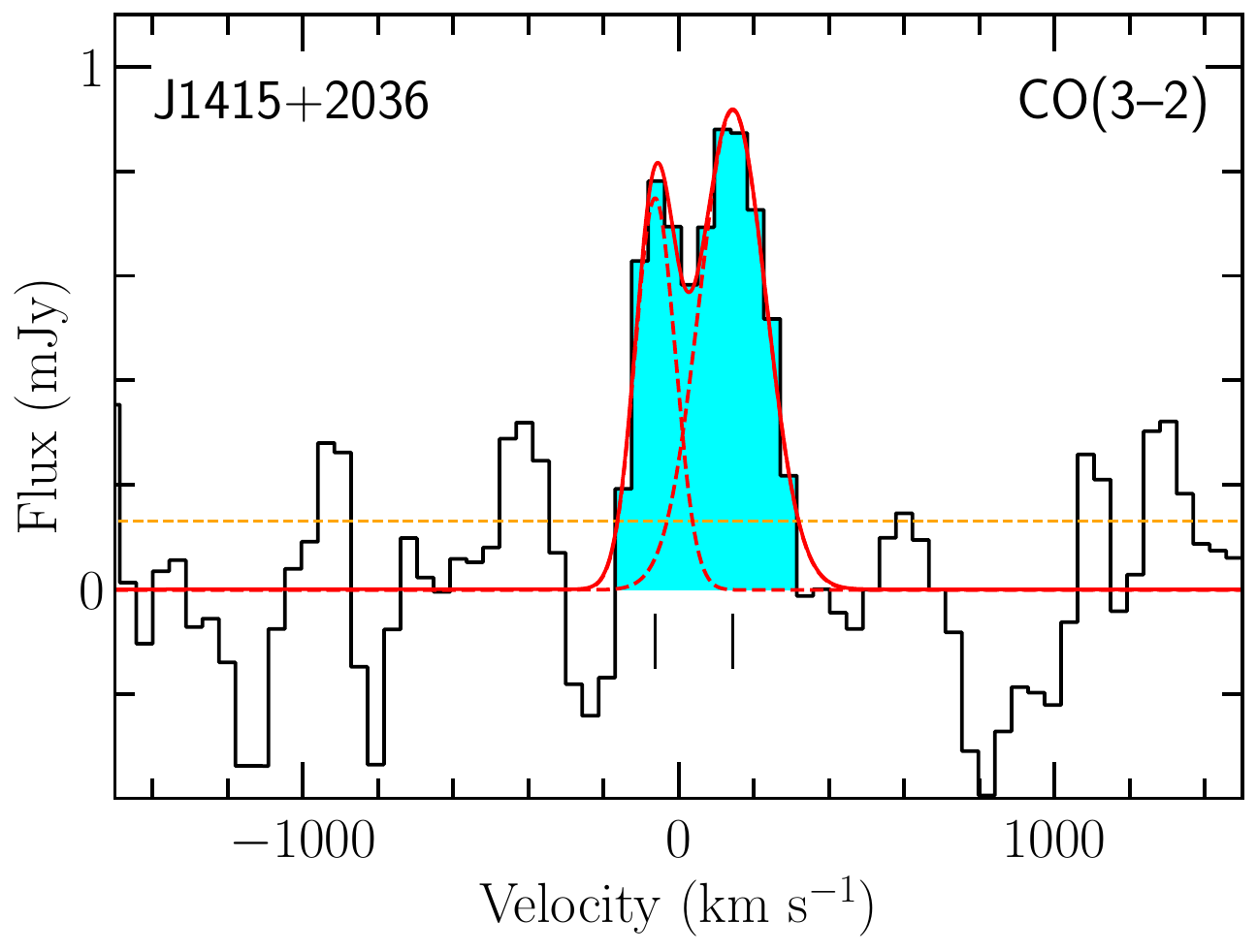}

\hspace{0.4cm}
\includegraphics[width=0.22\textwidth,clip]{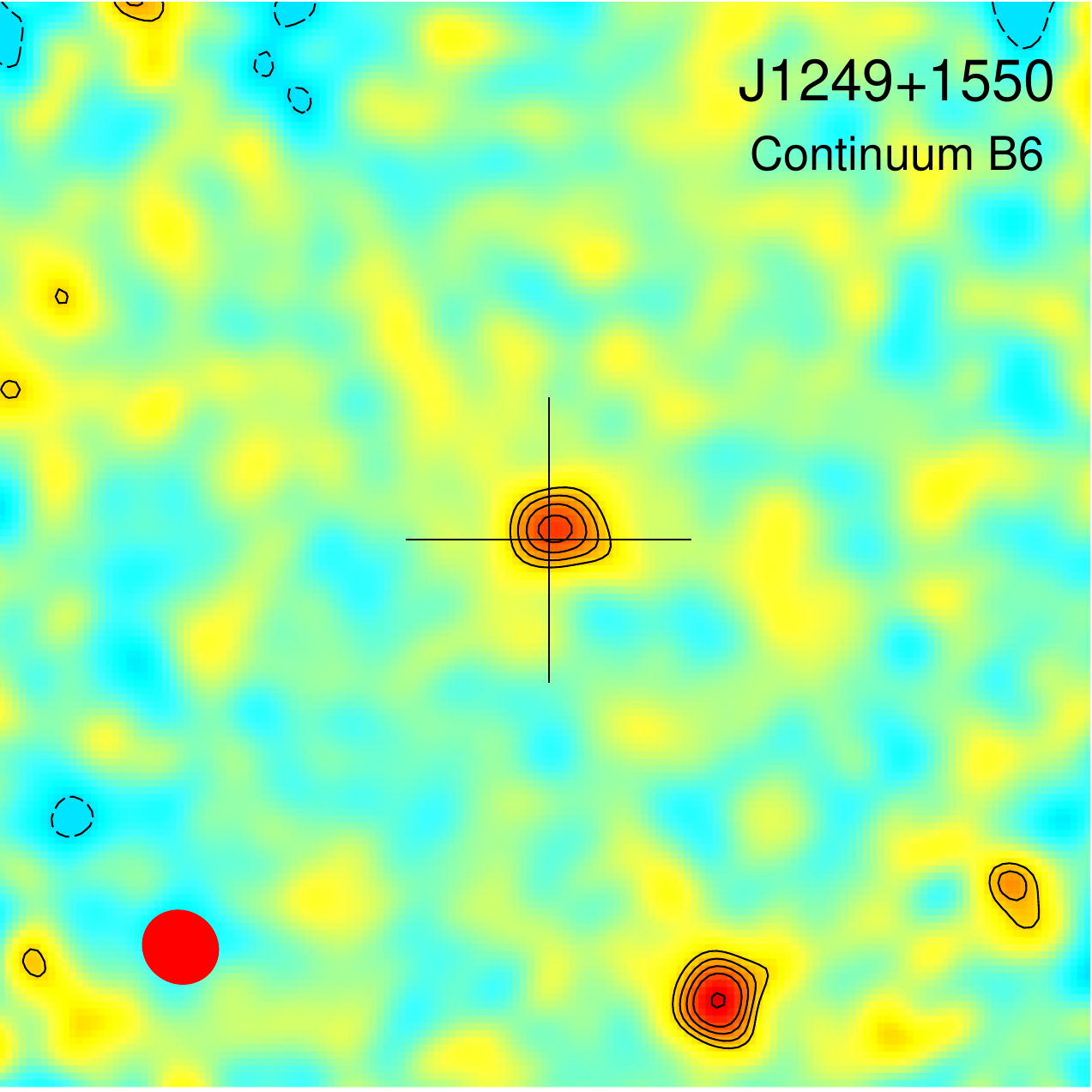}
\includegraphics[width=0.22\textwidth,clip]{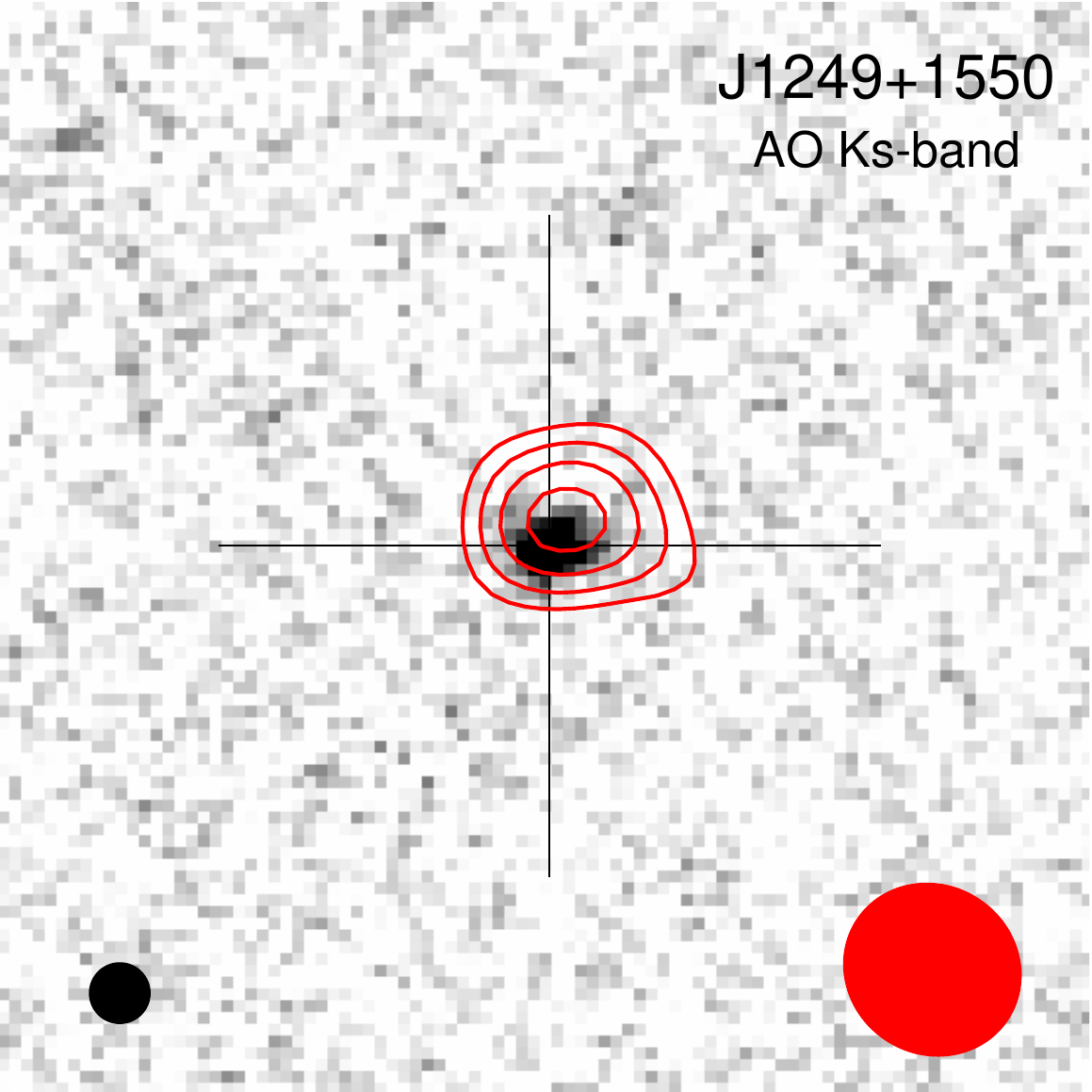}
\includegraphics[width=0.22\textwidth,clip]{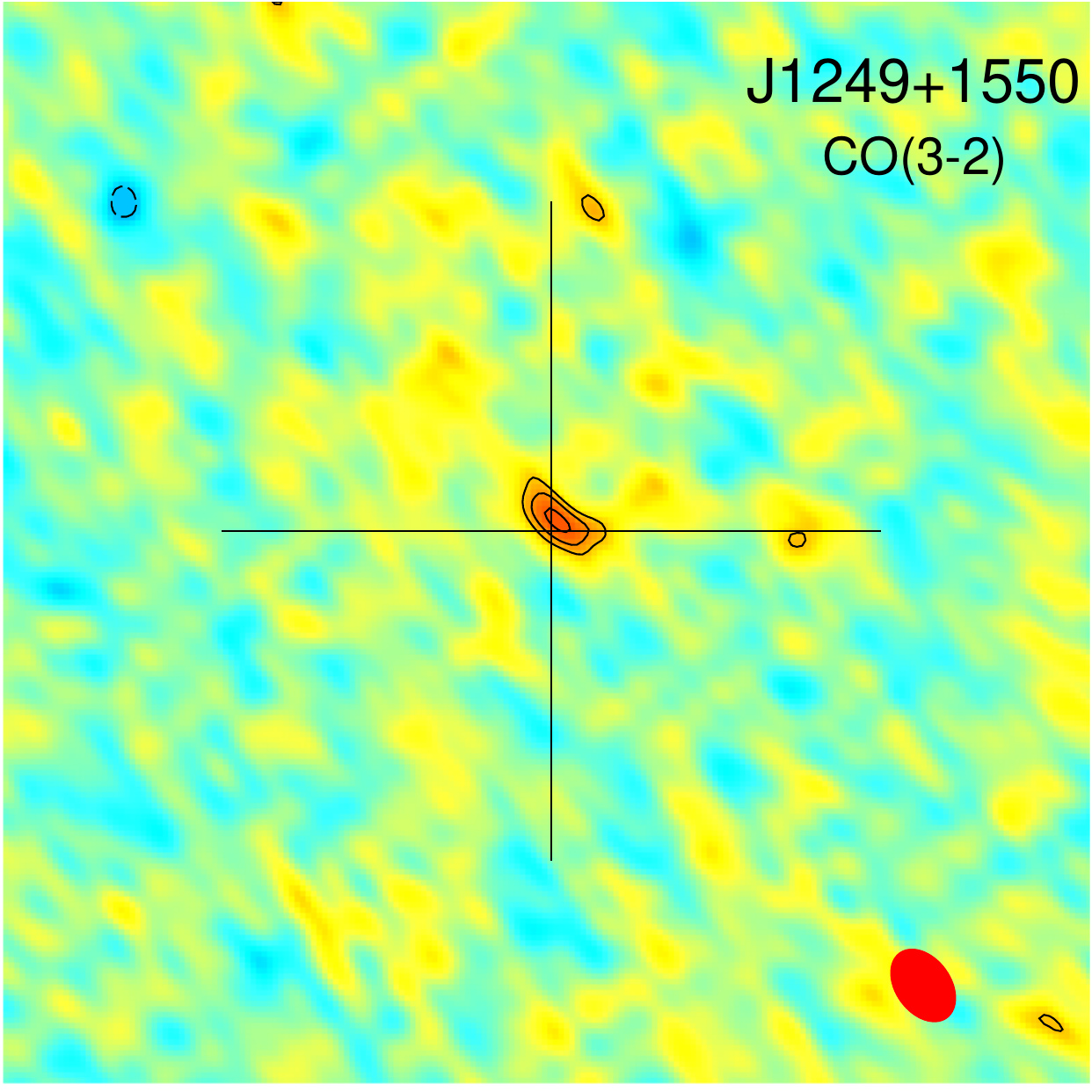}
\includegraphics[width=0.28\textwidth,clip]{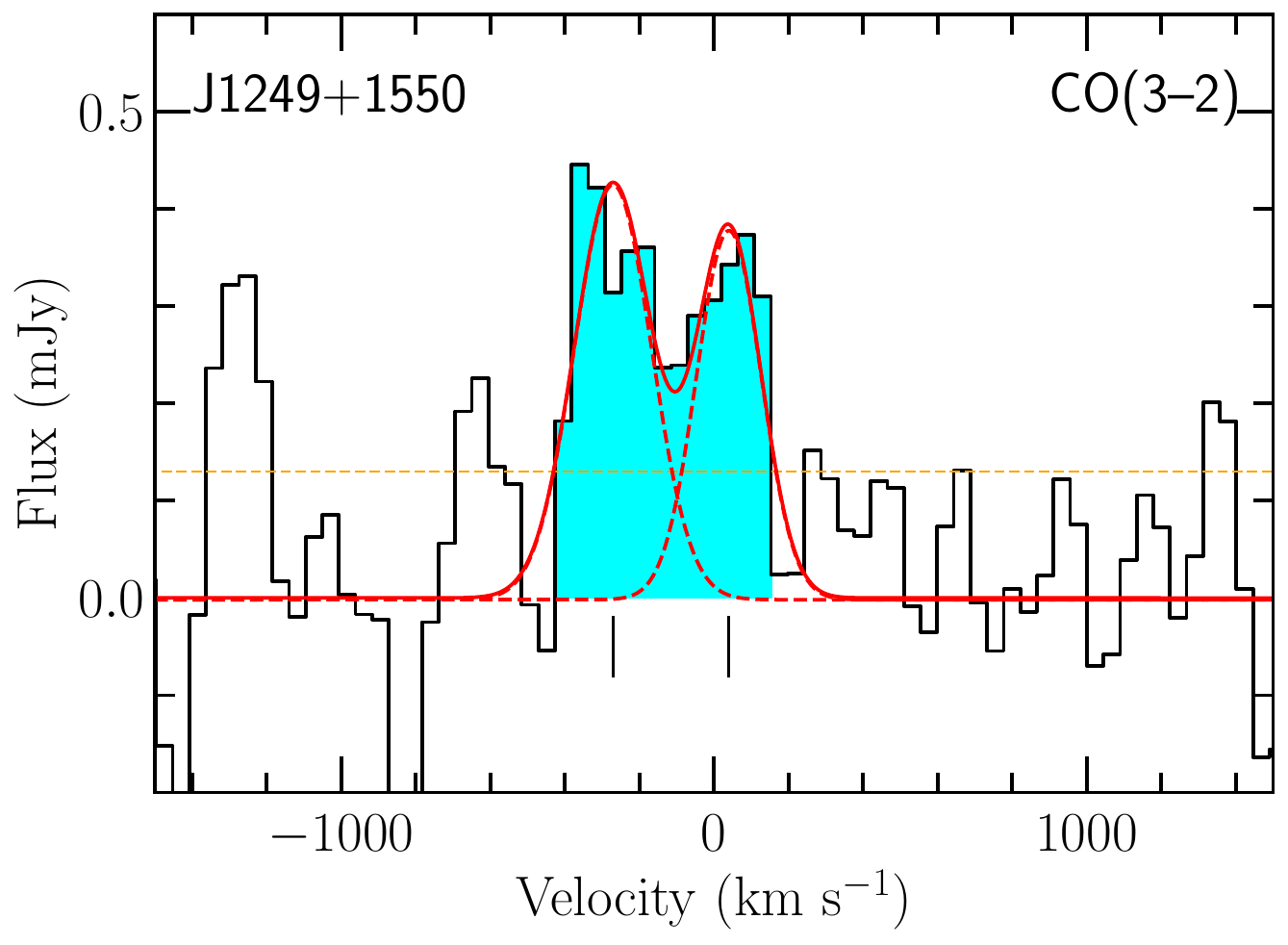}
\caption{
Images showing the FIR dust continuum emission, the $H$- or $K_{\rm s}$-band continuum emission, the CO(3--2) or CO(4--3) emission, and the related spectra (if CO is detected) of our 12 extreme starburst galaxies ordered by increasing redshift from top to bottom.
Left panels: $23\arcsec \times 23\arcsec$ ALMA band 6 1.3~mm (240~GHz) FIR dust continuum images of our targeted galaxies. Contour levels start at $\pm 4\sigma$ and are in steps of $1\sigma$ up to $10\sigma$ and in larger steps above. The respective synthesised beam size and orientation are indicated by the red filled ellipse in the bottom-left corner. The cross in each panel corresponds to the coordinates of the HAWK-I continuum emission peak (except for J0850+1549 and J0121+0025 for which we consider, respectively, the CFHT MegaCam and Subaru continuum emission peak) and is $\pm 3\arcsec$ in size.
Middle-left panels: $10\arcsec \times 10\arcsec$ VLT HAWK-I seeing-limited $H$-band or AO $K_{\rm s}$-band images of our starburst galaxies in greyscale with the ALMA band 6 FIR dust continuum contours overlaid in red. For J0850+1549 and J0121+0025, we show, respectively, the CFHT MegaCam and Subaru seeing-limited $R$-band images. Contour levels start at $\pm 4\sigma$ and are in steps of $2\sigma$, except for J0146--0220, J1249+1550, and J0850+1549 where contour levels are the same as in the left panels. The PSF is shown by the black filled circle in bottom-left corner, and the ALMA synthesised beam by the red filled ellipse in the bottom-right corner. The cross is the same as in the left panels.
Middle-right panels:  $10\arcsec \times 10\arcsec$ ALMA CO(3--2) or CO(4--3)  velocity-integrated intensity moment-0 maps of our galaxies. The maps were integrated over the cyan-shaded spectral channels shown in the right panels. Contour levels start at $\pm 4\sigma$ and are in steps of $1\sigma$. The respective synthesised beam size and orientation are indicated by the red filled ellipse in the bottom-right corner. The cross is the same as in the other panels.
Right panels: ALMA CO(3--2) or CO(4--3) emission line spectra of our galaxies, plotted when detected, in steps of $\sim 45~\rm km~s^{-1}$ and with the zero velocity centred on the redshifts derived from optical nebular emission lines (Table~\ref{tab:restUV-properties}). The cyan-shaded regions correspond to the velocity channels optimising the CO detections, as described in Sect.~\ref{sect:imaging}. The dashed orange lines correspond to the RMS noise level of spectra. The solid red lines are the multi-component Gaussian best-fits to the observed CO line profiles. The vertical bars mark the positions of the fitted Gaussian components.
The rest of the figure is available in Appendix~A.
}
\label{fig:contB6-CO_1}
\end{figure*}
%-----------------------------------------------------------------------

\section{Observations and data analysis}
\label{sect:observations}

\subsection{ALMA data reduction and imaging}
\label{sect:imaging}

The 12 targets are part of two ALMA observing programmes. The first programme 2018.1.00932.S (PI: R.~Marques-Chaves) includes J0146--0220, J0850+1549, J1220+0842, and J1157+0113, and was observed in Cycle 6 in bands 3 and 6 in the configurations C43-4 and C43-3, respectively, with short on-source integration times per target between 5 and 35 minutes. The second programme 2021.1.01438.S (PI: R.~Marques-Chaves) includes J1322+0423, J1415+2036, J1249+1550, J0006+2452, J1220--0051, J0950+0523, J0121+0025, and J1316+2614, and was observed in Cylcle 8 also in bands 3 and 6 in the configurations C43-5 and  C43-1 (C43-2 for J0950+0523), respectively, with on-source integration times of about 40 minutes per target. The band 3 observations were tuned to the CO(3--2) emission line, except for the 2 highest redshift targets at $z>3$, J1316+2614 and J0121+0025, that were tuned to the CO(4--3) emission line. The band 6 observations were aimed for the 240~GHz (1.3~mm) FIR dust continuum solely and therefore were not tuned to a specific frequency. The spectral resolution was set to $31.25~\rm MHz$ (i.e. $\sim 45~\rm km~s^{-1}$) for both band 3 and 6 observations.

The ALMA data were calibrated with the standard observatory pipeline, and imaged using the Common Astronomy Software Application \citep[CASA, version 6.2.1.7;][]{McMullin+07}. We imaged the band 6 calibrated visibilities of continuum over the four spectral windows with the multi-frequency synthesis, applying a pixel size of $0.1\arcsec$ (for the 2018.1.00932.S targets) or $0.15\arcsec$ (for the 2021.1.01438.S targets), and the natural weighting. The clean was repeated down to the threshold of $4\times$ the RMS noise level of the dirty images using the \texttt{tclean} routine in CASA. We then applied the primary beam correction on the cleaned continuum emission maps. The corresponding FIR continuum maps are shown in Figs.~\ref{fig:contB6-CO_1}, \ref{fig:contB6-CO_2}, and \ref{fig:contB6-CO_3} (left panels). 

%-----------------------------------------------------------------------
\begin{table*}
\caption{CO emission line observations (ALMA band 3).}             
\label{tab:COdata}      
\centering          
\begin{tabular}{cccccccc} 
\hline\hline      
Target & $z_{\rm CO}$\tablefootmark{a} & CO & Synthesised beam & RMS & $I_{\rm CO}$\tablefootmark{c} & FWHM\tablefootmark{d} & $L^{\prime}_{\rm CO}$ \\
 &  & line & Size $(\arcsec)$ / PA $(^\circ)$ & (\, \tablefootmark{b}) & $(\rm mJy~km~s^{-1})$ & $(\rm km~s^{-1})$ & $(10^9~\rm K~km~s^{-1}~pc^2)$ \\
\hline 
J1322+0423 & 2.0800\tablefootmark{e} & 3--2 & $1.86\times 1.22$ / $-50$ & 59 & $<236$ & -- & $< 5.82$ \\
J0146--0220 & $2.1596\pm 0.0002$    & 3--2 & $1.01\times 0.63$ / $+59$ & 71 & $373\pm 25$ & $358\pm 61$ & $9.82\pm 0.66$ \\
J1415+2036 & $2.2448\pm 0.0004$    & 3--2 & $0.87\times 0.62$ / $+17$ & 25 & $<304\pm 62$\tablefootmark{g} & $332\pm 47$ & $<8.57\pm 1.75$ \\
J1249+1550 & $2.2926\pm 0.0009$    & 3--2 & $0.74\times 0.51$ / $+36$\tablefootmark{f} & 28 & $193\pm 12$ & $446\pm 115$ & $5.64\pm 0.35$ \\
J0006+2452 & $2.3800\pm 0.0003$    & 3--2 & $0.81\times 0.52$ / $+30$\tablefootmark{f} & 45 & $906\pm 51$ & $546\pm 107$ & $28.3\pm 1.6$ \\
J0850+1549 & 2.4235\tablefootmark{e} & 3--2 & $0.98\times 0.76$ / $+30$ & 45 & $<180$ & -- & $< 5.78$ \\
J1220--0051 & $2.4271\pm 0.0001$    & 3--2 & $0.77\times 0.61$ / $+34$ & 28 & $240\pm 25$ & $295\pm 46$ & $7.74\pm 0.81$ \\
J0950+0523 & $2.4538\pm 0.0006$    & 3--2 & $0.50\times 0.36$ / $+34$\tablefootmark{f} & 36 & $470\pm 29$ & $524\pm 138$ & $15.4\pm 1.0$ \\
J1220+0842 & 2.4698\tablefootmark{e} & 3--2 & $0.95\times 0.77$ / $-35$ & 65 & $<260$ & -- & $< 8.63$ \\
J1157+0113 & $2.5459\pm 0.0003$    & 3--2 & $0.83\times 0.80$ / $-71$ & 25 & $237\pm 21$ & $394\pm 95$ & $8.29\pm 0.73$ \\
J0121+0025 & 3.2445\tablefootmark{e} & 4--3 & $0.60\times 0.57$ / $-45$ & 30 & $<225\pm 65$\tablefootmark{g} & -- & $<6.63\pm 1.92$ \\
J1316+2614 & 3.6122\tablefootmark{e} & 4--3 & $0.94\times 0.61$ / $+10$ & 27 & $<108$ & -- & $< 3.79$ \\
\hline                  
\end{tabular}
\tablefoot{
\tablefoottext{a}{Redshifts derived from Gaussian fits of the detected CO emission line profiles (Sect.~\ref{sect:CO}). For multiple components identified in the CO line profiles, we considered the component with the highest intensity to compute $z_{\rm CO}$, unless the multiple components have comparable intensities, then we considered the mean $z_{\rm CO}$ of the components.}
\tablefoottext{b}{RMS of the CO moment-0 maps in units of $\rm mJy~beam^{-1}~km~s^{-1}$.}
\tablefoottext{c}{CO velocity-integrated intensities derived from CO moment-0 maps. For the CO non-detections, we considered $4\sigma$ upper limits based on the RMS noise level of moment-0 maps integrated over $\sim 350~\rm km~s^{-1}$ (Sect.~\ref{sect:CO}).}
\tablefoottext{d}{Full-width half maximum derived from Gaussian fits of the detected CO emission line profiles (Sect.~\ref{sect:CO}), and corrected for the final channel spacing of $\sim 45~\rm km~s^{-1}$ (see the ALMA Technical Handbook). For multiple blended components identified in the CO line profiles, we give the sum of the FWHM of all components.}
\tablefoottext{e}{Redshifts derived from optical nebular emission lines instead of a CO line that is either undetected or has a low S/N.}
\tablefoottext{f}{For these high S/N CO detections, we applied the Briggs weighting with the robust factor of 1.0 instead of the natural weighting.}
\tablefoottext{g}{J1415+2036 and J0121+0025 show patchy CO emission peaks at $4-5\sigma$ significance level spread within $< 3\arcsec$ (i.e. $< 25$~kpc) around the phase centre in the CO moment-0 maps, but their respective integrated CO line detections seem to be relatively robust (Figs.~\ref{fig:contB6-CO_1}, \ref{fig:contB6-CO_2}, and \ref{fig:contB6-CO_3}, middle-right and right panels).}
}
\end{table*}
%-----------------------------------------------------------------------
\begin{table*}
\caption{Far-infrared dust continuum observations (ALMA band 6) and rest-frame UV/optical ground-based observations.}             
\label{tab:FIRdata}      
\centering          
\begin{tabular}{ccccccccc} 
\hline\hline      
Target & Synthesised beam & RMS & $S_{\rm 1.3mm}$\tablefootmark{b} & $L_{\rm IR}$\tablefootmark{c} & $R_{\rm eff,FIR}$\tablefootmark{d} & HAWK-I & PSF & $R_{\rm eff,UV/opt}$\tablefootmark{e} \\                   
 & Size $(\arcsec)$ / PA $(^\circ)$ & (\, \tablefootmark{a}) & $(\mu \rm Jy)$ & $(10^{11}~L_{\odot})$ & (kpc) & band & ($\arcsec$) & (kpc) \\
\hline 
J1322+0423 & $1.59\times 1.39$ / $-89$ & 12 & $<48$ &$< 2.59$ &  -- & $H$ & 0.49 & $2.01\pm 0.10$ \\
J0146--0220 & $0.63\times 0.56$ / $-74$ & 15 & $148\pm 10$ & $8.06\pm 2.90$ &  -- & $K_s$ & 0.37 & $2.59\pm 0.10^{\dagger}$ \\
J1415+2036 & $1.71\times 1.38$ / $+24$ & 15 & $<64$ & $<3.40$ & -- & $K_s$ & 0.38 & $1.24\pm 0.68$ \\
J1249+1550 & $1.65\times 1.54$ / $+56$ & 17 & $155\pm 10$ & $8.54\pm 3.01$ & $5.0\pm 2.0$ & $K_s$ & 0.28 & $1.04\pm 0.05$ \\
J0006+2452 & $1.70\times 1.30$ / $+9$ & 18 & $510\pm 30$ & $28.3\pm 9.8$ & $3.5\pm 0.9$ & $K_s$ & 0.39 & $2.23\pm 0.35^{\dagger}$ \\
J0850+1549 & $0.72\times 0.59$ / $-3$ &         13 & $106\pm 20$ & $5.90\pm 2.04$ & $1.8\pm 0.5$ & $R^f$ & 0.71 & $1.46\pm 0.18$ \\
J1220--0051 & $1.60\times 1.33$ / $-80$ & 16 & $172\pm 12$ & $9.57\pm 3.30$ & $2.4\pm 1.0$ & $H$ & 0.61 & $1.51\pm 0.22$ \\
J0950+0523 & $1.07\times 0.76$ / $+57$ & 14 & $331\pm 20$ & $18.5\pm 6.3$ & $2.0\pm 0.7$ & $K_s$ & 0.36 & $<0.74$ \\
J1220+0842 & $0.62\times 0.61$ / $-27$ & 25 & $<102$ & $<5.69$ & -- & $H$ & 0.44 & $1.64\pm 0.24$ \\
J1157+0113 & $0.63\times 0.58$ / $-89$ & 11 & $251\pm 40$ & $14.1\pm 4.8$ & -- & $H$ &  0.38 & $1.07\pm 0.14$ \\
J0121+0025 & $1.57\times 1.14$ / $+63$ & 10 & $122\pm 30$ & $7.12\pm 2.15$ & $3.6\pm 0.8$ & $R^g$ & 0.55 & $0.79\pm 0.15$ \\
J1316+2614 & $1.70\times 1.33$ / $+8$ & 11 & $124\pm 20$ & $7.37\pm 2.09$ & $1.7\pm 0.8$ & $K_s$ & 0.30 & $<0.55$ \\
\hline                  
\end{tabular}
\tablefoot{
\tablefoottext{a}{RMS of the 1.3~mm (240~GHz) band~6 continuum emission maps in units of $\mu \rm Jy~beam^{-1}$.}
\tablefoottext{b}{Far-infrared fluxes; for the non-detections we considered $4\sigma$ upper limits (Sect.~\ref{sect:continuum}).}
\tablefoottext{c}{Infrared luminosities derived from $S_{\rm 1.3mm}$ assuming the MBB function (Sect.~\ref{sect:continuum}).}
\tablefoottext{d}{Far-infrared dust continuum effective radii measured in the \textit{uv} plane only for the resolved galaxies using the 1.3~mm band 6 visibilities (Sect.~\ref{sect:continuum}).}
\tablefoottext{e}{Rest-frame UV or optical effective radii measured, respectively, from HAWK-I seeing-limited $H$-band or AO $K_{\rm s}$-band observations (Sect.~\ref{sect:HAWKI}).}
\tablefoottext{$\dagger$}{For J0146--0220 and J0006+2452, the HAWK-I AO $K_{\rm s}$-band observations reveal a main bright and compact (unresolved) component and a more extended diffuse component. The rest-frame optical effective radii of the bright compact component, respectively, are $< 0.78$~kpc and $<0.80$~kpc.}
\tablefoottext{f}{CFHT MegaCam seeing-limited $R$-band observations were used to derive the rest-frame UV effective radius of J0850+1549 (Sect.~\ref{sect:HAWKI}).}
\tablefoottext{g}{Subaru seeing-limited $R$-band observations were used to derive the rest-frame UV effective radius of J0121+0025 as described in \citet{Marques+21}.}
}
\end{table*}
%-----------------------------------------------------------------------

For the CO(3--2) and CO(4--3) emission imaging we used the pixel size of $0.1\arcsec$ (for the 2018.1.00932.S targets) or $0.05\arcsec$ (for the 2021.1.01438.S targets). The Briggs weighting with the robust factor of 1.0 was applied for the few galaxies with a high S/N CO detection, otherwise we applied the natural weighting. Similarly to the continuum maps, we cleaned all channels with the \texttt{tclean} routine down to the threshold of $4\times$ the RMS noise level of the dirty cubes, and a primary beam correction was applied. Finally, we also imaged the band 3 calibrated visibilities of continuum over the four spectral windows, excluding channels contaminated by the CO emission. No band 3 (100~GHz) continuum was detected in any target. The resulting band 3 (CO emission) and band 6 (continuum emission) synthesised beam sizes and RMS noise levels for the 12 targets are listed in Tables~\ref{tab:COdata} and \ref{tab:FIRdata}, respectively.

The CO moment-0 maps, that is, the velocity-integrated CO line intensity maps, were obtained by averaging the cleaned cube over the spectral channels where the CO emission is detected, using the \texttt{immoments} routine in CASA. We adopted the optimum channel range as the one that is maximising the S/N of the 1D spectrum of the CO emission line by iteratively testing different channel ranges \citep[following the method of, e.g.][]{Daddi+15,Zanella+18}. For each channel range tested, the CO emission was extracted within a custom (polygonal) aperture including all the flux above the RMS noise level and typically bigger than the synthesised beam size. No band 3 continuum was subtracted beforehand, since undetected. The corresponding CO moment-0 maps and the CO line spectra are shown in Figs.~\ref{fig:contB6-CO_1}, \ref{fig:contB6-CO_2}, and \ref{fig:contB6-CO_3} (middle-right and right panels).

%-----------------------------------------------------------------------
\begin{table*}
\caption{Dust and molecular gas properties of the galaxy sample.}             
\label{tab:COdust-properties}      
\centering          
\begin{tabular}{c c c c c c c c} 
\hline\hline       
Target & $z_{\rm nebular}$\tablefootmark{a} & $\rm SFR_{IR}$\tablefootmark{b} &
$M_{\rm molgas}$\tablefootmark{c} & $M_{\rm dust}$\tablefootmark{d} & $t_{\rm depl}$\tablefootmark{e,f} & $f_{\rm molgas}$\tablefootmark{f} & $\epsilon_{\rm SF}$\tablefootmark{f} \\
 & & $(\rm M_{\odot}~yr^{-1})$ & $(10^{10}~M_{\odot})$ & $(10^7~M_{\odot})$ & (Myr) & & (\%) \\
\hline
J1322+0423  &   2.0800 & $< 26$ & $<0.76$ & $<1.09$ & $<42$ & $<0.69$ & $>31$ \\
J0146--0220  & 2.1595 & $81\pm 29$ & $1.28\pm 0.09$ & $3.32\pm 1.78$ & $61\pm 23$ & $0.79\pm 0.11$ & $21\pm 3$ \\
J1415+2036 & 2.2435 & $<34$ & $<1.11\pm 0.23$ & $<1.38$ & $<50\pm 11$ & $<0.86\pm 0.20$ & $>14\pm 3$ \\
J1249+1550 & 2.2928 & $85\pm 30.$ & $0.73\pm 0.05$ & $3.42\pm 1.79$ & $36\pm 13$ & $0.63\pm 0.09$ & $37\pm 5$ \\
J0006+2452 & 2.3796 & $283\pm 98$ & $3.67\pm 0.21$ & $11.1\pm 5.7$ & $71\pm 25$ & $0.92\pm 0.11$ & $8\pm 0.9$ \\
J0850+1549  & 2.4235 & $59\pm 20$ & $<0.75$ & $2.30\pm 1.18$ & $<30$ & $<0.80$ & $>20$ \\
J1220--0051  & 2.4269 & $96\pm 33$ & $1.01\pm 0.10$ & $3.73\pm 1.90$ & $42\pm 15$ & $0.87\pm 0.13$ & $13\pm 2$ \\
J0950+0523  & 2.4548 & $185\pm 63$ & $2.01\pm 0.12$ & $7.15\pm 3.63$ & $55\pm 19$ & $0.91\pm 0.15$ & $9\pm 1$ \\ 
J1220+0842  & 2.4698 & $<57$ & $<1.12$ & $<2.20$ & $<31$ & $<0.76$ & $>24$ \\
J1157+0113 & 2.5450 & $141\pm 48$ & $1.08\pm 0.10$  & $8.97\pm 2.68$ & $44\pm 16$ & $0.85\pm 0.31$ & $15\pm 6$ \\
J0121+0025      & 3.2445 & $ 71\pm 22$ & $<1.11\pm 0.32$ & $3.75\pm 1.08$ & $<32\pm 14$ & $<0.71\pm 0.25$ & $>29\pm 10$ \\ 
J1316+2614      & 3.6122 & $74\pm 21$ & $<0.63$ & $3.53\pm 1.01$ & $<13$ & $<0.60$ & $>40$ \\
\hline                  
\end{tabular}
\tablefoot{
\tablefoottext{a}{Redshifts derived from optical nebular emission lines.}
\tablefoottext{b}{Obscured SFRs derived from the IR luminosities (Table~\ref{tab:FIRdata}) following the calibration of \citet{Kennicutt98} for the \citet{Chabrier03} IMF, ${\rm SFR_{IR}} = L_{\rm IR}/10^{10}$. For $\rm SFR_{IR}$ non-detections, we considered the $L_{\rm IR}$ $4\sigma$ upper limits.}
\tablefoottext{c}{Molecular gas masses computed from the CO(3--2) or CO(4--3) luminosities (Table~\ref{tab:COdata}) following Eq.~(1) and assuming here the CO-to-H$_2$ conversion factor derived for nearby starbursts, $\alpha_{\rm CO}^{\rm SB} = 1~M_{\odot}~\rm (K~km~s^{-1}~pc^2)^{-1}$ (Sect.~\ref{sect:CO}).}
\tablefoottext{d}{Dust masses derived from the FIR fluxes (Table~\ref{tab:FIRdata}) following Eq.~(2) (Sect.~\ref{sect:continuum}).}
\tablefoottext{e}{Molecular gas depletion timescales derived using the total $\rm SFR_{UV+IR}$.}
\tablefoottext{f}{Molecular gas depletion timescales, molecular gas mass fractions, and star-formation efficiencies defined in Sects.~\ref{sect:Mgas-tdepl} and \ref{sect:SFE}, and all computed here with $\alpha_{\rm CO}^{\rm SB}$.}
}
\end{table*}
%-----------------------------------------------------------------------

\subsection{From the CO emission line to molecular gas masses}
\label{sect:CO}

The CO emission is successfully detected close to the phase centre and at the expected frequency for eight targets. J0146--0220, J1249+1550, J0006+2452, J1220--0051, J0950+0523, and J1157+0113 show robust CO detections above $5\sigma$, while J1415+2036 and J0121+0025 show patchy CO emission peaks at $4-5\sigma$ significance level spread over $< 3\arcsec$ (i.e. $< 25~\rm kpc$)  around the phase centre (Figs.~\ref{fig:contB6-CO_1}, \ref{fig:contB6-CO_2}, and \ref{fig:contB6-CO_3}, middle-right panels). Four of the robust CO detections in J0006+2452, J1220--0051, J0950+0523, and J1157+0113 are spatially resolved, while J0146--0220 and J1249+1550 are unresolved/marginally resolved with their CO emission being comparable to the ALMA synthesised beam size. All the CO line profiles are complex, most are characterised by double-peaks and J1157+0113 even shows a triple-peak (Figs.~\ref{fig:contB6-CO_1}, \ref{fig:contB6-CO_2}, and \ref{fig:contB6-CO_3}, right panels). They are difficult to interpret (between rotation/mergers) with the current data which do not allow us to perform a detailed kinematic analysis via velocity (moment-1) and dispersion (moment-2) maps given that all our galaxies are not sufficiently resolved.

We measured the CO velocity-integrated intensities ($I_{\rm CO}$) from the CO moment-0 maps integrating all the signal located around the phase centre, above the surrounding RMS noise level. No aperture correction was needed as the customised apertures used were all bigger than the synthesised beam.
%using customized apertures, centred on the phase center, large enough to include for each emission all the signal above the surrounding RMS noise level (no aperture correction was needed as the apertures used were all larger than the synthesized beam). 
The derived $I_{\rm CO}$ agree very well with those determined from the Gaussian fitting of the CO line profiles. To fit the complex CO line profiles, we performed multi-component Gaussian fitting using the nonlinear $\chi^2$ minimisation and the Levenberg-Marquardt algorithm. Errors on the values of CO redshifts ($z_{\rm CO}$), full-width half maximum ($\rm FWHM_{CO}$), and  $I_{\rm CO}$ were estimated using the Monte Carlo approach by perturbing the observed spectrum with 1000 random realisations. For the CO non-detections, upper limits on $I_{\rm CO}$ were derived from the $4\sigma$ RMS noise level of moment-0 maps integrated over $\sim 350~\rm km~s^{-1}$,  the typical FWHM measured for the detected CO emission lines of our targets. We then used Eq.~(3) from \citet{Solomon+97} to derive the CO luminosities ($L^{\prime}_{\rm CO}$) from the respective $I_{\rm CO}$. The resulting measurements can be found in Table~\ref{tab:COdata}.

The molecular gas mass ($M_{\rm molgas}$) is then expressed as
\begin{equation}
M_{\rm molgas} = \alpha_{\rm CO} \frac{L^{\prime}_{{\rm CO}~J\rightarrow J-1}}{r_{J,1}},
\end{equation}
where $L^{\prime}_{{\rm CO}~J\rightarrow J-1}$ in $\rm K~km~s^{-1}~pc^2$ is the luminosity of a given CO transition, $r_{J,1}$ is the CO luminosity correction for this transition to the fundamental CO(1--0) transition, and $\alpha_{\rm CO}$ in $M_{\odot}~\rm (K~km~s^{-1}~pc^2)^{-1}$ is the conversion factor between the CO(1--0) luminosity and the H$_2$ gas mass. 

To convert the CO(3--2) and CO(4--3) luminosities to the fundamental CO(1--0) luminosity, which ultimately gives the total H$_2$ molecular gas mass, we applied the CO luminosity correction factors $r_{3,1} = 0.77\pm 0.14$ and $r_{4,1} = 0.61\pm 0.13$, respectively, derived by \citet{Boogaard+20} from the stacking of CO-flux-limited star-forming galaxies at $z\sim 2.5$ from the ALMA SPECtroscopic Survey (ASPECS). Overall, the adopted CO excitation (i.e. CO spectral line energy distribution) is comparable to the one of a lensed star-forming galaxy at $z\sim 3.6$ with a similar IR luminosity to our targets \citep{Dessauges+17} and $z\sim 2.5$ starburst galaxies \citep{Xiao+22}. It is also comparable to the recently derived CO excitation of IR-luminous sub-mm galaxies (SMGs) at $z=2-5$ \citep[$r_{3,1}^{\rm SMG} = 0.75\pm 0.39$ and $r_{4,1}^{\rm SMG} = 0.63\pm 0.44$;][]{Castillo+23}, in contrast with the initially higher CO excitation derived for SMGs at $z=1-4$  \citep{Bothwell+13}. 

As for the CO-to-H$_2$ conversion factor, hereafter we favour the value derived for nearby starburst galaxies, $\alpha_{\rm CO}^{\rm SB} = 1~M_{\odot}~\rm (K~km~s^{-1}~pc^2)^{-1}$, given the starburst nature of our targets (Sect.~\ref{sect:targets}). 
%the range of $\alpha_{\rm CO}$ factors between the nearby starburst value $\alpha_{\rm CO}^{\rm SB} = 1~M_{\odot}~\rm (K~km~s^{-1}~pc^2)^{-1}$ and the Milky Way value $\alpha_{\rm CO}^{\rm MW} = 4.36~M_{\odot}~\rm (K~km~s^{-1}~pc^2)^{-1}$ \citep[including the factor of 1.36 to account for heavy elements, primarily Helium;][]{Bolatto+13}. 
The corresponding $M_{\rm molgas}$ measurements are summarised in Table~\ref{tab:COdust-properties}. In Sects.~\ref{sect:Mgas-tdepl} and \ref{sect:SFE} we discuss the impact of $\alpha_{\rm CO}$ values between $\alpha_{\rm CO}^{\rm SB}$ and the Milky Way value $\alpha_{\rm CO}^{\rm MW} = 4.36~M_{\odot}~\rm (K~km~s^{-1}~pc^2)^{-1}$ \citep[including the factor of 1.36 to account for heavy elements, primarily Helium;][]{Bolatto+13}.

%-----------------------------------------------------------------------

\subsection{From the FIR dust continuum to IR luminosities and dust masses}
\label{sect:continuum}

The 1.3~mm (240~GHz) band 6 FIR dust continuum emission is robustly detected at the phase centre for nine targets: J0146--0220, J1249+1550, J0006+2452, J0850+1549, J1220--0051, J0950+0523, J1157+0113, J0121+0025, and J1316+2614 (Figs.~\ref{fig:contB6-CO_1}, \ref{fig:contB6-CO_2}, and \ref{fig:contB6-CO_3}, left panels). All the continuum detections are spatially resolved, except in J0146--0220 whose continuum emission is comparable to the ALMA synthesised beam size. 

We measured the FIR fluxes ($S_{\rm 1.3mm}$) from the 1.3~mm (240~GHz) continuum emission maps using customised apertures big enough to include for each emission all the signal located around the phase centre, above the surrounding RMS noise level (no aperture correction was needed as the apertures used were all bigger than the synthesised beam). The derived $S_{\rm 1.3mm}$ agree very well with those determined from the fitting of the continuum emission in the \textit{uv} plane by adopting the Fourier transform of the elliptical Gaussian 2D model. The fits were done using the \texttt{UV\_FIT} routine from the GILDAS software package \citep{Guilloteau+00}, leaving the centre coordinates, flux, $\rm FWHM_{major}$, $\rm FWHM_{minor}$, and position angle of the Gaussian 2D model as free parameters. The fit in the \textit{uv} plane did not converge for one target J1157+0113, certainly because of the complex shape of its emission in the continuum map that could not be approximated by the elliptical Gaussian 2D model. For the continuum non-detections, we considered $4\sigma$ upper limits estimated from the RMS noise level of the continuum dirty images. The resulting $S_{\rm 1.3mm}$ measurements are listed in Table~\ref{tab:FIRdata}.

We used the results of the \textit{uv} fits to derive the deconvolved FIR dust continuum sizes of the spatially resolved targets (except for J1157+0113). We computed the effective radius ($R_{\rm eff,FIR}$), defined as the radius enclosing half of the total emission, following $R_{\rm eff,FIR} = \sqrt{\rm FWHM_{major} \times FWHM_{minor}}/2$, where the square root of the product of the deconvolved full-width half maximum along the major and minor axes of the Gaussian 2D most precise fits of the FIR continuum emission corresponds to the circularised $\rm FWHM$ radius. The resulting $R_{\rm eff,FIR}$ are listed in Table~\ref{tab:FIRdata}. 

Since our galaxies have a low dust attenuation $E(B-V)\leq 0.1$ (Sect.~\ref{sect:targets}), we assumed that the emission models for these galaxies should be optically thin. Therefore, to determine the IR luminosity ($L_{\rm IR}$) we scaled the optically thin modified black-body (MBB) function \citep{Casey12} to the measured 1.3~mm (240~GHz) continuum flux of each target, and integrated over the wavelength range between $8~\mu \rm m$ and $1000~\mu \rm m$. We derived the dust mass ($M_{\rm dust}$) from $S_{\rm 1.3mm}$ in Jy by following \citet{Casey12}
\begin{equation}
M_{\rm dust} = \frac{D_{\rm L} S_{\rm 1.3mm}}{\kappa B_{\nu}(T_{\rm dust})},
\end{equation}
where $D_{\rm L}$ in m is the luminosity distance, and $B_{\nu}(T_{\rm dust})$ in $\rm Jy~sr^{-1}$ is the Planck function at the dust temperature ($T_{\rm dust}$) and the observer-frame frequency ($\nu = 240~\rm GHz$). We assumed $T_{\rm dust} = 40~\rm K$ for the dust associated with the starburst regions of our targets at $z\sim 2.1-3.6$ \citep[e.g.][]{Schreiber+18,Sommovigo+22,Viero+22,Witstok+23}. For the dust mass absorption coefficient $\kappa = \kappa_0 (\frac{\nu}{\nu_0})^{\beta}$, where $\beta$ is the dust emissivity index and $\kappa_0$ the dust opacity at $\nu_0$ ($\lambda_0$), we assumed $\beta = 1.4$, $\kappa_0 = 40~\rm cm^2~g^{-1}$, and $\lambda_0 = 100$~$\mu$m \citep{BianchiSchneider07}. We then derived the uncertainties on $L_{\rm IR}$ and $M_{\rm dust}$ as the median absolute deviations (mad) obtained when averaging over $T_{\rm dust} = 30$~K, 35~K, 40~K, and  45~K and seven different $\kappa$ models of dust composition listed in \citet[][Table~2]{Ginolfi+19}, each of them providing a combination of parameters $\beta$, $\kappa_0$, and $\lambda_0$ selected from the literature \citep{WeingartnerDraine01,Bertoldi+03,Robson+04,Beelen+06,
BianchiSchneider07,Galliano+11,Jones+17}. The resulting uncertainties are big but allow us to place our $L_{\rm IR}$ and $M_{\rm dust}$ estimates on the conservative side given the poorly constrained FIR SED with one single band continuum measurement. They are also big enough to encompass the $\gtrsim 5-10\%$ correction factor potentially needed to derive the intrinsic dust emission detected against the cosmic microwave background at the redshift of our galaxies (which we have not applied) computed by \citet{daCunha+13}.
%what fraction of the intrinsic dust emission can be detected against the cosmic microwave background at the redshift of our galaxies.
In Table~\ref{tab:FIRdata} we summarize the derived $L_{\rm IR}$, and $M_{\rm dust}$ can be found in Table~\ref{tab:COdust-properties}.

%-----------------------------------------------------------------------

\subsection{VLT HAWK-I data reduction and analysis}
\label{sect:HAWKI}

The near-IR imaging was obtained in $H$- and $K_{\rm s}$-bands for 10 targets with HAWK-I on the VLT UT4. These observations were conducted between March 2023 and February 2024 in service mode as part of the programme 111.251K.001 (PI: R.~Marques-Chaves). The $H$-band observations were taken under very good seeing conditions of $0.4\arcsec - 0.6\arcsec$ (FWHM). The $K_{\rm s}$-band observations were obtained with the GRound layer Adaptive optics system Assisted by Lasers (GRAAL), enhancing the final image quality down to $0.3\arcsec - 0.4\arcsec$ (FWHM). For each target, the total on-source exposure times were 560 seconds and 1350 seconds in the $H$- and $K_{\rm s}$-bands, respectively. Data were reduced using the standard ESO pipeline version 2.4.12\footnote{\url{https://www.eso.org/sci/software/pipelines/hawki/hawki-pipe-recipes.html}} and were flux calibrated against 2MASS stars in the field. The astrometry was calibrated using the GAIA DR3 catalogue \citep{Gaia+23} yielding an RMS precision of $0.10\arcsec - 0.13\arcsec$, which is roughly similar to the HAWK-I native pixel-scale ($0.107\arcsec$ per pixel). 

Given the very good HAWK-I image quality, we investigated the morphology and sizes of the galaxies\footnote{However, one needs to keep in mind that the HAWK-I images at the redshift of our galaxies are likely dominated by the strong rest-frame optical lines, [O\,{\sc iii}]+H$\beta$ or H$\alpha$.}. We used GALFIT \citep{Peng+02} to fit the light distribution of each galaxy with the 2D Gaussian function convolved to the instrumental point spread function (PSF) that was measured from bright stars in the HAWK-I field-of-view. The fitting process was performed on $10\arcsec \times 10\arcsec$ background-subtracted cutouts centred on each target, shown in Figs.~\ref{fig:contB6-CO_1}, \ref{fig:contB6-CO_2}, and \ref{fig:contB6-CO_3} (middle-left panels). Overall, a single 2D Gaussian profile fitted well the observed light profiles for most of our galaxies, except for J0146--0220 and J0006+2452 that required an additional component. We find that eight galaxies are resolved in the HAWK-I images, for which we derived the rest-frame UV or optical effective radius ($R_{\rm eff,UV/opt}$) measurements. 
%ranging from $R_{\rm eff,UV/opt} = 1.04$~kpc to 2.59~kpc. 
The two remaining galaxies, J0950+0523 and J1316+2614, are unresolved with effective radii as small as $R_{\rm eff,UV} < 0.55$~kpc for J1316+2614. J0146--0220 and J0006+2452 show a bright and unresolved component beside the extended and resolved component.

For the two targets with no HAWK-I observations, J0850+1549 and J0121+0025, we used, respectively, the public CFHT MegaCam and Subaru seeing-limited $R$-band images to infer their sizes, following the same methodology as described for HAWK-I images. The rest-frame UV or optical effective radii of our 12 UV-bright galaxies are summarised in Table~\ref{tab:FIRdata}. 
%Following the same methodology as described for HAWK-I images, we found a very compact morphology for J0121+0025 with $R_{\rm eff,UV} = 0.79 \pm 0.15$~kpc \citep[see][]{Marques+21}, whereas J0850+1549 shows $R_{\rm eff,UV} = 1.46\pm 0.18$~kpc. The rest-frame UV or optical effective radii of our 12 UV-bright galaxies are summarised in Table~\ref{tab:FIRdata}. 

%-----------------------------------------------------------------------

\section{Results}
\label{sect:results}

\subsection{Dust-obscured star formation}
\label{sect:obscured-SFR}

%As summarized in Sect.~\ref{sect:targets}, our galaxies were selected because of their extreme brightness in the rest-frame UV with absolute magnitudes of $M_{\rm UV} = -24.7$ to $-23.4$ (Table~\ref{tab:CO-dust-properties}), comparable and brighter than the UV-bright galaxies recently unveiled by JWST at much higher redshifts $z>7$ \citep[e.g.,][]{Casey+23,Bunker+23,Castellano+24}. These galaxies appear to be 
As summarised in Sect.~\ref{sect:targets}, our galaxies are intense starburst galaxies with very high unobscured star-formation rates for their $M_{\rm stars}$ and redshifts, ranging between $\rm SFR_{UV} = 104~M_{\odot}~\rm yr^{-1}$ and $415~M_{\odot}~\rm yr^{-1}$ (uncorrected for dust attenuation; Table~\ref{tab:restUV-properties}). The measured $L_{\rm IR}$ (Table~\ref{tab:FIRdata}), moreover, show that most of these UV-bright galaxies reside in the regime of luminous IR galaxies (LIRGs; $10^{11} < L_{\rm IR}/L_{\odot} < 10^{12}$) and 3 galaxies reside at the faint end of ultra-luminous IR galaxies (ULIRGs; $10^{12} < L_{\rm IR}/L_{\odot} < 10^{13}$). They provide a measure of the obscured star-formation rates, defined as ${\rm SFR_{IR}}  =  L_{\rm IR} /10^{10}$, following the calibration of \citet{Kennicutt98} for the \citet{Chabrier03} IMF. They reach $\rm SFR_{IR} = 59~M_{\odot}~\rm yr^{-1}$ to $283~M_{\odot}~\rm yr^{-1}$, with three upper limits below $57~M_{\odot}~\rm yr^{-1}$ (Table~\ref{tab:COdust-properties}). The resulting total SFR ($\rm SFR_{UV+IR} = SFR_{UV}+SFR_{IR}$) place our galaxies well above the MS of star-forming galaxies at similar $z\sim 2.5$ and with similar stellar masses, with very high MS offsets of $\rm \Delta MS = SFR_{UV+IR}/\langle SFR_{\rm MS}\rangle \sim 30$ \citep[e.g.][]{Speagle+14,Leslie+20}. Obviously, the offsets could be lower in presence of old, yet currently poorly evidenced, stellar populations (Sect.~\ref{sect:targets}).

%-----------------------------------------------------------------------
\begin{figure}
\includegraphics[width=0.48\textwidth,clip]{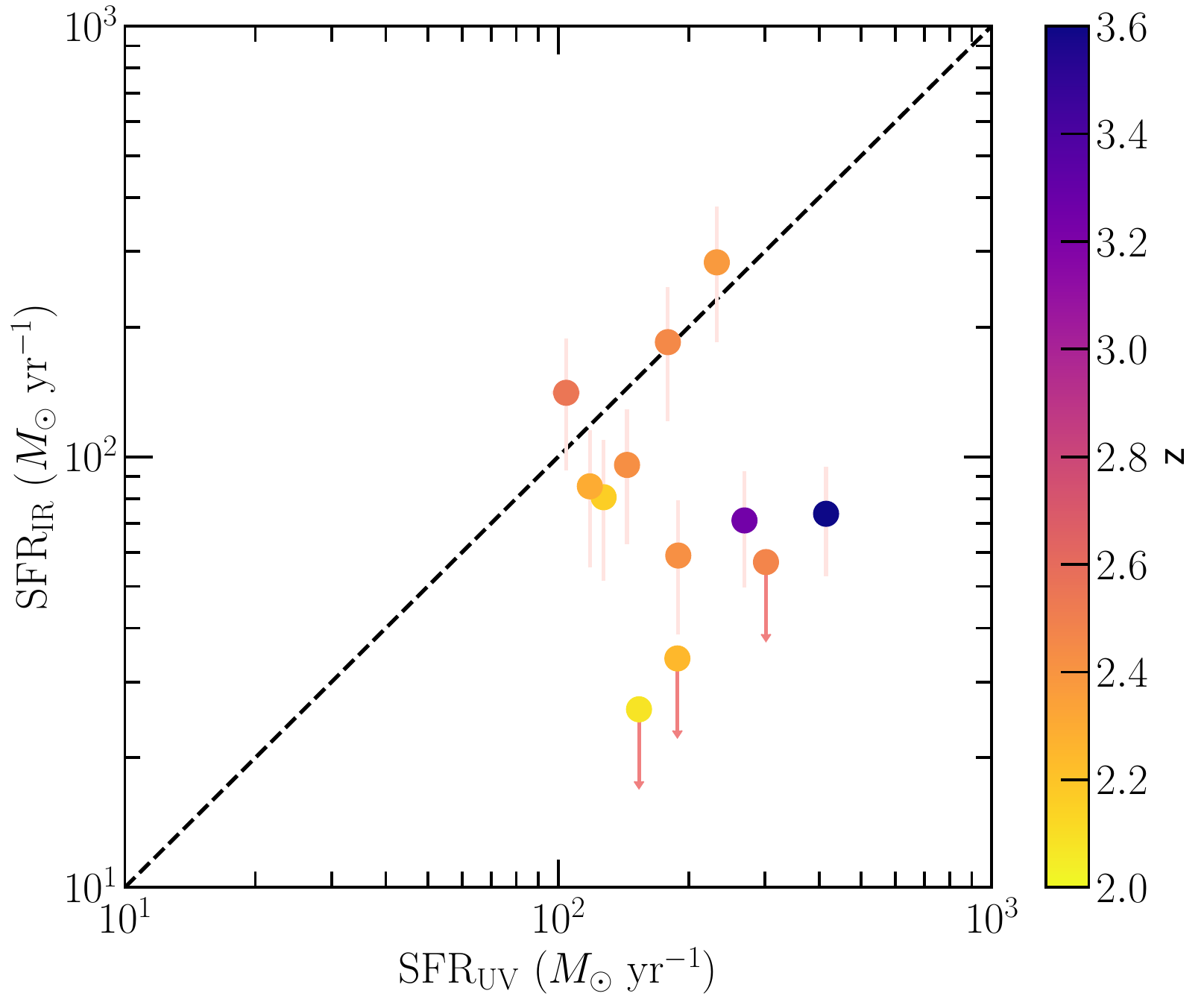}
\caption{Comparison of the unobscured $\rm SFR_{UV}$ (uncorrected for dust attenuation) and obscured $\rm SFR_{\rm IR}$ of our galaxies, colour-coded by redshift. The majority of our galaxies is dominated by the unobscured $\rm SFR_{UV}$, such that the corresponding obscured fractions of star formation range between $f_{\rm obscured} < 14$\% and 42\%, and is above 50\% for three galaxies only. The dashed line traces the one-to-one relationship.}
\label{fig:SFRIR-SFRUV}
\end{figure}
%-----------------------------------------------------------------------

In Fig.~\ref{fig:SFRIR-SFRUV} we show the comparison of the unobscured $\rm SFR_{UV}$ and obscured $\rm SFR_{IR}$ of our galaxies. The bulk of these extreme starbursts are dominated by the unobscured $\rm SFR_{UV}$ and have obscured fractions of star formation ($f_{\rm obscured} = \rm SFR_{IR}/SFR_{UV+IR}$) ranging between $\lesssim 14$\% and 42\%. Only three galaxies, J0006+2452, J0950+0523, and J1157+0113, have $f_{\rm obscured}$ moderately above 50\%, with at most 57\%. Based on different galaxy samples and exploring the possible effects of the assumed FIR SED template in the determination of $\rm SFR_{IR}$ values, \citet{Whitaker+17} showed that $f_{\rm obscured}$ is highly mass dependent but redshift independent for galaxies at $0<z<2.5$. Half of our extreme starburst galaxies follows their $f_{\rm obscured}$--$M_{\rm stars}$ relation, while the other half of galaxies with $f_{\rm obscured} < 25$\% is located a factor of two below the empirical relation. The latter galaxies have among the lowest $f_{\rm obscured}$ measurements known at their $M_{\rm stars}$, and this even in comparison to stacks of UV-selected galaxies at $z\sim 5$ \citep{Fudamoto+20b} and $z\sim 7$ \citep{Algera+23}. The deviation from the mean $f_{\rm obscured}$--$M_{\rm stars}$ relation would be all the more significant if the stellar masses of our galaxies are underestimated (see Sect.~\ref{sect:targets}). Among these low $f_{\rm obscured}$ galaxies, we find the two strong LyC leaking galaxies, J0121+0025 and J1316+2614, at $z>3$ \citep{Marques+21,Marques+22}, both with FIR dust continuum detections. 
%and two newly confirmed LyC leakers, J1415+2036 and J0850+1549 (Marques-Chaves et~al.\ in prep.), with three of them having a FIR dust continuum emission detection.

\begin{figure}
\includegraphics[width=0.48\textwidth,clip]{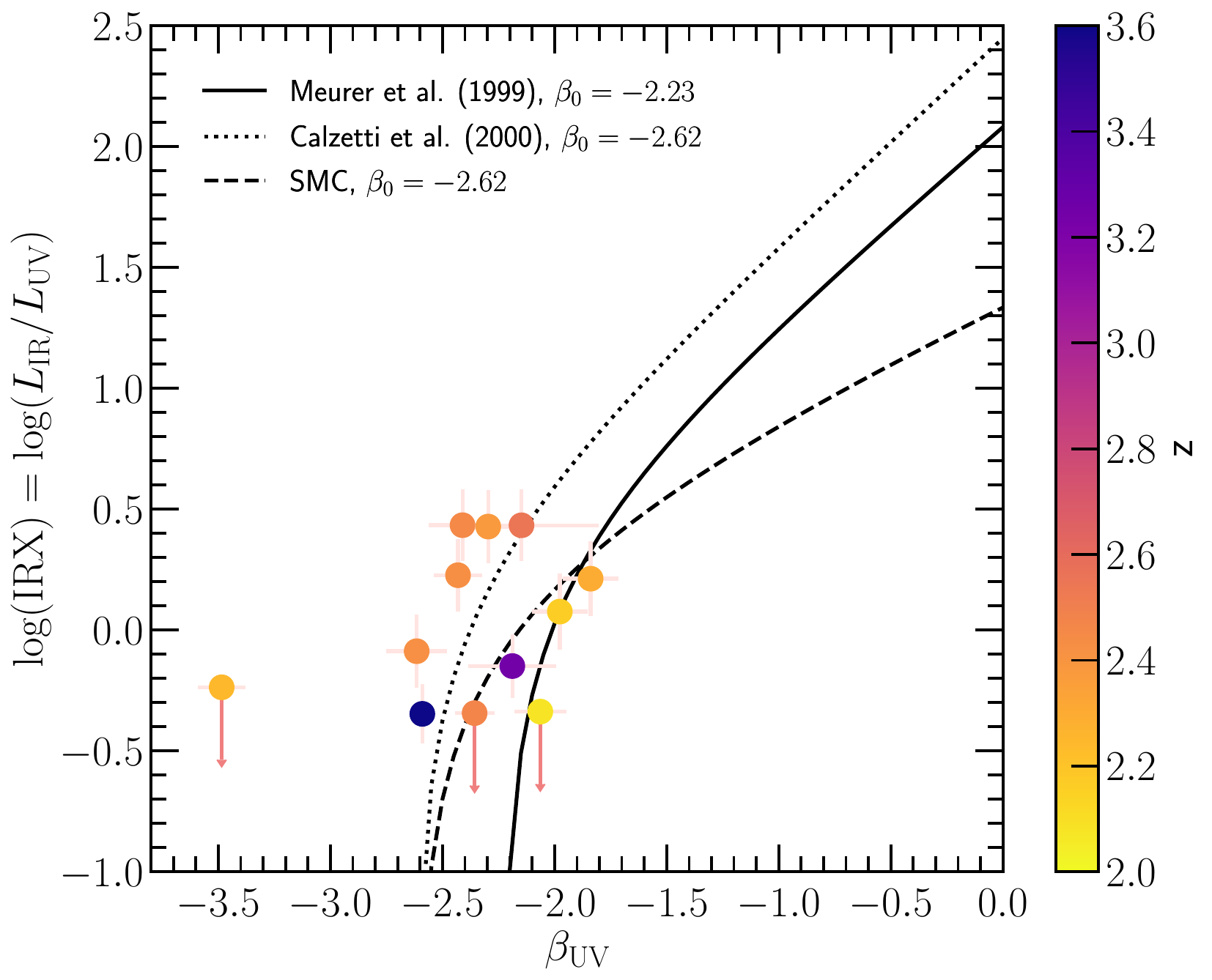}
\caption{Infrared excess as a function of the UV spectral slope of our galaxies, colour-coded by redshift. Our galaxies are all characterised by very blue $\beta_{\rm UV} < -1.8$ and low $\rm IRX < +0.5$. Most of them deviate from the canonical IRX--$\beta_{\rm UV}$ relation \citep[solid line;][]{Meurer+99}, and are better reproduced with the bluer intrinsic UV continuum slope $\beta_0 = -2.62$ as computed by \citet{Reddy+18} for the SMC (dashed line) and \citet{Calzetti+00} (dotted line) extinction curves.}
\label{fig:IRX-betaUV}
\end{figure}

In Fig.~\ref{fig:IRX-betaUV} we show the IR excess (${\rm IRX} = L_{\rm IR}/L_{\rm UV}$) as a function of the UV spectral slope of our galaxies. This is another empirical relation commonly studied for star-forming galaxies used to evaluate their dust attenuation, which does not evidence significant redshift evolution at least for galaxies at $z\lesssim 4$ \citep[e.g.][]{Meurer+99,Whitaker+14,Bouwens+16,Fudamoto+17,Alvarez+16,Alvarez+19}. In comparison to the existing galaxy samples around the cosmic noon era, our galaxies are characterised by very blue $\beta_{\rm UV} < -1.8$ and low $\rm IRX < +0.5$ because of their particularly high $L_{\rm UV}$ dominating over the LIRG/ULIRG regime. These two characteristics are more typical of very high redshift UV-selected galaxies at $z\sim 4.5-7.7$ \citep[e.g.][]{Fudamoto+20b,Inami+22,Bowler+24}; in particular, mean $\beta_{\rm UV}$ between $-2.2$ and $-2.6$ were derived from about a thousand of $5 < z <13$ galaxies \citep{RobertsBorsani+24,Heintz+24}. As a result, similarly to the $z\sim 4.5-7.7$ galaxies, most of our galaxies deviate from the canonical IRX--$\beta_{\rm UV}$ relation derived by \citet{Meurer+99} for nearby starburst galaxies with an intrinsic UV continuum slope $\beta_0 = -2.23$ (solid line in Fig.~\ref{fig:IRX-betaUV}). \citet{Reddy+18} computed IRX--$\beta_{\rm UV}$ relations assuming a constant star formation with an age of 100~Myr and including nebular continuum emission for a low stellar metallicity of $Z= 0.14~Z_{\odot}$, assumed to better characterise high redshift galaxies expected to have different physical conditions from nearby starburst galaxies with younger stellar populations and lower metallicities, yielding bluer $\beta_0 = -2.62$ relative to the canonical relation of \citet{Meurer+99}. The resulting IRX--$\beta_{\rm UV}$ predictions with $\beta_0 = -2.62$ were found to agree with stacks of UV-selected galaxies at $z>4.5$ \citep[e.g.][]{Fudamoto+20b}. In Fig.~\ref{fig:IRX-betaUV} we show the corresponding IRX--$\beta_{\rm UV}$ predictions obtained for the Small Magellanic Cloud \citep[SMC;][]{Gordon+03} extinction curve (dashed line) and the \citet{Calzetti+00} attenuation curve (dotted line). For three of our galaxies we observe a good match with the SMC and $\beta_0 = -2.62$ prediction, although the \citet{Meurer+99} IRX--$\beta_{\rm UV}$ relation remains consistent within the measurement uncertainties. Six galaxies are clearly offset to bluer $\beta_{\rm UV}$ values, and are most precisely reproduced by the \citet{Calzetti+00} and $\beta_0 = -2.62$ prediction. 
%with the exception of J1415+2036 given its extremely blue $\beta_{\rm UV} = -3.49\pm 0.11$ (Table~\ref{tab:CO-dust-properties}). 

%-----------------------------------------------------------------------
%-----------------------------------------------------------------------

Another well-studied empirical relation that links dust attenuation and stellar masses of galaxies is the IRX--$M_{\rm stars}$ relation, which is expected because $M_{\rm stars}$ is the outcome of past star-formation activity responsible for producing dust in supernovae (SNe) and pulsating moderate-mass asymptotic giant branch (AGB) stars, and is also suggested by simulations \citep{Graziani+20}. Different parametrisations are reported for samples of MS galaxies at $z\sim 1.5-4$, showing, on average, that MS galaxies follow a relatively tight and shallow IRX--$M_{\rm stars}$ correlation with small variations from one sample to another \citep[e.g.][]{Heinis+14,Bouwens+16,Alvarez+16,Koprowski+18}, with the exception of the study of \citet{Fudamoto+20a} where they report a significantly steeper IRX--$M_{\rm stars}$ relation for their sample of $z=2.5-4.0$ MS galaxies. Starburst galaxies at similar redshifts are found above the IRX--$M_{\rm stars}$ relation of MS galaxies with $\sim +0.5$~dex higher IRX values, indicating that starbursts are more dust extinct at a fixed $M_{\rm stars}$ \citep{Fudamoto+20a}. Our extreme starburst galaxies are among the few galaxies known with individual IRX measurements at stellar masses as low as $\log(M_{\rm stars}/M_{\odot}) = 9.17-9.66$. Half of them follows more or less the shallow IRX--$M_{\rm stars}$ parametrisation, and the other half lies in between, that is, above the steep \citet{Fudamoto+20a} parametrisation and below the shallow parametrisation, and even below the parametrisation from \citet{Heinis+14} specifically derived from stacks of high $L_{\rm UV}$ MS galaxies ($\log L_{\rm UV}/L_{\odot} = 10.64-10.94$, which, nevertheless, are lower than the UV luminosities of our galaxies). Consequently, the currently still debated IRX--$M_{\rm stars}$ relation of MS galaxies, particularly for low stellar masses ($M_{\rm stars}/M_{\odot}<10^{10}$), makes it difficult to bring definitive conclusions on the dust attenuation of our extreme starburst galaxies with respect to MS galaxies with similar $M_{\rm stars}$. Globally, our galaxies populate the IRX and $M_{\rm stars}$ parameter space encompassed by the relations extrapolated from individual measurements and stacks of different samples of more massive MS galaxies at comparable redshifts. Nevertheless, they seem to agree particularly well with the new IRX--$M_{\rm stars}$ parametrisation recently derived by \citet{Bowler+24} for the very high redshift galaxies ($z\sim 4.5-7.7$).

%-----------------------------------------------------------------------

\subsection{Molecular gas mass content and depletion timescale}
\label{sect:Mgas-tdepl}

\begin{figure}
\includegraphics[width=0.48\textwidth,clip]{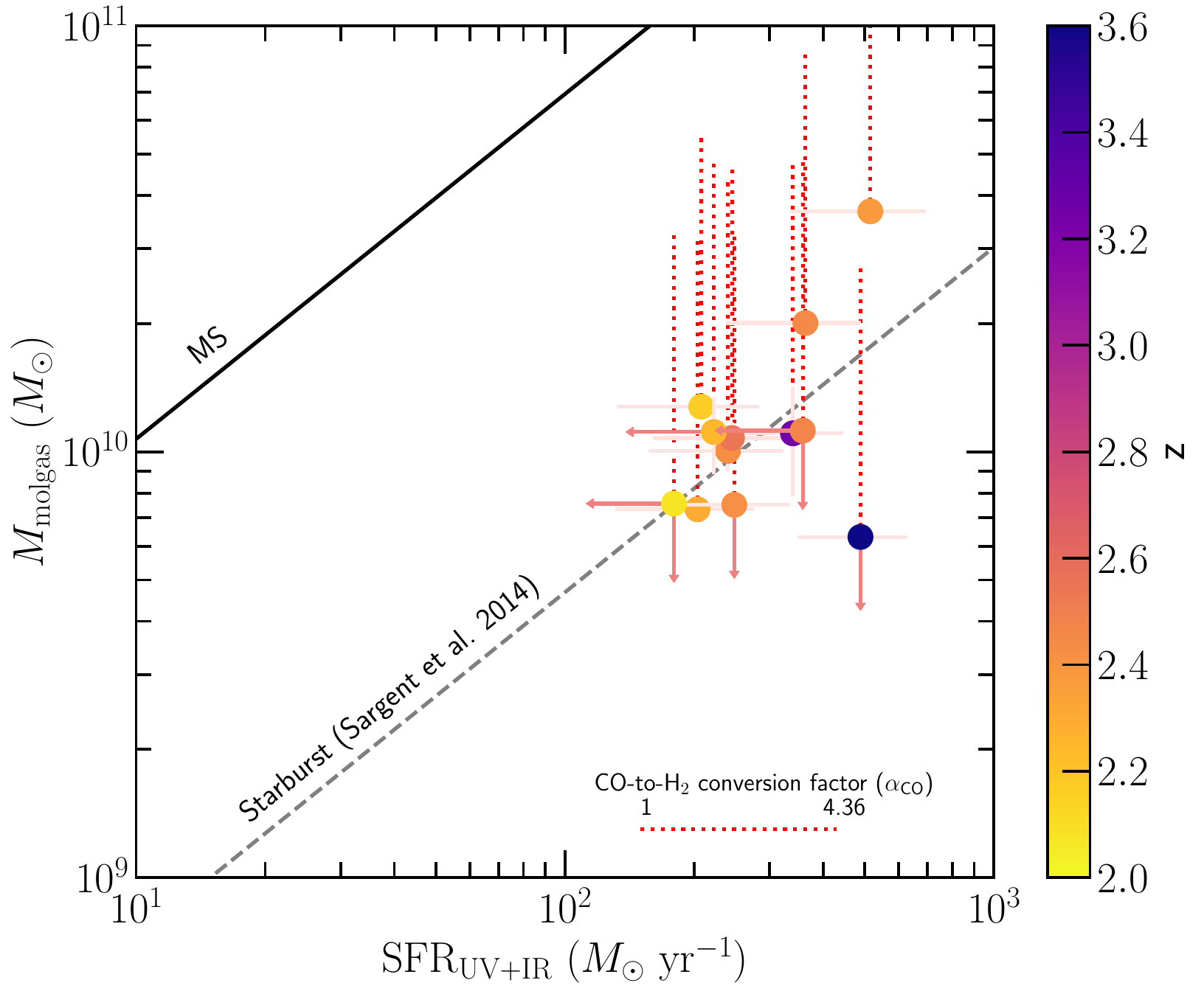}
\caption{Molecular gas mass as a function of the total (unobscured plus obscured) star-formation rate of our galaxies, colour-coded by redshift. The filled circles correspond to $M_{\rm molgas}$ determined with $\alpha_{\rm CO}^{\rm SB} = 1~M_{\odot}~\rm (K~km~s^{-1}~pc^2)^{-1}$, and the dotted red segments show the range of possible $M_{\rm molgas}$ with higher CO-to-H$_2$ conversion factors up to $\alpha_{\rm CO}^{\rm MW} =  4.36~M_{\odot}~\rm (K~km~s^{-1}~pc^2)^{-1}$. Whatever $\alpha_{\rm CO}$, our extreme starburst galaxies have significantly higher $\rm SFR_{UV+IR}$ than MS galaxies at comparable $M_{\rm molgas}$ (see the solid line as derived by \citet{Sargent+14}). For $\alpha_{\rm CO}^{\rm SB}$ our galaxies are the first galaxies known at high redshift to reach the starburst $M_{\rm molgas}$--SFR relation (dashed line) offset by $>1$~dex from the MS relation, according to the parametrisation of \citet{Sargent+14} based on nearby and $z<0.1$ ULIRGs. }
\label{fig:Mmol-SFR}
\end{figure}

The measurements (and upper limits) of the molecular gas masses of our galaxies, derived from the CO(3--2) or CO(4--3) luminosities, range between $M_{\rm molgas} <0.63\times 10^{10}~M_{\odot}$ and $3.67\times 10^{10}~M_{\odot}$ when derived with $\alpha_{\rm CO}^{\rm SB}$ (Table~\ref{tab:COdust-properties}).
%are spread over 1.4~dex between $M_{\rm molgas} <6.3\times 10^9~M_{\odot}$ and $1.6\times 10^{11}~M_{\odot}$, depending on the assumed $\alpha_{\rm CO}^{\rm SB}$ or $\alpha_{\rm CO}^{\rm MW}$. 
As shown in Fig.~\ref{fig:Mmol-SFR}, whatever $\alpha_{\rm CO}$, our extreme starburst galaxies have significantly higher $\rm SFR_{UV+IR}$ than MS galaxies at comparable $M_{\rm molgas}$, or, inversely, have significantly lower $M_{\rm molgas}$ than MS galaxies at comparable $\rm SFR_{UV+IR}$ (if these latter exist). This is in line with the integrated Kennicutt-Schmidt (KS) star-formation law and its $M_{\rm molgas}$--SFR parametrisations determined by \citet{Sargent+14} for MS galaxies (solid line in Fig.~\ref{fig:Mmol-SFR} derived from nearby and cosmic noon galaxies) and starburst galaxies (dashed line in Fig.~\ref{fig:Mmol-SFR} derived from nearby and $z<0.1$ ULIRGs), as also reported by \citet{Daddi+10} and \citet{Genzel+10}. Our galaxies actually represent the first sample of high redshift galaxies that shows an offset from MS galaxies high enough, when assuming $\alpha_{\rm CO}^{\rm SB}$, to lie on the starburst $M_{\rm molgas}$--SFR relation offset by more than 1~dex from the MS relation. Both the sample of massive starburst galaxies at $z\sim 1.6$ studied by \citet{Silverman+15,Silverman+18} and SMGs at $z\sim 1-5$ \citep[e.g.][]{Bothwell+13,CalistroRivera+18,Liu+19,Castillo+23} do not reach the starburst star-formation regime even with $\alpha_{\rm CO}^{\rm SB}$ . On the other hand, our galaxies still satisfy the empirical relation defined between $L_{\rm IR}$ and $L^{\prime}_{\rm CO(1-0)}$ established for diverse galaxy types from nearby galaxies (disks, dwarfs, starbursts) to ULIRGs, MS galaxies, SMGs, and quasars at $z\sim 0-5$ \citep[e.g.][]{CarilliWalter13,Dessauges+15}. This means that, for their IR luminosities, our starburst galaxies have the expected CO luminosities. They thus end up in the extreme starbursting regime mostly because of their very high unobscured $\rm SFR_{UV}$ (see also Sect.~\ref{sect:obscured-SFR}).

%-----------------------------------------------------------------------
%-----------------------------------------------------------------------
\begin{figure*}
\centering
\includegraphics[width=0.43\textwidth,clip]{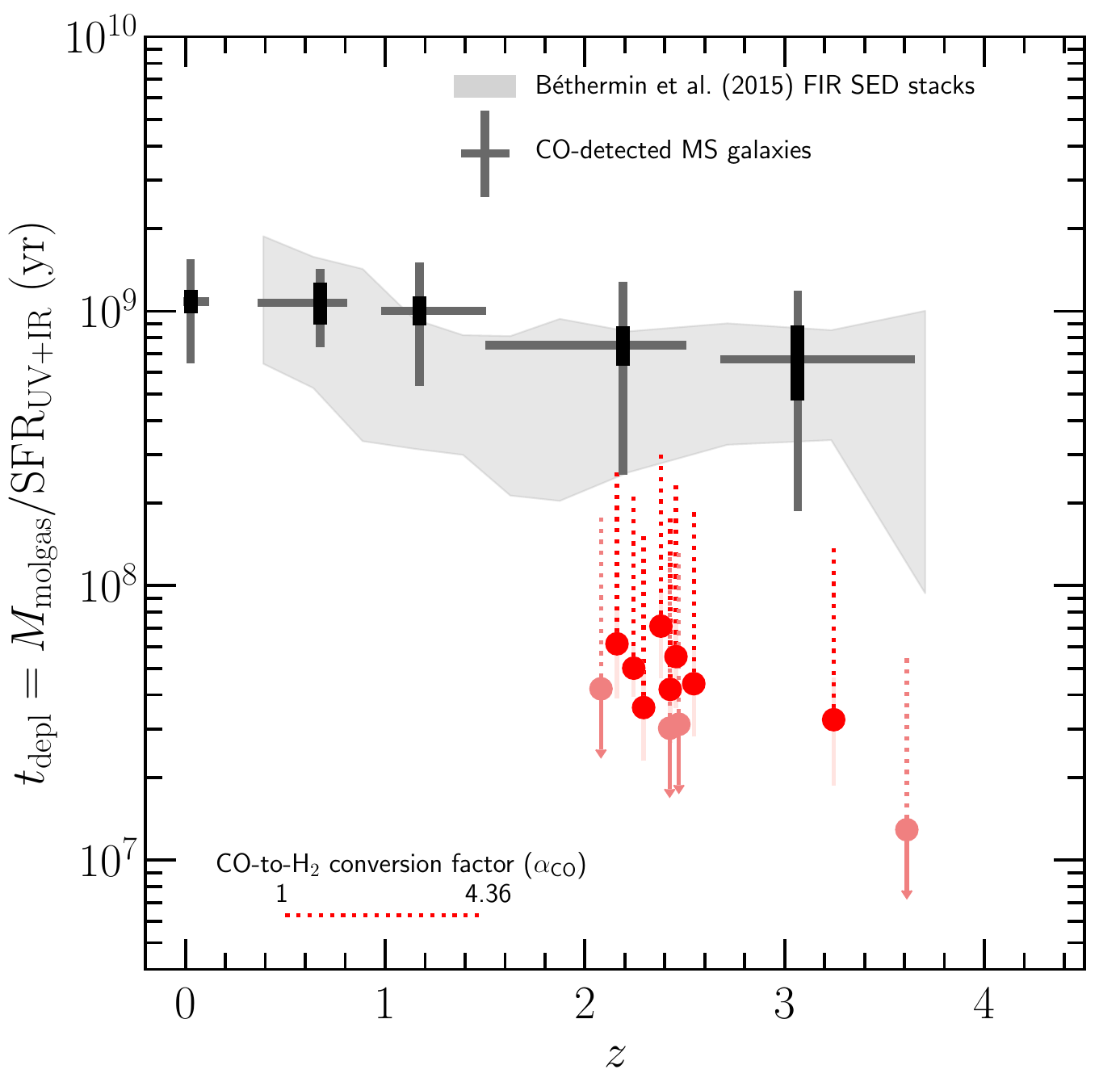}\hspace{1cm}
\includegraphics[width=0.43\textwidth,clip]{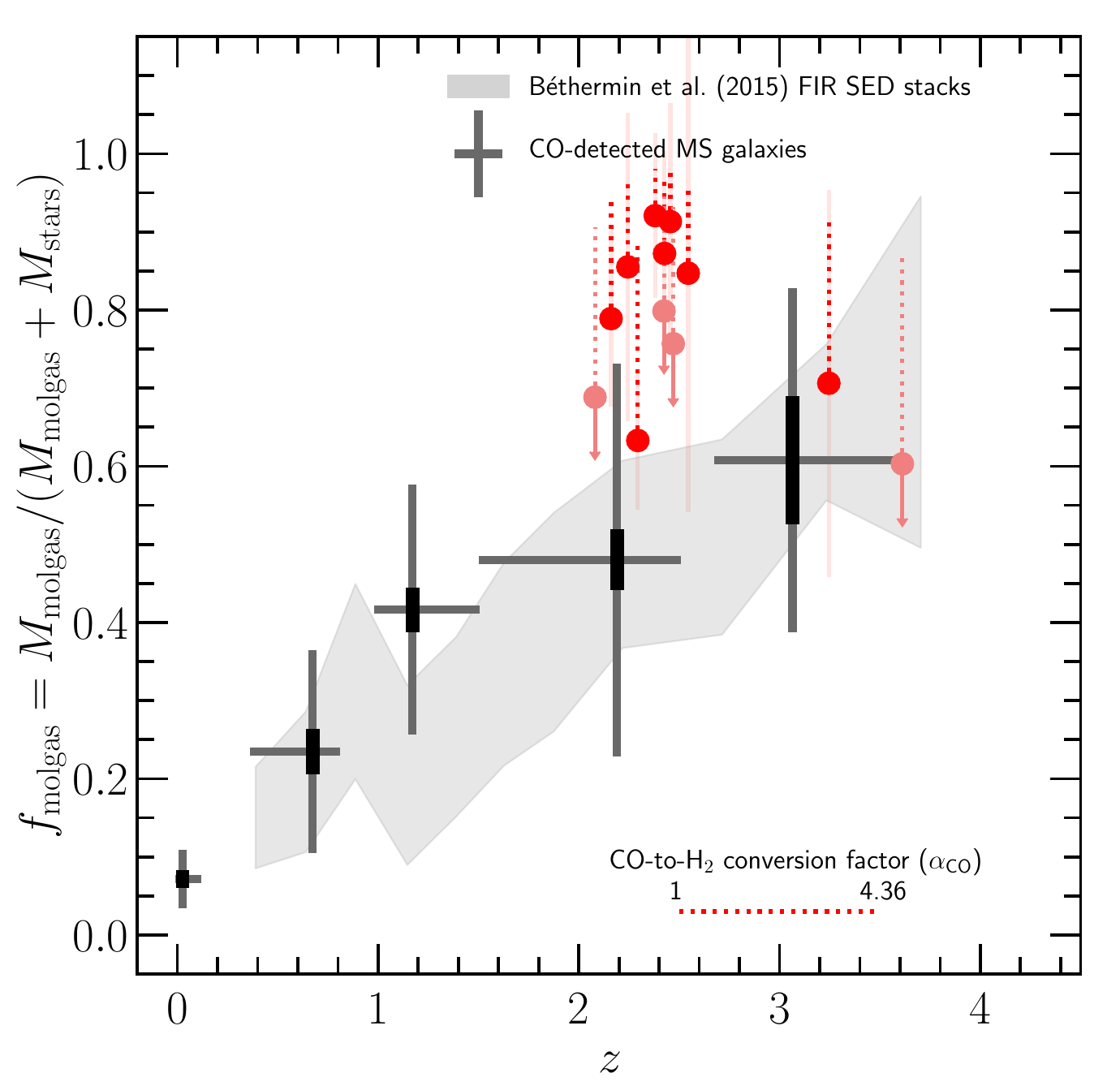}
\caption{Molecular gas depletion timescale (left) and molecular gas mass fraction (right) as a function of the redshift of our galaxies shown by the red filled circles. Similarly to Fig.~\ref{fig:Mmol-SFR}, the dotted red segments represent the range of possible $t_{\rm depl}$ and $f_{\rm molgas}$ measurements of our galaxies as derived with CO-to-H$_2$ conversion factors sampling values from $\alpha_{\rm CO}^{\rm SB} = 1~M_{\odot}~\rm (K~km~s^{-1}~pc^2)^{-1}$ to $\alpha_{\rm CO}^{\rm MW} = 4.36~M_{\odot}~\rm (K~km~s^{-1}~pc^2)^{-1}$. The $t_{\rm tepl}$ and $f_{\rm molgas}$ means, errors on the means, and standard deviations obtained for the CO-detected MS galaxies from the compilation of \citet{Dessauges+20} are shown by the black/grey crosses in five redshift bins $0<z<0.1$, $0.1<z<1$, $1<z<1.5$, $1.5<z<2.5$, and $2.5<z<3.7$.
%The crosses represent the CO-detected MS galaxies from the compilation of \citet{Dessauges+20} in five redshift bins $0<z<0.1$, $0.1<z<1$, $1<z<1.5$, $1.5<z<2.5$, and $2.5<z<3.7$; the respective means, errors on the means, and standard deviations per redshift bin are indicated by the black/grey crosses. 
The light-grey shaded area corresponds to $t_{\rm tepl}$ and $f_{\rm molgas}$ derived by \citet{Bethermin+15} from the FIR SED stacks of MS galaxies. For all MS galaxy measurements compiled here, the same $\alpha_{\rm CO}^{\rm MW}$ conversion factor is assumed. Our extreme starburst galaxies have significantly shorter $t_{\rm depl}$ than those measured in MS galaxies at any redshift (left). This indicates that they are vigorously consuming their molecular gas mass reservoir to build up their stellar mass. On the other hand, they have very high $f_{\rm molgas}$ (right), in excess with respect to MS galaxies at similar redshifts, showing that their mass remains dominated by the molecular gas mass over the stellar mass.}
\label{fig:tdepl-fmol-z}
\end{figure*}
%-----------------------------------------------------------------------

The high $\rm SFR_{UV+IR}$ of our galaxies 
%The deficit in $M_{\rm molgas}$ of our galaxies for their high $\rm SFR_{UV+IR}$ 
yield very short molecular gas depletion timescales ($t_{\rm depl} = M_{\rm molgas}/\rm SFR_{UV+IR}$) with a mean of $\rm 49~Myr \pm 12~Myr$ for $\alpha_{\rm CO}^{\rm SB}$ (Table~\ref{tab:COdust-properties}).
%with mean values of $49~\rm Myr$ and $214~\rm Myr$ ($\pm 91~\rm Myr$), when derived with $\alpha_{\rm CO}^{\rm SB}$ and $\alpha_{\rm CO}^{\rm MW}$, respectively (see Table~\ref{tab:CO-dust-properties}). 
As shown in the left panel of Fig.~\ref{fig:tdepl-fmol-z}, these $t_{\rm depl}$ are significantly shorter by more than one order of magnitude than those measured in MS galaxies at any redshift \citep[e.g.][]{Bethermin+15,Liu+19,Tacconi+20,Dessauges+20}. They are also shorter than those reported for the massive starburst galaxies at $z\sim 1.6$ with their mean $t_{\rm depl} \sim 60~\rm Myr$ \citep{Silverman+15,Silverman+18} and for the SMGs with their mean $t_{\rm depl} \sim 200~\rm Myr$ \citep[e.g.][]{Bothwell+13,CalistroRivera+18,Castillo+23}, considering the same $\alpha_{\rm CO}^{\rm SB}$. The upper limit on $t_{\rm depl}$ derived for J1316+2614, our highest redshift galaxy, even suggests that all its molecular gas mass ($M_{\rm molgas} < 6.3\times 10^9~M_{\odot}$) is depleted in $<13$~Myr! This remains true despite the uncertain $L_{\rm IR}$ measurements (Sect.~\ref{sect:continuum}) affecting $\rm SFR_{UV+IR}$: if $\rm SFR_{UV+IR}$ are underestimated the $t_{\rm depl}$ values would be even shorter, and if, on the other hand, $\rm SFR_{UV+IR}$ are overestimated they would need to be overestimated by a factor of more than 10 to bring our galaxies to the $t_{\rm depl}$ values of MS galaxies at similar redshifts, which is implausible as $\rm SFR_{UV+IR}$ are dominated by $\rm SFR_{UV}$ and not by the uncertain $\rm SFR_{IR}$. Consequently, our extreme starburst galaxies are really vigorously consuming their molecular gas mass reservoir and are rapidly building up their stellar mass. 

In the right panel of Fig.~\ref{fig:tdepl-fmol-z} we show the molecular gas mass fractions ($f_{\rm molgas} = M_{\rm molgas}/(M_{\rm molgas}+M_{\rm stars})$) of our galaxies as a function of redshift that we compare to MS galaxies. We observe that our galaxies have very high $f_{\rm molgas}$ with a mean of $82\%\pm 10\%$ for $\alpha_{\rm CO}^{\rm SB}$ (Table~\ref{tab:COdust-properties}). This reveals that their baryonic mass is dominated by the cold molecular gas mass over the stellar mass, with the mean molecular gas mass to stellar mass ratio ($\mu_{\rm molgas} = M_{\rm molgas}/M_{\rm stars}$) of $6\pm 3$, 
%very large mean $\langle f_{\rm molgas}\rangle$ values of 82\% and 95\% ($\pm 10$\%), derived with $\alpha_{\rm CO}^{\rm SB}$ and $\alpha_{\rm CO}^{\rm MW}$, respectively (Table~\ref{tab:CO-dust-properties}). This reveals that their baryonic mass is dominated by the cold molecular gas mass over the stellar mass (Table~\ref{tab:CO-dust-properties}), with the mean molecular gas mass to stellar mass ratios ($\mu_{\rm molgas} = M_{\rm molgas}/M_{\rm stars}$) reaching 6 and 26 when derived with $\alpha_{\rm CO}^{\rm SB}$ and $\alpha_{\rm CO}^{\rm MW}$, respectively, 
and this despite the observed very rapid molecular gas consumption timescales (Fig.~\ref{fig:tdepl-fmol-z}, left panel). This suggests that our extreme starburst galaxies are caught at the very beginning of their stellar mass build-up, or that cosmic gas inflows are actively feeding these galaxies. The measured $f_{\rm molgas}$ of our galaxies are in significant excess with respect to those of MS galaxies at similar redshifts, and still higher than those of MS galaxies with similar $M_{\rm stars}$ (and $z$) despite the existing steep $f_{\rm molgas}$--$M_{\rm stars}$ anticorrelation \citep[e.g.][]{Bethermin+15,Liu+19,Tacconi+20,Dessauges+20}. The stellar masses of our galaxies would need to be underestimated by, on average, a factor of $\sim 3-4$ to reach the upper end of the $f_{\rm molgas}$ distribution of MS galaxies at $z\sim 2.5$. The same is true for them to reach the mean $f_{\rm molgas} \sim 50\%$ of the massive starburst galaxies at $z\sim 1.6$ of \citet{Silverman+15,Silverman+18} and even more to reach the mean $f_{\rm molgas} \sim 40\%$ of SMGs \citep[e.g.][]{Bothwell+13,CalistroRivera+18,Castillo+23}, considering the same $\alpha_{\rm CO}$.

%------------------------------------------------------------------------------------

\subsection{Star-formation efficiency}
\label{sect:SFE}

Our extreme starburst galaxies are dominated by very young stellar populations of $\sim 10$~Myr, with the bulk of their stellar mass being assembled within this short timescale given the absence of relevant old stellar populations, as described in Sect.~\ref{sect:targets}. \citet{Upadhyaya+24}, moreover, showed that their rest-frame UV spectra resemble those of nearby young star clusters (R136--\citet{Crowther+16} and SB179 \citet{Senchyna+17}) and the Sunburst star cluster at $z\simeq 2.4$ \citep{Mestric+23}. As a result, in contrast to most of the galaxies at cosmic noon whose stellar populations are a mixed bag of young and old populations dominated by old populations, our galaxies represent rare objects whose stellar populations are dominated by young populations representing $> 30$\% to 62\% of the total stellar mass of our galaxies \citep{Marques+20b,Marques+21,Marques+22,Marques+24}.

We can thus measure the star-formation efficiency ($\epsilon_{\rm SF}$) of our galaxies, defined as the fraction of molecular gas mass converted into the recently formed stellar mass, namely here over the past $\sim 10$~Myr, following $\epsilon_{\rm SF} = M_{\rm stars}^{\rm young}/(M_{\rm molgas}+M_{\rm stars}^{\rm young})$ \citep{Evans+09,Dessauges+19,Dessauges+23}. We find $\epsilon_{\rm SF}$ ranging from 8\% to 37\% (Table~\ref{tab:COdust-properties}), with a mean of  $18\% \pm 9\%$ for $\alpha_{\rm CO}^{\rm SB}$. J1249+1550 has the highest measured $\epsilon_{\rm SF}$ approaching 40\%, together with J1316+2614 with its lower limit $\epsilon_{\rm SF} > 40$\%. They both are associated with very short molecular gas depletion timescales of 36~Myr and $<13$~Myr, respectively, among the shorter in our galaxy sample (Table~\ref{tab:COdust-properties}). One needs to keep in mind that the derived $\epsilon_{\rm SF}$ likely are lower limits, as we do not know whether all the measured $M_{\rm molgas}$ is associated with the star formation taking place in the compact starburst UV-bright component of our galaxies revealed by the HST and HAWK-I rest-frame UV/optical images (see Sect.~\ref{sect:UVmorphology}).
%since spatial offsets between UV and CO emissions are observed in some galaxies (see Sect.~\ref{sect:UVmorphology}).
%at disposal for the star formation linked to the compact starburst UV-bright emission in our galaxies or if part of $M_{\rm molgas}$ feeds an underlying galactic star formation (see Sect.~\ref{sect:UVmorphology}).

The derived $\epsilon_{\rm SF}$ of our extreme starburst galaxies are higher in comparison to those of a few percent only ($\lesssim 5$\%) measured in nearby galaxies \citep[e.g.][]{Schruba+19,Utomo+18,Kim+23}. On the other hand, they agree with the $\epsilon_{\rm SF}$ values of $\sim 30$\% inferred for two strongly lensed MS galaxies at $z\simeq 1$ from the analysis of giant molecular clouds and their association with star-forming regions \citep{Dessauges+19,Dessauges+23}. These enhanced $\epsilon_{\rm SF}$ are comparable to those proposed in ultra-massive ($M_{\rm stars} > 10^{11}~M_{\odot}$) optically dark and quenched galaxies recently discovered at $z\sim 5-6$ \citep{Xiao+24,deGraaff+24}, whose existence can be explained by invoking very high star-formation efficiencies ($\gtrsim20$\% to 100\%). Similarly, \citet{Weibel+24} suggested that an increasing $\epsilon_{\rm SF}$ with redshift, reaching values of $\epsilon_{\rm SF}\sim 30$\% at $z\sim 7-8$, is required to explain the high-mass end of the stellar mass functions at $z\geq 4-9$ as determined with JWST observations. On the simulation side, MillenniumTNG simulations also find that $\epsilon_{\rm SF}$ of about $10\%-30$\% are necessary in the early Universe to solve the observed excess of luminous/massive galaxies at $z\gtrsim 8$ discovered with JWST \citep{Kannan+23}. Values of $\epsilon_{\rm SF}\geq 57$\% are even invoked by \citet{Boylan23} to reconcile the standard $\Lambda$CDM cosmological model with the observed excess. 

%-----------------------------------------------------------------------

\subsection{Dust mass content}
\label{sect:Mdust}

The detected dust continuum emission in the Rayleigh-Jeans tail of the FIR SED ($\lambda_{\rm rest}>230~\mu$m) was used to estimate the dust mass content of our galaxies, as described in Sect.~\ref{sect:continuum}. However, the derived $M_{\rm dust}$ ranging from $<1.1\times 10^7~M_{\odot}$ to $1.1\times 10^8~M_{\odot}$ (Table~\ref{tab:COdust-properties}) have to be considered with some caution given their large uncertainties coming from the poorly constrained FIR SED with one single band continuum measurement and the large number of free parameters entering in the dust composition models. The corresponding dust-to-gas mass ratios ($\delta_{\rm DGR} = M_{\rm dust}/M_{\rm molgas}$) are spread out between $\log (\delta_{\rm DGR}) < -2.9$ and $-2.1$ for $\alpha_{\rm CO}^{\rm SB}$.
%$\log (\delta_{\rm DGR}) < -3.5$ and $-2.1$, depending on the assumed $\alpha_{\rm CO}^{\rm MW}$ or $\alpha_{\rm CO}^{\rm SB}$. This more than 1~dex spread 
This spread is comparable to the dispersion of $\delta_{\rm DGR}$ measurements reported in the literature at a given metallicity and the various extrapolated $\delta_{\rm DGR}$--metallicity relations that diverge even more at $12+\log(\rm O/H) < 8$ \citep[e.g.][]{Magdis+12,RemyRuyer+14,DeVis+19,Popping+23,Valentino+24}. The observed $\delta_{\rm DGR}$--metallicity correlation is also found in semi-analytic models of galaxy formation that include the tracking of dust formation and destruction over cosmic time \citep[e.g.][]{Popping+17}. Globally, the inferred $\delta_{\rm DGR}$ of our galaxies lower than $10^{-2}$ (the typical $\delta_{\rm DGR}$ value measured around solar metallicity) favour sub-solar metallicities in agreement with the metallicity measurements of our galaxies ($12+\log(\rm O/H) = 8.13-8.49$; Sect.~\ref{sect:targets}).

\begin{figure}
\includegraphics[width=0.48\textwidth,clip]{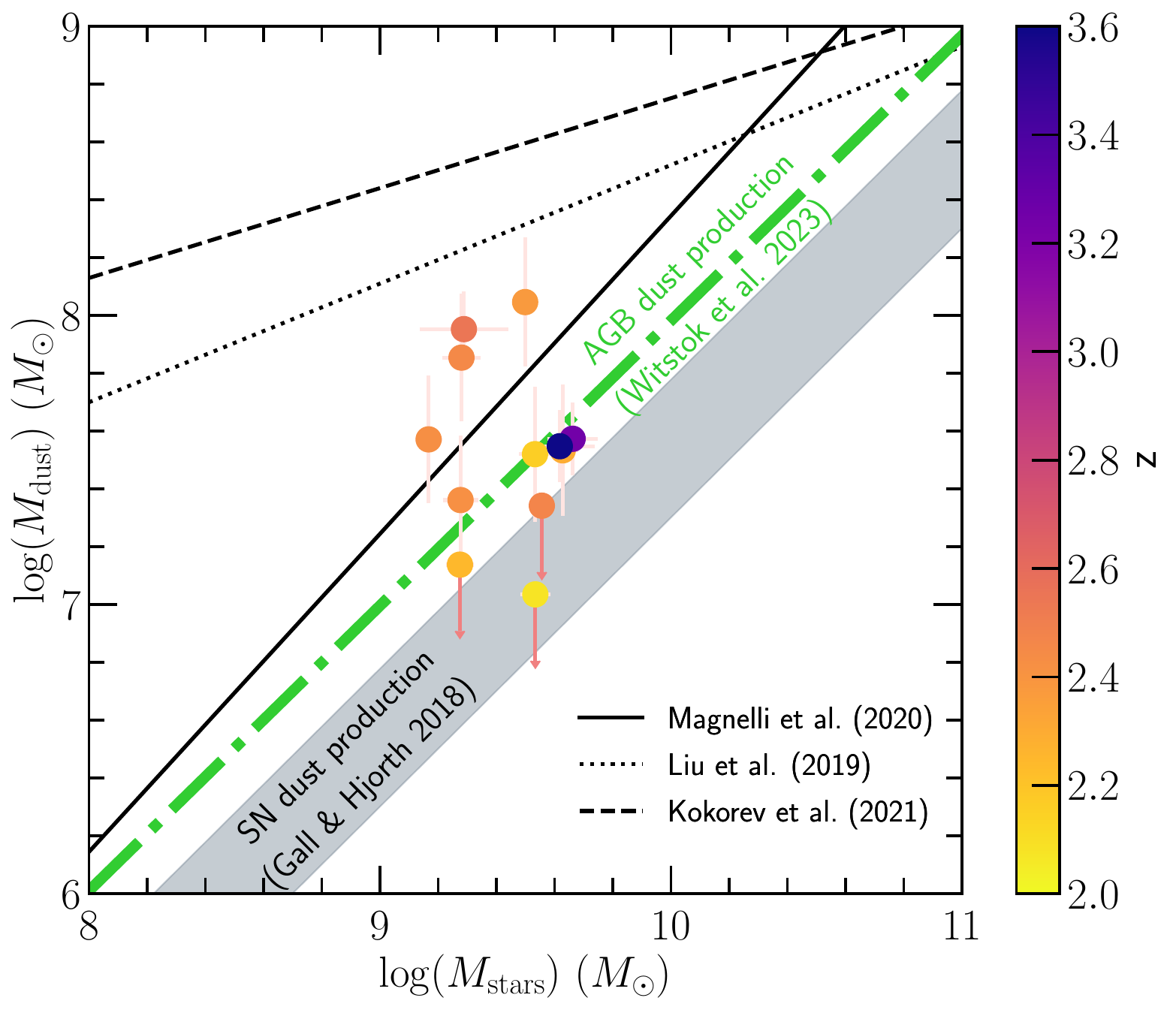}
\caption{Dust mass as a function of the stellar mass of our extreme starburst galaxies, colour-coded by redshift. Our galaxies agree with the $M_{\rm dust}$--$M_{\rm stars}$ relation derived by \citet{Magnelli+20} as shown by the solid black line, but are well below the relations of \citet{Liu+19} and \citet{Kororev+21} shown by the dotted and dashed black lines, respectively. The grey shaded zone defines the SN dust production as predicted by \citet{GallHjorth18}, which is barely sufficient to reproduce $M_{\rm dust}$ of most of our galaxies. 
%except four despite their conservative/large $M_{\rm dust}$ uncertainties}. 
The thick dashed-dotted green line defines the start of the AGB star contribution to the dust production \citep{Witstok+23}.}
\label{fig:Mdust-Mstars}
\end{figure}

In Fig.~\ref{fig:Mdust-Mstars} we compare the dust masses of our extreme starburst galaxies to different empirical $M_{\rm dust}$--$M_{\rm stars}$ relations presented in \citet[][their Fig.~12]{Kororev+21} for MS galaxies at similar redshifts but extrapolated from more massive galaxies ($\log(M_{\rm stars}/M_{\odot}) > 10$). Our galaxies agree with the $M_{\rm dust}$--$M_{\rm stars}$ relation derived by \citet{Magnelli+20} as shown by the solid black line, but are well below the relations of \citet{Liu+19} and \citet{Kororev+21} shown by the dotted and dashed black lines, respectively, which predict much higher $M_{\rm dust}$ for the low-$M_{\rm stars}$ MS galaxies at $z\sim 2.5$. The $M_{\rm dust}$ estimates of our extreme starburst galaxies in fact resemble those of $z\sim 4-7$ galaxies with comparable stellar masses \citep{Pozzi+21,Sommovigo+22,Witstok+23,Valentino+24}. 

%-----------------------------------------------------------------------
%-----------------------------------------------------------------------

The efficiency of dust production of our galaxies can be assessed by comparing their $M_{\rm dust}$ and $M_{\rm stars}$ with dust production models. In Fig.~\ref{fig:Mdust-Mstars} the grey shaded zone defines the SN dust production as predicted by the model of \citet{GallHjorth18}, which assumes that the dust is produced during a star-formation episode lasting over $\Delta t$ with a mean SFR and with SNe (or their massive star progenitors\footnote{The model of \citet{GallHjorth18} does not specify how exactly the dust is produced in SNe or in massive stars prior the SN explosion.}) producing a fraction $\eta$ of solar masses of dust, with the rate of SNe being proportional to SFR with the proportionality factor $\gamma$; such that the produced dust mass can be expressed as $M_{\rm dust} = \gamma \eta {\rm SFR} \Delta t$. \citet{GallHjorth18} computed the dust productivity $\mu_{\rm dust} = \gamma \eta$ and derived $\mu_{\rm dust} = 0.004\pm 0.002$ for the \citet{Chabrier03} IMF. The product ${\rm SFR} \Delta t$ gives the stellar mass assembled during the star-formation episode, which for our galaxies is computed over a burst age of $\sim 10~\rm Myr$ as described in Sect.~\ref{sect:targets}. The SN dust production can hence be expressed as $M_{\rm dust} = (0.004\pm 0.002) M_{\rm stars}$. This simple model indicates that SNe (or massive stars) alone barely produce enough dust (without considering dust destruction) during the short burst timescale of our extreme starburst galaxies (see the grey shaded zone in Fig.~\ref{fig:Mdust-Mstars}) to enable to reproduce their dust masses. For four galaxies, despite their big and conservative $M_{\rm dust}$ uncertainties (Sect.~\ref{sect:continuum}), their $M_{\rm dust}$ measurements even clearly step outside the predicted SN dust production, unless a top-heavy IMF is invoked, or the (young) stellar masses of these galaxies are underestimated by a factor of up to $\sim 4$. On longer timescales, but relatively rapidly (for $\Delta t\gtrsim 40~\rm Myr$), AGB stars start to contribute to the dust production in addition to SNe \citep[][their Fig.~1]{SchneiderMaiolino24}, as delimited by the thick dash-dotted green line in Fig.~\ref{fig:Mdust-Mstars} derived by \citet[][following the simulations of \citet{Cesare+23}]{Witstok+23}. With respect to other dust formation models, only those by \citet{Imara+18} and \citet{Vijayan+19} predict the formation of enough dust to reproduce $M_{\rm dust}$ of all the low-$M_{\rm stars}$ galaxies in our sample; models of \citet{Popping+17} at $z\sim 2-3$ globally follow the same trend as the SN dust production of \citet{GallHjorth18}. 
%The dust masses of our extreme starburst galaxies support a relatively efficient dust production, especially given their redshifts and low $M_{\rm stars}$.

%-----------------------------------------------------------------------

\subsection{FIR dust continuum and rest-frame UV/optical sizes}
\label{sect:sizes}

\begin{figure}
\includegraphics[width=0.48\textwidth,clip]{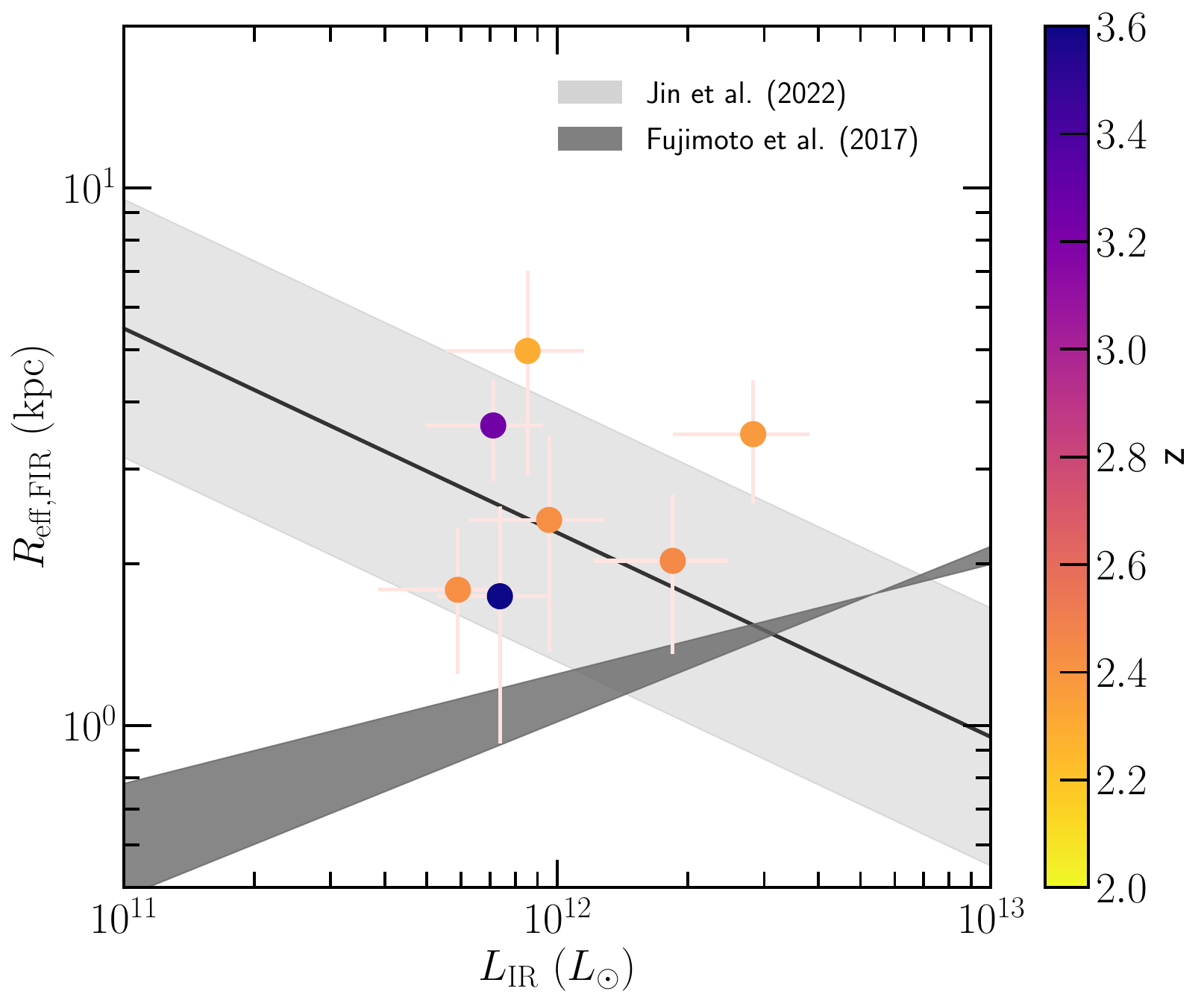}
\caption{Far-infrared dust continuum effective radius as a function of the IR luminosity of our galaxies, colour-coded by redshift. Our extreme starburst galaxies agree better with the $R_{\rm eff,FIR}$--$L_{\rm IR}$ anti-correlation (light-grey shaded area) found by \citet{Jin+22} than with the $R_{\rm eff,FIR}$--$L_{\rm IR}$ correlation (dark-grey shaded area) of \citet{Fujimoto+17}.}
\label{fig:R-LIR}
\end{figure}

The FIR dust continuum sizes were measured in the \textit{uv} plane for seven galaxies of our sample using the ALMA 1.3~mm band 6 observations (Sect.~\ref{sect:continuum}). They range between $R_{\rm eff,FIR} = 1.7$~kpc and 5.0~kpc (Table~\ref{tab:FIRdata}), and are shown as a function of $L_{\rm IR}$ in Fig.~\ref{fig:R-LIR}. The dependence of the FIR dust sizes of high redshift galaxies ($1<z<6$) on $L_{\rm IR}$ is still debated between, on one hand, the positive correlation (plotted as the dark-grey shaded area) reported by \citet[][based on a thousand of ALMA archival galaxies]{Fujimoto+17} that is in line with the stellar rest-frame UV size--luminosity correlation for star-forming galaxies \citep[e.g.][]{Shibuya+15}, and, on the other hand, the anti-correlation (light-grey shaded area) proposed by \citet[][based on a much smaller compilation of massive dusty galaxies, but with a better dynamical range towards lower $L_{\rm IR}$, from \citet{Valentino+20}, \citet{Franco+20}, and \citet{Gomez+22}]{Jin+22}. In both studies, the big $R_{\rm eff,FIR}$ $1\,\sigma$ scatter ($\sim 0.3-0.5~\rm dex$) at a given $L_{\rm IR}$ weakens the significance of the respective correlations. We observe that $R_{\rm eff,FIR}$ of our galaxies, on average, better agree with the relation of \citet{Jin+22}.

\begin{figure}
\includegraphics[width=0.48\textwidth,clip]{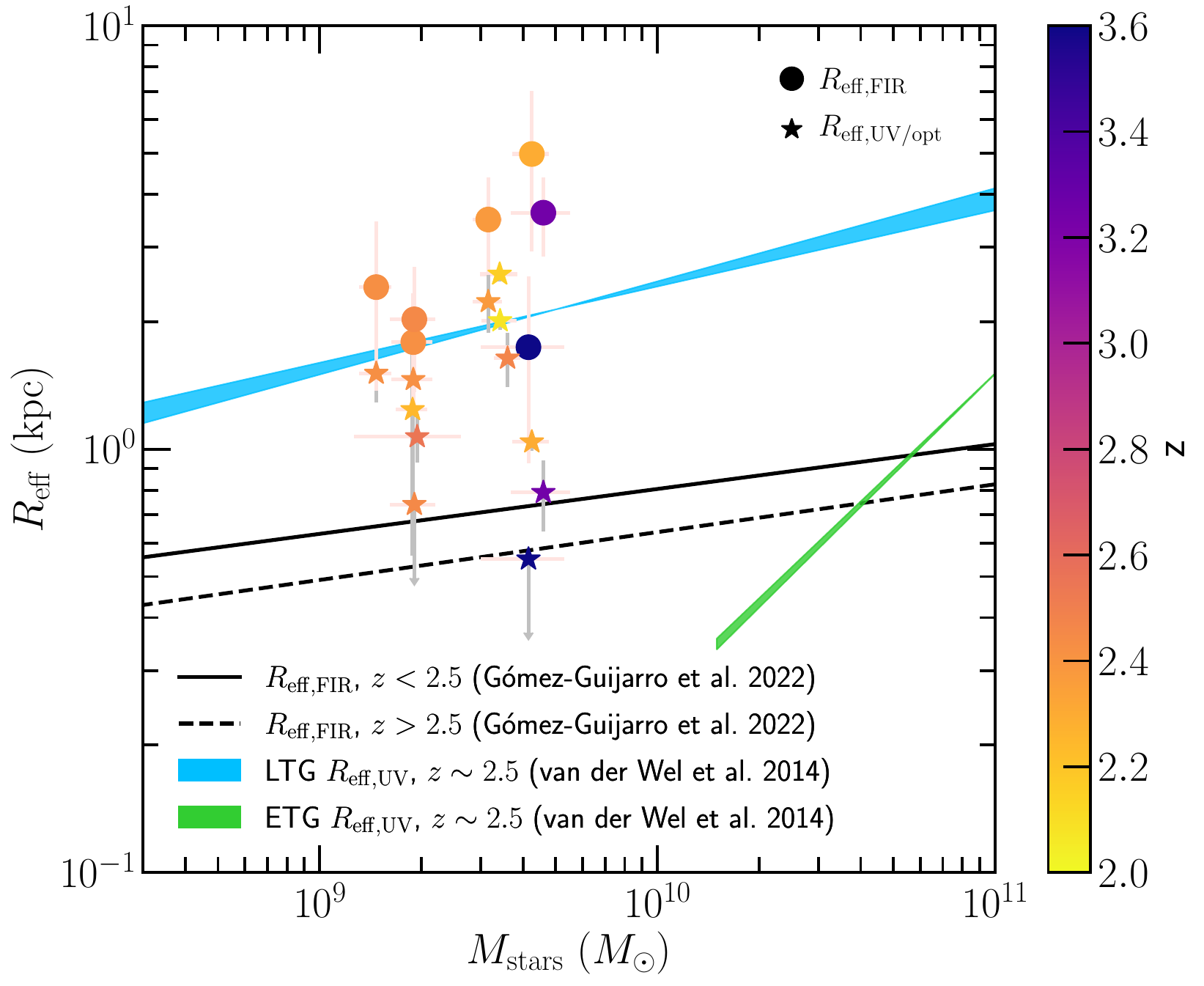}
\caption{Far-infrared dust continuum effective radius (filled circles) and rest-frame UV/optical effective radius (filled stars) as a function of the stellar mass of our galaxies, colour-coded by redshift. All our galaxies have systematically higher $R_{\rm eff,FIR}$ than $R_{\rm eff,UV/opt}$. As a result, $R_{\rm eff,FIR}$ of our galaxies strongly deviate from the $R_{\rm eff,FIR}$--$M_{\rm stars}$ relations of \citet{Gomez+22} at $z<2.5$ and $z>2.5$, shown by the solid and dashed black lines, respectively, as they are more extended. On the other hand, their $R_{\rm eff,UV/opt}$ agree with the $R_{\rm eff,UV}$--$M_{\rm stars}$ relation (blue shaded area) derived for LTG at $z\sim 2.5$ by \citet{Wel+14}. The green shaded area corresponds to the $R_{\rm eff,UV}$--$M_{\rm stars}$ relation of $z\sim 2.5$ early-type galaxies \citep[ETG;][]{Wel+14}.
}
\label{fig:R-Mstars}
\end{figure}

A dependence of $R_{\rm eff,FIR}$ on stellar mass was also reported for massive galaxies at $1.5<z<4.5$ by \citet{Gomez+22} as shown in Fig.~\ref{fig:R-Mstars} by the solid and dashed black lines at $z<2.5$ and $z>2.5$, respectively. Its slope is comparable to the slope of the stellar rest-frame UV effective radius versus $M_{\rm stars}$ relation derived for late-type galaxies (LTG) at similar redshifts $z\sim 2.5$ (blue shaded area) but its normalisation is much lower by a factor of $\sim 3-4$ \citep{Wel+14}. \citet{Fujimoto+17} also found that the FIR dust continuum sizes of high redshift galaxies are more compact than those at UV/optical wavelengths, albeit with a less different normalisation factor. Our galaxies have much higher $R_{\rm eff,FIR}$ (filled circles) with respect to the \citet{Gomez+22} relations, and in fact very much agree with the stellar rest-frame UV sizes of LTG at $z\sim 2.5$. TNG50 simulations from \citet{Popping+22} predict FIR dust continuum (850~$\mu$m rest-frame) sizes as a function of the stellar mass of MS galaxies from $z=1$ to $z=5$, which at $z\sim 2-3$ approximately match $R_{\rm eff,FIR}$ of our galaxies, but clearly overpredict the $R_{\rm eff,FIR}$--$M_{\rm stars}$ relations proposed by \citet{Gomez+22}.
%They find larger 850~$\mu$m sizes towards lower redshifts at fixed $M_{\rm stars}$, and conclude that the evolution of FIR dust continuum sizes is linked to the build-up of the galaxy disk across cosmic time.

In Fig.~\ref{fig:R-Mstars} we furthermore compare $R_{\rm eff,FIR}$ (filled circles) with the rest-frame UV or optical effective radii (filled stars) derived from our HAWK-I seeing-limited $H$-band or AO $K_{\rm s}$-band observations\footnote{For J0850+1549 and J0121+0025 we used the CFHT MegaCam and Subaru seeing-limited $R$-band observations, respectively.}, respectively (Sect.~\ref{sect:HAWKI} and Table~\ref{tab:FIRdata}). We observe that half of our galaxies have $R_{\rm eff,UV/opt}$ in agreement with those of LTG at $z\sim 2.5$ \citep{Wel+14}, and the other half is considerably more compact. However, globally, all our extremely UV-luminous galaxies have $R_{\rm eff,FIR}$ systematically higher than their $R_{\rm eff,UV/opt}$. As a result, the mean $R_{\rm eff,UV/opt}/R_{\rm eff,FIR} =0.43\pm 0.17$ ratio of our galaxies contrasts with what is reported in the literature for galaxies at comparable redshifts $1<z<4$, $R_{\rm eff,HST}/R_{\rm eff,ALMA} \sim 1.6$ and $\sim 2.4$ for, respectively, ALMA archival galaxies from \citet{Fujimoto+17} and massive galaxies from \citet{Franco+20}. Our galaxies actually share one additional physical property in common with very high redshift UV-selected galaxies at $4.5<z<6$, because these latter are also characterised by relatively extended FIR dust continuum sizes with respect to the compact UV sizes ($\lesssim 1.5$~kpc) with $R_{\rm eff,HST}/R_{\rm eff,ALMA} = 0.39\pm 0.15$ as recently measured by \citet[][see also \citet{Mitsuhashi+24}]{Pozzi+24}.

%-----------------------------------------------------------------------
%-----------------------------------------------------------------------

\subsection{Rest-frame UV/optical morphology and spatial offsets}
\label{sect:UVmorphology}

As shown in the middle-left panels of Figs.~\ref{fig:contB6-CO_1}, \ref{fig:contB6-CO_2}, and \ref{fig:contB6-CO_3}, most of our extremely UV-luminous galaxies are characterised by a simple `roundish' rest-frame UV or optical morphology with one main bright component with $R_{\rm eff,UV/opt}$ ranging from 0.79~kpc to 2.01~kpc (Table~\ref{tab:FIRdata}). Two galaxies, J0950+0523 and J1316+2614, are particularly compact with $R_{\rm eff,opt} < 0.74$~kpc and $< 0.55$~kpc, respectively, unresolved in the HAWK-I AO $K_{\rm s}$-band images. Only two galaxies, J0146--0220 and J0006+2452, show more complex rest-frame optical morphologies with one main bright and compact component with $R_{\rm eff,opt} < 0.8$~kpc (unresolved in HAWK-I AO $K_{\rm s}$-band observations), centred on the phase centre, and a more diffuse component extended over $>2$~kpc in radius and offset in one direction with respect to the main component. A second tiny component seems also to be present on top of the diffuse component in J0146--0220. The detected rest-optical diffuse  emission can be interpreted as a possible signature of either a diffuse underlying old stellar population or diffuse outflowing gas as the HAWK-I $K_{\rm s}$-band emission can instead trace the nebular H$\alpha$ emission at the redshift of these two galaxies.
%These features are possible signatures of either a diffuse underlying old stellar population, or diffuse outflowing gas as the HAWK-I $K_{\rm s}$-band is contaminated by the nebular H$\alpha$ emission at the redshifts of these 2 galaxies, or a merger companion specifically in J0146--0220. 
%both HAWK-I $H$- and $K_{\rm s}$-bands are contaminated by the nebular [O\,{\sc III}]+H$\beta$ and H$\alpha$ emission, respectively, at the redshifts of our galaxies, 

The recently acquired HST images of J1415+2036, J0850+1549, and J1316+2614 confirm that the rest-frame UV morphology of our extreme starburst galaxies is a mixed bag between one single component for J1316+2614 \citep{Marques+24} and multi-components with J1415+2036 and J0850+1549 showing, respectively, two and three very compact components (Marques-Chaves et~al.\ in prep.) that are blended in ground-based observations (Figs.~\ref{fig:contB6-CO_1} and \ref{fig:contB6-CO_2}, middle-left panels). 
%This suggests that the measured ground-based $R_{\rm eff,UV/opt}$ (Table~\ref{tab:FIRdata}) represent the total galaxy sizes, while 
The HST images also reveal that the extreme $L_{\rm UV}$ we measure mostly comes from a very compact component, which is barely resolved even at the HST resolution. 
%with $R_{\rm eff,UV}^{\rm HST} \lesssim 360$~pc (at the average $z=2.5$). 
The galaxy J1316+2614 appears as the most extreme case in our sample, because it is composed of one single very compact HST component with an effective radius as small as $R_{\rm eff,UV}^{\rm HST} = 220\pm 12$~pc, co-spatial with the extended FIR dust continuum emission, and without any sign of underlying HST diffuse emission \citep{Marques+24}. 
%With its $M_{\rm UV} = -24.65$, this component outshines by far the absolute UV magnitudes of the brightest stellar clusters known in the Sunburst ($M_{\rm UV} \sim -19$) at $z\simeq 2.4$ and Sunrise ($M_{\rm UV} \sim -18$) at $z\simeq 6$ galaxies \citep{Vanzella+22,Vanzella+23}.

%-----------------------------------------------------------------------
%-----------------------------------------------------------------------

In both J0146--0220 and J0006+2452 galaxies, the FIR dust continuum emission is co-spatial with the CO emission, but is offset from the main rest-optical component, whereas it aligns with the extended diffuse rest-optical emission. The offsets between the main rest-optical component and the FIR emission are $0.66\arcsec$ and $0.59\arcsec$, that is, about 5.5~kpc and 4.9~kpc, in J0146--0220 and J0006+2452, respectively. The possible origin of this offset is discussed in Sect.~\ref{sect:discussion}. J1249+1550 and J0850+1549 also seem to have their FIR emission slightly offset by $\lesssim 0.25-0.3\arcsec$ from the bright rest-optical and rest-UV emission, respectively, but these offsets remain within the synthesised beam and PSF uncertainties and need to be confirmed.
%J1249+1550 also has its FIR and CO emissions slightly offset from the bright rest-optical emission, although an underlying diffuse component is barely detected at the sensitivity of the available HAWK-I AO $K_{\rm s}$-band observations in the direction of the FIR emission. The FIR emission of J0850+1549 as well seems to present a small offset ($\lesssim 0.24\arcsec$) from the rest-UV bright emission although observed at the limited CFHT MegaCam resolution of $0.71\arcsec$. The measured respective offsets between the main rest-optical component and the FIR emission are $0.66\arcsec$, $0.59\arcsec$, and $0.28\arcsec$, i.e., about 5.5~kpc, 4.9~kpc, and 2.3~kpc, respectively, in J0146--0220, J0006+2452, and J1249+1550. 
Thus, for seven out of nine galaxies one can assume that the UV/optical bright component is co-spatial with the more extended FIR emission and the CO emission (when detected) as shown in the middle-left and middle-right panels of Figs.~\ref{fig:contB6-CO_1}, \ref{fig:contB6-CO_2}, and \ref{fig:contB6-CO_3}.

J1415+2036 and J0121+0025 show patchy CO emission peaks at $4-5\sigma$ significance level spread within $< 3\arcsec$ (i.e. $< 25$~kpc) around the phase centre (Figs.~\ref{fig:contB6-CO_1} and \ref{fig:contB6-CO_3}, middle-right panels). Their integrated CO line detections nevertheless seem relatively robust, especially for J1415+2036, as shown by their respective CO line spectra plotted in the right panels of Figs.~\ref{fig:contB6-CO_1} and \ref{fig:contB6-CO_3}. However, how this patchy CO emission with its relatively big spatial extent relates to the rest-frame UV/optical emission remains unclear, especially in J0121+0025 where the escaping LyC radiation \citep{Marques+21} needs to be reconciled with the presence of cold molecular gas. The other leaker, J1316+2614 \citep{Marques+22}, is undetected in the CO emission with a stringent upper limit on its very depleted $M_{\rm molgas}<6.3\times 10^9~M_{\odot}$ (Fig.~\ref{fig:Mmol-SFR}).

%and, this even more, as both J1415+2036 and J0121+0025 have been proven to be LyC leakers \citep[Marques-Chaves et~al.\ in prep.;][]{Marques+21}, and that cold/molecular gas cannot readily be reconciled with LyC radiation. The two other leakers, J0850+1549 (Marques-Chaves et~al.\ in prep.) and J1316+2614 \citep{Marques+22}, are undetected in the CO emission with stringent upper limits on their very depleted $M_{\rm molgas}< 7.5\times 10^9~M_{\odot}$ and $<6.3\times 10^9~M_{\odot}$, respectively (Table~\ref{tab:CO-dust-properties} and Fig.~\ref{fig:Mmol-SFR}).

%-----------------------------------------------------------------------

\section{Discussion}
\label{sect:discussion}

The studied sample consists of the 12 most UV-bright galaxies known at $z\simeq 2.1-3.6$ with $M_{\rm UV} = -23.40$ to $-24.65$ (Sect.~\ref{sect:targets} and Table~\ref{tab:restUV-properties}), surpassing by $2-3$ magnitudes the recently   detected JWST UV-luminous galaxies at $7<z<16$ brighter than $M^{\star}_{\rm UV}$ \citep[e.g.][]{Bouwens+23,Atek+23,Bunker+23,Casey+23,Castellano+24,Carniani+24}. They also outshine by far the brightest stellar clusters known in the Sunburst ($M_{\rm UV} \sim -19$) at $z\simeq 2.4$ and Sunrise ($M_{\rm UV} \sim -18$) at $z\simeq 6$ galaxies \citep{Vanzella+22,Vanzella+23}. They are dominated by very young stellar populations of $\sim 10$~Myr (as assessed from the rest-frame UV spectroscopy) and are powerful starbursts with a mean specific SFR (${\rm sSFR = SFR}/M_{\rm stars}$) of $\rm 112~Gyr^{-1}$ ($\pm 45~\rm Gyr^{-1}$), placing them $\sim 1.5$~dex above the MS for their redshifts and stellar masses ($M_{\rm stars} \sim (1.47-4.59)\times 10^9~M_{\odot}$). They show unattenuated starlight with blue UV spectral slopes between $\beta_{\rm UV} = -2.62$ and $-1.84$, similar to those observed in several UV-bright galaxies at $z>7$ \citep{RobertsBorsani+24,Heintz+24}. They present complex gas kinematics showing signatures of outflows and inflows. Furthermore, they are not only very efficient at producing ionising radiation given their luminous $M_{\rm UV}$, but some of them leak high amounts of LyC photons \citep{Marques+21}, including the record-holder J1316+2614 at $z=3.6122$ with a LyC escape fraction of $f_{\rm esc}{\rm (LyC)} \simeq 90\%$ \citep{Marques+22}. In summary, our galaxies possibly experience very early stages of vigorous starbursts with a fast mode of stellar mass build-up, likely resembling the major bursts of star formation taking place in the UV-luminous galaxies recently discovered with JWST at the EoR. 

The ALMA observations, analysed in this paper, furthermore reveal the relatively bright FIR dust continuum emission for nine of our galaxies, placing these extremely UV-luminous galaxies in the LIRG regime, and even the ULIRG regime for three galaxies, with $L_{\rm IR} = (5.9-28.3)\times 10^{11}~L_{\odot}$ (Table~\ref{tab:FIRdata}). All galaxies in the LIRG regime remain nevertheless dominated by the unobscured $\rm SFR_{UV}$, and half of them even have $f_{\rm obscured} < 25$\%, much lower than UV-selected galaxies at $z\sim 4.5-7.7$ (Sect.~\ref{sect:obscured-SFR}). The resulting IRX ratios together with the blue $\beta_{\rm UV}$ best match the IRX--$\beta_{\rm UV}$ relations determined for young stellar populations (Fig.~\ref{fig:IRX-betaUV}), in line with their very young ages of $\sim 10$~Myr derived from rest-frame UV spectra. 
%On the other hand, the inferred relatively large dust masses of $M_{\rm dust} = (2.3-11.1)\times 10^7~M_{\odot}$ (Table~\ref{tab:CO-dust-properties} and Sect.~\ref{sect:Mdust}), contrast with their very blue unattenuated starlight. While the presence of dust cannot be contested in these extreme UV-bright galaxies, we, however, need to keep in mind that the derived $M_{\rm dust}$ remain relatively uncertain since constrained from single band continuum measurements at 1.3~mm. 
%and the large LyC leakage confirmed for some of these galaxies \citep{Marques+21,Marques+22}. 

Large amounts of CO molecular gas in eight of our galaxies are also shown by ALMA observations, with much higher (up to an order of magnitude) molecular gas masses than their stellar masses, yielding very high $f_{\rm molgas}$ between 63\% and 92\% (Table~\ref{tab:COdust-properties} and Sect.~\ref{sect:Mgas-tdepl}). In what follows we refer to the $M_{\rm molgas}$ measurements derived using the CO-to-H$_2$ conversion factor  $\alpha_{\rm CO}^{\rm SB}$, but our conclusions do not change when considering the $\sim 4$ times higher $\alpha_{\rm CO}^{\rm MW}$ instead. $\epsilon_{\rm SF}$ reliably estimated for our galaxies thanks to their dominant very young ($\sim 10$~Myr) stellar populations, show that very efficient star formation is taking place in our galaxies with $\epsilon_{\rm SF}$ values reaching up to $\sim 40$\% (Sect.~\ref{sect:SFE}). Those are likely even higher if only part of the detected $M_{\rm molgas}$ is associated with the star formation ongoing in the compact UV-bright component of our galaxies. This can, in particular, be the case for galaxies showing spatial offsets between the main UV-bright component and CO emission, currently evidenced in J1415+2036 and J0121+0025
%and then in J0146--0220, J0006+2452, and J1249+1550 
(Sect.~\ref{sect:UVmorphology}). The high $\epsilon_{\rm SF}$ are a unique physical property characterising our galaxies together with their amazingly short $t_{\rm depl}$ between 36~Myr and 71~Myr (Table~\ref{tab:COdust-properties} and Sect.~\ref{sect:Mgas-tdepl}), which highlight a very efficient and fast conversion of gas into stars that can only result from the gas collapse within a very short free-fall time. 

In their model, \citet{Dekel+23} suggest the formation of feedback-free starbursts (FFB) through the collapse of gas clouds within very short free-fall times, approaching the monolithic collapse model, in high density ($> 10^3~\rm cm^{-3}$ and $> 2\times 10^3~M_{\odot}~\rm pc^{-2}$) and low metallicity ($Z\leq 0.2~Z_{\odot}$) environments. This makes star formation very efficient, because the cloud collapse occurs before the onset of stellar winds and SNe feedback, allowing the formation of massive galaxies within a few millions of years; and this also makes galaxies appear blue with none to little dust attenuation. A significant fraction of gas mass ($\sim 80$\%) is predicted to be in massive gas clouds shielded against feedback and which can hence participate in the FFB star formation. 
%permitting high $\epsilon_{\rm SF}$. 
%both in line with the $f_{\rm molgas}$ and $\epsilon_{\rm SF}$ measurements of some of our extreme starburst galaxies. 
More specifically, \citet{Li+24} showed that globally the star-formation history (SFH) in an FFB galaxy occurs in several generations separated by about 10~Myr, which consist of a peak of nearly simultaneous FFB starbursts lasting for about $2-5$~Myr and ending in low amounts of gas left over, followed by a period of gas accumulation until the onset of the following generation of bursts. The dispersion of the $f_{\rm molgas}$ and $\epsilon_{\rm SF}$ measurements could be linked to the different SFH evolutionary stages of our galaxies following this model.

The stellar mass surface densities of our galaxies, defined as $\Sigma M_{\rm stars} = M_{\rm stars}/(2\pi R_{\rm eff}^2)$, range between $\Sigma M_{\rm stars} = 80~M_{\odot}~\rm pc^{-2}$ and $1170~M_{\odot}~\rm pc^{-2}$ when measured with the ground-based rest-frame UV/optical effective radii listed in Table~\ref{tab:FIRdata}. %representing the total sizes of the galaxies. 
However, as discussed in Sect.~\ref{sect:UVmorphology}, the extreme $L_{\rm UV}$ of our galaxies emerge from a compact component which is unresolved in the HAWK-I images (even with AO; $R_{\rm eff,opt} < 0.8$~kpc) and is barely resolved in the recently acquired HST images with $R_{\rm eff,UV}^{\rm HST} = 220\pm 12$~pc measured in J1316+2614 \citep{Marques+24}. 
%as well as in the recently acquired HST images with $R_{\rm eff,UV}^{\rm HST} \lesssim 360$~pc, on average (Marques-Chaves et~al.\ in prep.). 
The derived $M_{\rm stars}$ also come from these compact UV-bright components (see Sect.~\ref{sect:targets}), and represent $>30$\% to $>62$\% of the total stellar mass of our galaxies \citep{Marques+20b,Marques+21,Marques+22,Marques+24}. As a result, the starbursting UV-bright components\footnote{The corresponding $\rm SFR_{UV}$ surface densities of the UV-bright components are also huge with $\Sigma {\rm SFR_{UV}} \sim (0.27-1.1)\times 10^3~M_{\odot}~\rm yr^{-1}~kpc^{-2}$. Only a few massive stellar clusters at sub-10 pc scales, hosted in the Sunburst and Sunrise high redshift galaxies, have such extreme $\Sigma {\rm SFR_{UV}}$ \citep{Vanzella+22,Vanzella+23,Messa+24}.}, for an average $R_{\rm eff,UV}^{\rm HST} \sim 250$~pc, have very high $\Sigma M_{\rm stars} \sim (0.37-1.2)\times 10^4~M_{\odot}~\rm pc^{-2}$\footnote{Such high $\Sigma M_{\rm stars}$ were only reported for very few star-forming galaxies with similar $M_{\rm stars}$ and redshifts, and are more common for more massive compact star-forming galaxies \citep{Barro+17}, as well as for young star-forming clumps at $1<z<3$ yet typically having two orders of magnitude lower stellar masses \citep[e.g.][]{Claeyssens+23,Messa+24}.} well in line with the surface density threshold required for an FFB to occur\footnote{For one of our galaxies, J1220+0842, \citet{Marques+20b} could even measure its density from the [O\,{\sc ii}]\,$\lambda$3729/$\lambda$3726 and C\,{\sc iii}] $\lambda$1906/$\lambda$1908 line ratios that is $>10^3~\rm cm^{-3}$, namely in agreement with the FFB density threshold.} \citep{Dekel+23}. Most of the physical properties of our extremely UV-bright galaxies at $z\sim 2.5$ actually seem to be naturally explained in the context of the FFB model, and this despite their $Z \sim 0.5~Z_{\odot}$ metallicities \citep{Upadhyaya+24} while the FFB model was so far validated at $Z\leq 0.2~Z_{\odot}$. However, the measured metallicities should reflect the chemical enrichment of the starbursts over the past $\sim 10$~Myr, potentially differing from the gas metallicity in the pre-FFB phase.

What furthermore needs to be understood is the relatively bright FIR dust continuum emission detected in nine of our galaxies (Sect.~\ref{sect:Mdust}), and which we find co-spatial with the main rest-frame UV/optical component in seven galaxies (Sect.~\ref{sect:UVmorphology}) in contrast with their very blue ($\beta_{\rm UV} < -1.84$) unattenuated starlight. A possible simple explanation could be geometrical, such that the UV-bright emission is located in front of the dust layer, hence appearing unaffected by dust obscuration. However, this specific geometrical alignment appears surprisingly too common given its high occurrence for seven out of nine galaxies. 

Another interpretation could follow the \citet{Ferrara+23} model \citep[see also][]{Ziparo+23,Ferrara24} which suggests a temporary removal of dust (and gas) cleared by radiation-driven outflows that develop when galaxy luminosities become super-Eddington, making galaxies appear UV brighter and bluer with very low dust attenuation ($A_{\rm V} \lesssim 0.02$), and having an impact on the occurrence of huge LyC leakage. All our galaxies  (see Table~\ref{tab:restUV-properties}) can potentially develop such radiation-driven outflows given their sSFR well above the super-Eddington $\rm sSFR > 25~Gyr^{-1}$ threshold\footnote{The super-Eddington limit from \citet{Ferrara24} was reassessed by \citet{Li+24}, who found that super-Eddington ejections potentially take place at $z>16$ only.} computed by \citet{Ferrara24}, in contrast to MS galaxies that reach this super-Eddington limit only at very high redshifts of $z>7$ \citep{Faisst+16,Bouwens+22b,Bradley+23,Stefanon+22,Stefanon+23}. 

Evidence of outflowing gas was reported in two of our galaxies so far, J1220+0842 \citep{Alvarez+21} and J0121+0025 \citep{Marques+21}, from ISM absorption lines showing blueshifted centroids with respect to the systemic redshift by $-400~\rm km~s^{-1}$ and $-450~\rm km~s^{-1}$, respectively, in line with their Ly$\alpha$ profiles. The Ly$\alpha$ profiles of two other galaxies, J1249+1550 and J0850+1549, also support outflows. And, evidence of dust being possibly blown away in our galaxies
%expelled out of the UV-bright component of our galaxies 
is maybe best highlighted in J0146--0220 and J0006+2452,
%J0146--0220, J0006+2452, and J1249+1550, where the FIR emission is spatially offset from the UV-bright component along with the diffuse rest-optical emission detected for 2 of them (Sect.~\ref{sect:UVmorphology}). 
where the FIR emission is spatially offset from the main compact rest-optical component along with the diffuse rest-optical emission (Sect.~\ref{sect:UVmorphology}). For the seven other galaxies with co-spatial FIR dust continuum emission, we can speculate that the dust is expelled out spherically creating a `hole' in the FIR emission at the location of the UV-bright component and with a size of at least the UV-bright component's radius ($R_{\rm eff,UV}^{\rm HST} \sim 250$~pc; see above), currently unresolved at the angular resolution of our ALMA observations. The recent discovery of the Ly$\alpha$ hole in J1316+2614 at the location of the stellar UV emission while the Ly$\alpha$ emission extends out to $\sim 6$~kpc \citep{Marques+24} provides support to a comparable FIR dust emission hole, because dust should follow the same distribution as the Ly$\alpha$ emission that is tracing neutral gas. Moreover, the particularly big $R_{\rm eff,FIR}>1.7$~kpc (higher than $R_{\rm eff,UV/opt}$) of the FIR dust emission of our galaxies (Table~\ref{tab:FIRdata}) are also in line with dust being pushed beyond the radius of $\sim 2-3$~kpc at which it becomes optically thin \citep[as predicted by][]{Ferrara24}. Same is likely true for the gas that is also expected to be removed by outflows, and the CO emission is indeed found co-spatial with the FIR dust emission (Figs.~\ref{fig:contB6-CO_1}, \ref{fig:contB6-CO_2}, and \ref{fig:contB6-CO_3}). Nevertheless, we do not observe any evidence of outflowing molecular gas in any of the currently available CO line profiles (no broad CO line component is detected; see Figs.~\ref{fig:contB6-CO_1}, \ref{fig:contB6-CO_2}, and \ref{fig:contB6-CO_3}, right panels).
%It is yet unclear in the currently available ALMA data if the observed complex CO line profiles (Sect.~\ref{sect:CO}) result from the outflowing molecular gas or not.

\citet{Menon+24} also demonstrated in their radiation hydrodynamical numerical simulations that localised starbursts can evacuate (local) surrounding dust/gas through radiation-driven outflows in very high stellar/gas surface density conditions ($\Sigma > 10^3~M_{\odot}~\rm pc^2$) in line with the $\Sigma M_{\rm stars}$ measured for our starbursting UV-bright components (see above). Their results are slightly different from the \citet{Ferrara+23} model which invokes a global galaxy-scale radiation-driven outflow, whereas their simulations suggest a picture where `localised' outflows (at tens of parsecs scale) enable the UV-emitting stellar populations to be visible, while other regions can still be obscured, resulting in co-spatial UV/FIR emission when viewed with large apertures.

The origin of the dust detected in our galaxies remains to be explained. Dust could be produced by SNe at the end of the FFB phase, or have an independent origin before the onset of the FFB. As discussed in Sect.~\ref{sect:Mdust}, the inferred dust masses (Table~\ref{tab:COdust-properties}) -- however relatively uncertain since derived from single band continuum measurements at 1.3~mm -- can barely be produced by SNe (or massive stars) during the $\sim 10$~Myr starburst phase \citep{GallHjorth18} for most of our galaxies and are clearly in excess in four galaxies,
%are larger {\bf for four galaxies} than what SNe (or massive stars) can produce during a $\sim 10$~Myr starburst phase \citep{GallHjorth18},
%and other contributions (e.g.~from AGB stars) are therefore needed 
at least with the standard \citet{Chabrier03} IMF. This suggests that some dust was probably already produced prior to the burst of star formation by an older stellar population (from AGB stars and previous SNe). However, the evidence of an old stellar population currently remains elusive. So far, the SED analysis supports the absence of significant old stellar populations in these galaxies \citep{Marques+20b,Marques+21,Marques+22,Marques+24}, and the nature of the underlying diffuse rest-frame optical emission detected with HAWK-I in at least two galaxies, J0146--0220 and J0006+2452, still needs to be understood (Sect.~\ref{sect:UVmorphology}). 
%between a stellar or nebular H$\alpha$ contribution. 

Other dust production mechanisms can be invoked, such as the rapid grain growth in the galaxy ISM \citep{Graziani+20,Dayal+22,SchneiderMaiolino24} that would bypass the need of an old stellar population. Moreover, recently \citet{Higgins+23} reported that VMSs with masses of some $200~M_{\odot}$ can eject 100 times more heavy elements in their winds than $50~M_{\odot}$ stars and hence possibly produce more dust. Evidence of VMSs in some of our extreme starburst galaxies is obtained from the detection of intense and broad He\,{\sc ii}\,$\lambda$1640 emission lines, with equivalent widths of $EW_0 = 3-5~\AA$ \citep{Upadhyaya+24,Marques+20b,Marques+21,Marques+22}, that is, much stronger than that typically found in normal star-forming galaxies \citep[$EW_0 \simeq 1~\AA$;][]{Shapley+03}. Such high $EW_0$ are difficult to explain by standard stellar models with an IMF upper mass cut-off $M_{\rm upp} = 100~M_{\odot}$, but are reproduced well by evolutionary models and atmospheres of VMSs with masses from $100~M_{\odot}$ to $400~M_{\odot}$ \citep{Martins+22}. If such populations including VMSs can produce sufficient amounts of dust as those observed in our galaxies remains to be worked out.

\citet{Marques+20b} speculated on the fate of these extreme starburst galaxies and proposed two scenarios. Either these are the progenitors of compact present-day ellipticals \citep[the so-called red nuggets; e.g.][]{DekelBurkert14} in case the star formation quenches right after the starburst phase, or these are ultra-/hyper-luminous dusty star-forming galaxies \citep[DSFGs;][]{Santini+10} caught in their very early evolutionary stage ($\sim 10$~Myr), before their dust-poor extremely UV-luminous phase gets obscured by dust, in case the intense SFR activity lasts over a few hundreds of millions of years if located in gas-rich environments. The $t_{\rm depl}$ measurements derived in this work now provide a constraint on the gas consumption timescales of our galaxies. They highlight that most of our galaxies have enough gas to sustain the SFR over a few dozens of millions of years, that is, a factor of around three to seven longer than the current $\sim 10$~Myr age of the burst. As a result, over these timescales they will more than triple their $M_{\rm stars}$, and produce large amounts of dust by more than tripling their $M_{\rm dust}$ just from the SN dust production \citep[reaching $M_{\rm dust}/M_{\odot} \gtrsim 10^8-10^9$ according to][]{GallHjorth18}. They could hence turn into FIR-bright DSFGs, heavily obscured in the UV. On the other hand, galaxies with $t_{\rm depl}$ comparable to the burst age ($\sim 10$~Myr), and especially J1316+2614 with $t_{\rm depl} < 13$~Myr, are expected to quench rapidly by starvation even before reaching the SMG phase, likely representing the very initial phases in the evolution of massive quiescent galaxies. 

\section{Summary and conclusions}
\label{sect:conclusions}

In this study, we performed an analysis of the ALMA FIR 1.3~mm (240~GHz) dust continuum and the CO(3--2) or CO(4--3) emission of 12 starburst galaxies at $z=2.08-3.61$ selected for their extreme brightness in the rest-frame UV with absolute magnitudes  $M_{\rm UV} = -23.4$ to $-24.7$. 
%even brighter than the UV-bright galaxies recently unveiled by JWST at much higher redshifts $7<z<16$ \citep[e.g.,][]{Casey+23,Bunker+23,Castellano+24,Carniani+24}. 
We also analysed the HAWK-I seeing-limited $H$-band and AO $K_{\rm s}$-band images acquired for ten targets. The rest-frame UV and optical properties of seven of these targets have already been analysed in detail in previous works \citep{Marques+20b,Marques+21,Marques+22,Marques+24,Alvarez+21,Upadhyaya+24}. All the galaxies are characterised by steep UV spectral slopes $\beta_{\rm UV}=-2.62$ to $-1.84$ and dominated by very young stellar populations of $\sim 10$~Myr, indicating that the bulk of their stellar mass $M_{\rm stars} = (1.47-4.59)\times 10^9~M_{\odot}$ was formed in an intense starburst phase over $\sim 10$~Myr given the absence of relevant old stellar populations. They have sub-solar metallicities ranging between $12+\log({\rm O/H}) = 8.13$ and 8.49. Signatures of very massive stars with masses from $100~M_{\odot}$ to $400~M_{\odot}$ are revealed in their rest-frame UV spectra \citep{Upadhyaya+24} from the strong and broad He\,{\sc ii}\,$\lambda$1640 emission line \citep[e.g.][]{Crowther+16,Martins+23,Schaerer+24}, suggesting an IMF with possibly a higher upper mass cut-off and likely also boosting their $L_{\rm UV}$ \citep{Schaerer+24}. Several targets in our sample present complex gas kinematics showing evidence of outflows and even inflows for one galaxy (J1316+2614). They are all strong producers of ionising radiation given their very luminous $M_{\rm UV}$, and two of them are identified as strong LyC leakers (J0121+0025 and J1316+2614). Placed at the EoR, these galaxies alone would be able to ionise their environment.

Detected in nine galaxies, the ALMA FIR dust continuum emission provides constraints on their IR luminosities, obscured $\rm SFR_{IR}$, and dust masses, and the CO emission detected in eight galaxies provides constraints on their molecular gas masses:

\begin{itemize}
\item With $L_{\rm IR} = (5.9-28.3)\times 10^{11}~L_{\odot}$ our galaxies populate the LIRG regime, and even the ULIRG regime for three galaxies. All the LIRG-type galaxies remain dominated by the unobscured $\rm SFR_{UV}$. For half of the galaxies, the obscured fractions of star formation are less than $25\%$, placing them a factor of two below the empirical $f_{\rm obscured}$--$M_{\rm stars}$ relation of \citet{Whitaker+17}, that is, even below the UV-selected galaxies at $z\sim 4.5-7.7$ \citep{Fudamoto+20b,Algera+23}. 

\item The resulting total $\rm SFR_{UV+IR}$ bring these UV-bright galaxies considerably above (a factor of $\sim 30$) the main sequence of star-forming galaxies for similar redshifts and stellar masses \citep[e.g.][]{Speagle+14}, confirming they are powerful starbursts with specific SFRs as high as $\rm sSFR = 48-191~Gyr^{-1}$.

\item The blue UV spectral slopes and luminous $L_{\rm UV}$, implying $\rm IRX<+0.5$, shift our extreme starburst galaxies away from the canonical IRX--$\beta_{\rm UV}$ relation of \citet{Meurer+99} to relations expected for younger and lower metallicity stellar populations \citep{Reddy+18}, in agreement with their $\sim 10$~Myr ages derived from rest-frame UV spectroscopy and sub-solar metallicities. Our galaxies are among the few with individual IRX measurements at their low stellar masses. They populate the IRX--$M_{\rm stars}$ parameter space between the relations found for main sequence galaxies at $z\sim 1.5-4$ when extrapolated to lower $M_{\rm stars}$ \citep[e.g.][]{Bouwens+16,Fudamoto+20a} and agree best with the relation derived for very high redshift UV-selected galaxies at $z\sim 4.5-7.7$ \citep{Bowler+24}, but they definitely disagree with starburst galaxies at comparable redshifts that have a much higher IRX at fixed $M_{\rm stars}$ \citep{Fudamoto+20a}. 

\item The high dust masses $M_{\rm dust} = (2.3-11.1)\times 10^7~M_{\odot}$ of our galaxies (to be confirmed with a better FIR SED sampling) support an efficient dust production given their $\sim 10$~Myr ages. Indeed, SNe alone barely produce such dust amounts during this short timescale \citep{GallHjorth18} and even fail for four galaxies unless a top-heavy IMF is invoked. Other dust production mechanisms can be invoked, for instance the rapid grain growth in the ISM, VMSs, or AGB stars in the presence of an old stellar population that remains elusive \citep{SchneiderMaiolino24,Higgins+23}. 

\item The dust-to-gas mass ratios of our UV-bright galaxies range between $<10^{-2.9}$ and $10^{-2.1}$ (for $\alpha_{\rm CO}^{\rm SB}$) and are consistent with different $\delta_{\rm DGR}$--$12+\log({\rm O/H})$ relations when extrapolated to their sub-solar metallicities \citep[e.g.][]{Popping+23}.

\item Our extreme starburst galaxies are offset from the integrated KS $M_{\rm molgas}$--$\rm SFR_{UV+IR}$ relation of main sequence galaxies and match the starburst regime as defined by \citet{Sargent+14}. Their high $\rm SFR_{UV+IR}$ with respect to their $M_{\rm molgas}$ give unprecedentedly short molecular gas depletion timescales between $t_{\rm depl} <13$~Myr and 71~Myr (for $\alpha_{\rm CO}^{\rm SB}$). On the other hand, their baryonic masses remain dominated by the cold molecular gas over stars, which results in very high molecular gas mass fractions between $f_{\rm molgas}  <60\%$ and 92\% (for $\alpha_{\rm CO}^{\rm SB}$), in significant excess with respect to those of main sequence galaxies \citep[e.g.][]{Tacconi+20}. This suggests that these extreme starburst galaxies are caught at the very beginning of their intense starburst phase while vigorously consuming their still high molecular gas mass reservoir to build up their stellar mass.

\item Dominated by very young stellar populations of $\sim 10$~Myr, we obtained direct star-formation efficiency measurements of our galaxies ranging from $\epsilon_{\rm SF} = 8$\% to $>40\%$ (for $\alpha_{\rm CO}^{\rm SB}$); these are much higher than $\epsilon_{\rm SF} < 5$\% measured in nearby galaxies \citep[e.g.][]{Kim+23}. The derived $\epsilon_{\rm SF}$ should even be considered as lower limits given that only a percentage of the detected CO gas may be involved in the ongoing star formation. These high $\epsilon_{\rm SF}$, together with very short $t_{\rm depl}$, highlight the amazingly efficient and fast conversion of gas into stars of these UV-bright galaxies.

\item The FIR dust continuum effective radii of our galaxies range between $R_{\rm eff,FIR}=1.7$~kpc and 5~kpc, and are bigger than their $R_{\rm eff,UV/opt}$. The corresponding mean $R_{\rm eff,UV/opt}/R_{\rm eff,FIR} = 0.43\pm 0.17$ is in contrast to what is typically observed for galaxies at comparable redshifts \citep{Fujimoto+17} but in agreement with the size properties of very high redshift UV-selected galaxies at $4.5<z<6$ \citep{Mitsuhashi+24,Pozzi+24}.

\item Our extremely UV-luminous galaxies are mostly characterised by a simple rest-frame UV/optical morphology, as assessed from HAWK-I observations, composed of one main bright component with $R_{\rm eff,UV/opt} < 0.55$~kpc to 2.01~kpc. 
%J0950+0523 and J1316+2614 are even more compact (unresolved) with $R_{\rm eff,opt} < 0.74$~kpc and $<0.55$~kpc, respectively. 
Only two galaxies, J0146--0220 and J0006+2452, have more complex rest-frame optical morphologies with one main bright compact component ($R_{\rm eff,opt} < 0.8$~kpc) and a diffuse component extended over $R_{\rm eff,opt} > 2$~kpc that is offset in one direction with respect to the main component. A second tiny component is also present in J0146--0220. The diffuse emission could either trace an underlying old stellar population or outflowing gas as the HAWK-I $K_{\rm s}$-band is contaminated by the nebular H$\alpha$ emission at the redshift of these two galaxies. HST images acquired for three of our galaxies confirmed the multi-component morphology (blended in the ground-based images) for two other galaxies, J1415+2036 and J0850+1549, whereas J1316+2614 remains composed of one single extremely UV-bright and compact component. All of these components are barely resolved even at the HST resolution \citep[$R_{\rm eff,UV}^{\rm HST} = 220\pm 12$~pc for J1316+2614;][]{Marques+24}. 

\item For seven galaxies, the main rest-UV/optical bright component is co-spatial with the more extended FIR dust continuum emission and the CO emission (when detected). This is quite surprising given their very blue unattenuated starlight. Only two galaxies, J0146--0220 and J0006+2452, have their FIR and CO emissions offset from the main rest-optical component but aligned with the extended diffuse rest-optical emission.
% We speculate that in all galaxies the dust is possibly being expelled from the rest-UV/optical bright component by outflows.
%Three galaxies (plus another tentatively), J0146--0220, J0006+2452, and J1249+1550, have their FIR dust continuum and CO emissions offset from the main rest-UV/optical bright component, while they are aligned with the extended diffuse rest-optical emission detected for 2 of them.

\end{itemize}

The extremely UV-luminous galaxies studied in this paper are recognised as being among the brightest UV galaxies known with $M_{\rm UV} = -23.4$ to $-24.6$ and among the most powerful starbursts, harbouring very young stellar populations of about 10~Myr. Their high star-formation efficiencies and very short molecular gas depletion timescales, together with their still very high molecular gas masses dominating over their stellar masses, strongly suggest that these galaxies are caught at the very beginning of their stellar mass build-up in an incredibly intense, rapid, and efficient gas-to-stars conversion phase. We find that the feedback-free starburst model of \citet{Dekel+23} seems to be able to explain the unique physical properties of our galaxies through the collapse of gas clouds within very short free-fall times before the onset of stellar winds and SN feedback. However, in addition to this, our galaxies are characterised by a relatively bright FIR dust continuum emission that is found to be co-spatial with the main rest-UV/optical bright component in seven out of nine galaxies. This is in strong contrast with their blue unattenuated starlight, as one would rather expect the galaxy to be dust obscured. To reconcile these physical properties, the \citet{Ferrara+23} model of dust removal by radiation-driven outflows taking place above the super-Eddington threshold at the location of the starburst appears to be a viable option and would also explain the big FIR effective radii of our galaxies in comparison to their stellar effective radii \citep[see also][]{Menon+24}. Finally, the measured dust masses seem to require dust production mechanisms that are more rapid and efficient than SNe given the short 10~Myr timescale available for their formation.

While these extremely UV-luminous starburst galaxies are rare at $z = 2.1-3.6$, with only about 70 of them found in the $\sim 9000~\rm deg^2$-wide eBOSS survey, corresponding to a number density of $10^{-9}~\rm Mpc^{-3}$ (Marques-Chaves et~al.\ in prep.), intrinsically they are certainly not so rare. The detection of some of them at $z>6$ \citep[e.g.][]{Sobral+15,Matsuoka+18,Hashimoto+19} and, more recently, their unexpected overabundance discovered at $7<z<16$ with JWST, implying steeper UV luminosity functions than predicted \citep[e.g.][]{Bouwens+23,Atek+23,Bunker+23,Casey+23,Castellano+24,Carniani+24}, show that galaxies towards the EoR are more often experiencing such UV-luminous and still unobscured phases, with high sSFRs. As discussed in Sect.~\ref{sect:introduction}, different scenarios were invoked to explain the UV-luminous starburst galaxies at the EoR and their number excess. Among these scenarios are the very efficient star formation following the \citet{Dekel+23} model and the \citet[][see also \cite{Ziparo+23}, \cite{Ferrara24}]{Ferrara+23} model of temporary dust removal as a consequence of radiation-driven outflows, for which the molecular gas and FIR dust continuum analysis of our galaxies %which we may consider as their lower redshift `analogues', 
provides some support. Alternatively, a modified IMF (towards a top-heavy one) boosting the UV radiation and the luminosity-to-mass ratio of galaxies was also proposed \citep[e.g.][]{BekkiTsujimoto23,Trinca+24} as well as a stochastic variability of the SFR \citep[e.g.][]{Mirocha+23,Gelli+24}. Currently, we have no evidence supporting a modified IMF except possibly an IMF with a higher upper mass cut-off given that our galaxies host VMSs \citep{Upadhyaya+24} that can also lead to high UV luminosity increases \citep{Schaerer+24}. 
%we cannot exclude the stochastic variability of the SFR if the replenishment in molecular gas is sufficiently efficient (in quantity and time) to allow it.

In summary, these galaxies show several similarities with the galaxies recently unveiled at the EoR with JWST. They hence possibly represent a sample of lower redshift `analogues' that benefit from the advantage of a rich dataset and data to come that will be difficult to acquire for galaxies at $z>7$. 
%These are still essential to understanding what triggers extreme UV luminosities and starbursts in galaxies in general. Our UV-bright galaxies at cosmic noon can thus serve as ideal laboratories to test different scenarios proposed to explain the extremely UV-luminous galaxies and their number excess at the EoR.
Our UV-bright galaxies at cosmic noon thus are ideal laboratories to test different scenarios proposed to explain what triggers the extreme UV luminosities and starbursts of galaxies and their number excess at the EoR. 

%-----------------------------------------------------------------------
\begin{acknowledgements}
This paper makes use of the following ALMA data: ADS/JAO.ALMA\#2018.1.00932.S and 2021.1.01438.S. ALMA is a partnership of ESO (representing its member states), NSF (USA) and NINS (Japan), together with NRC (Canada), NSC and ASIAA (Taiwan), and KASI (Republic of Korea), in cooperation with the Republic of Chile. The Joint ALMA Observatory is operated by ESO, AUI/NRAO and NAOJ. This paper is also based on observations collected at the European Organisation for Astronomical Research in the Southern Hemisphere under ESO programme 111.251K.001.
\end{acknowledgements}

%-----------------------------------------------------------------------

%-----------------------------------------------------------------------
\begin{appendix}
\section{Figure~1 continued}
\label{appendix}

\begin{figure*}[h]
\hspace{0.4cm}
\includegraphics[width=0.22\textwidth,clip]{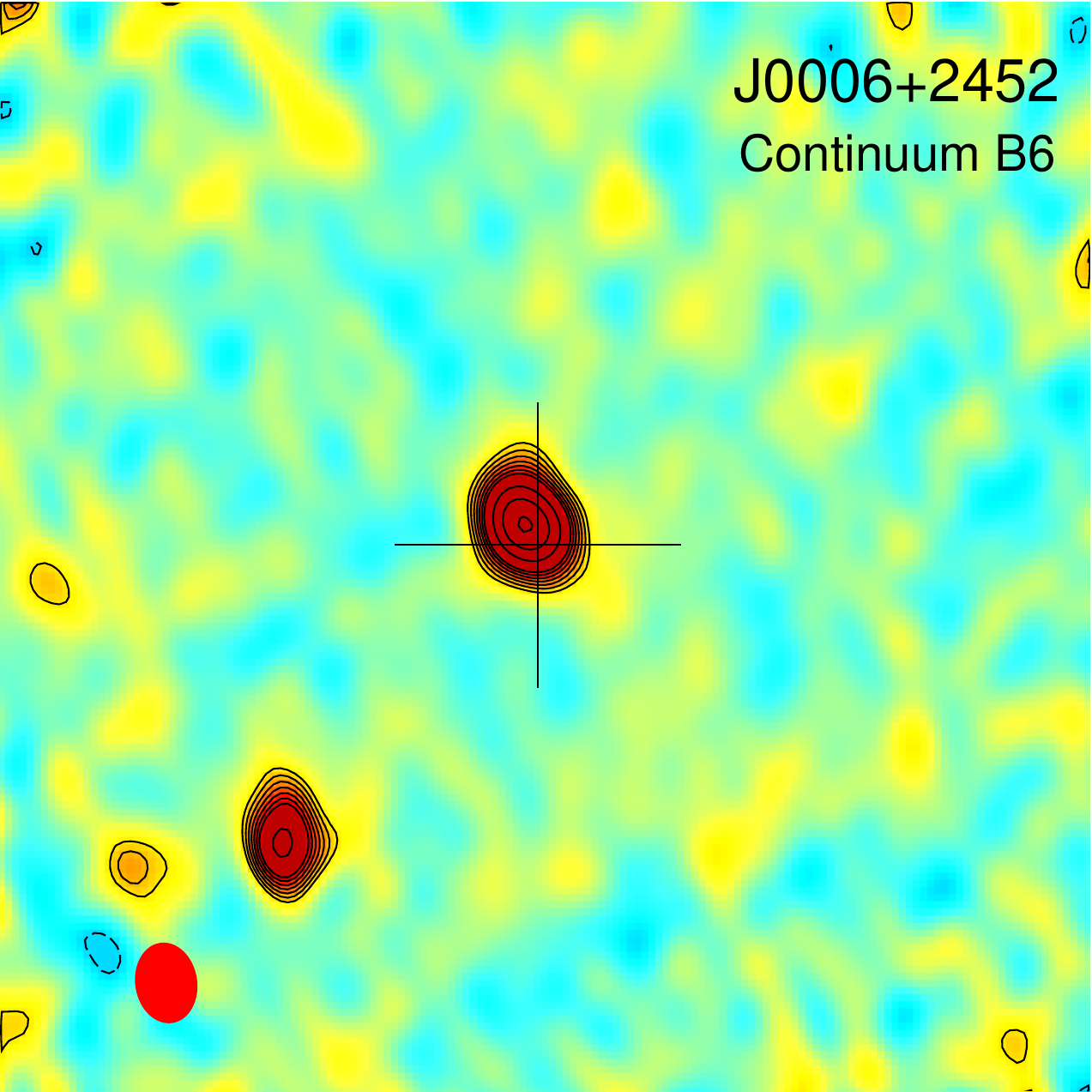}
\includegraphics[width=0.22\textwidth,clip]{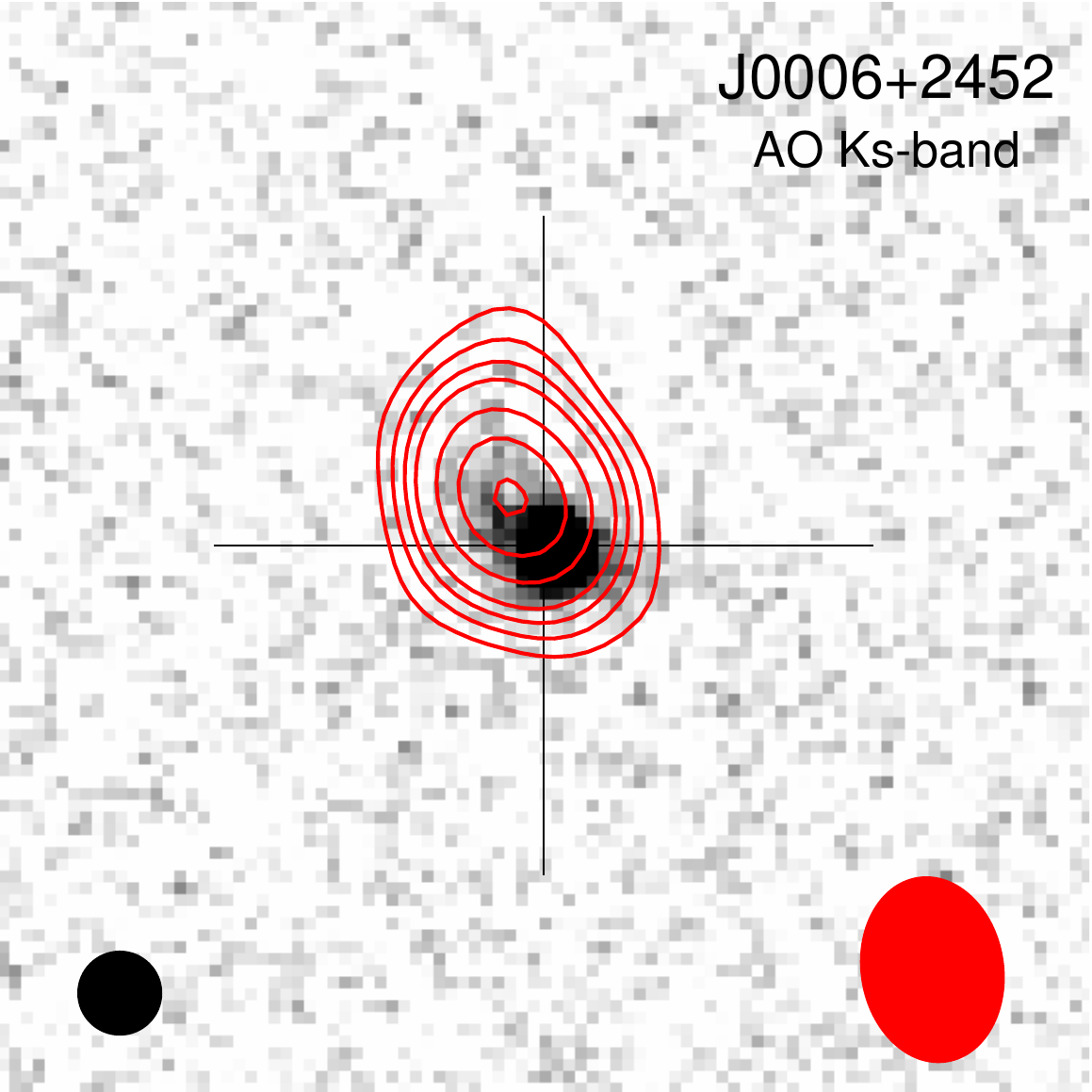}
\includegraphics[width=0.22\textwidth,clip]{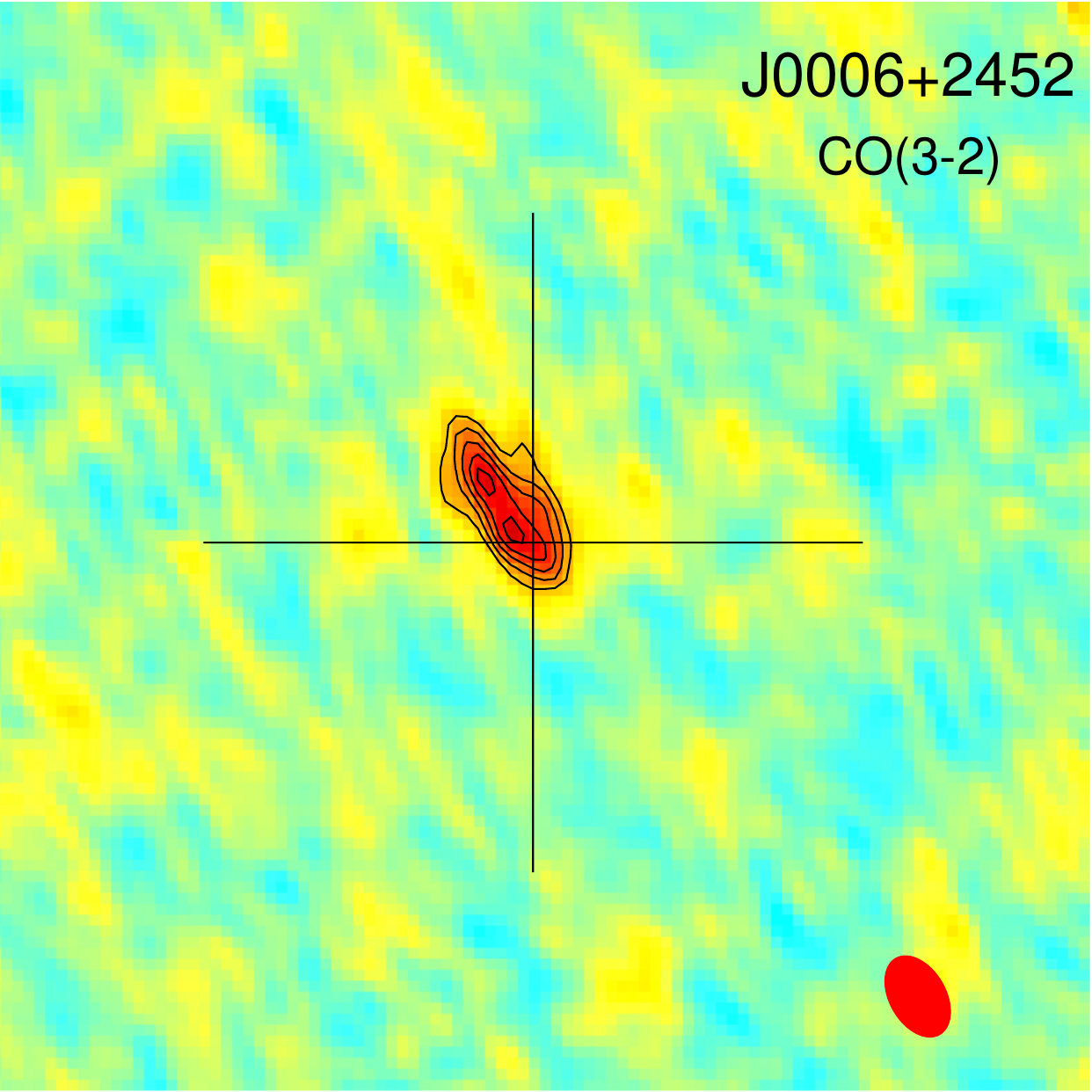}
\includegraphics[width=0.28\textwidth,clip]{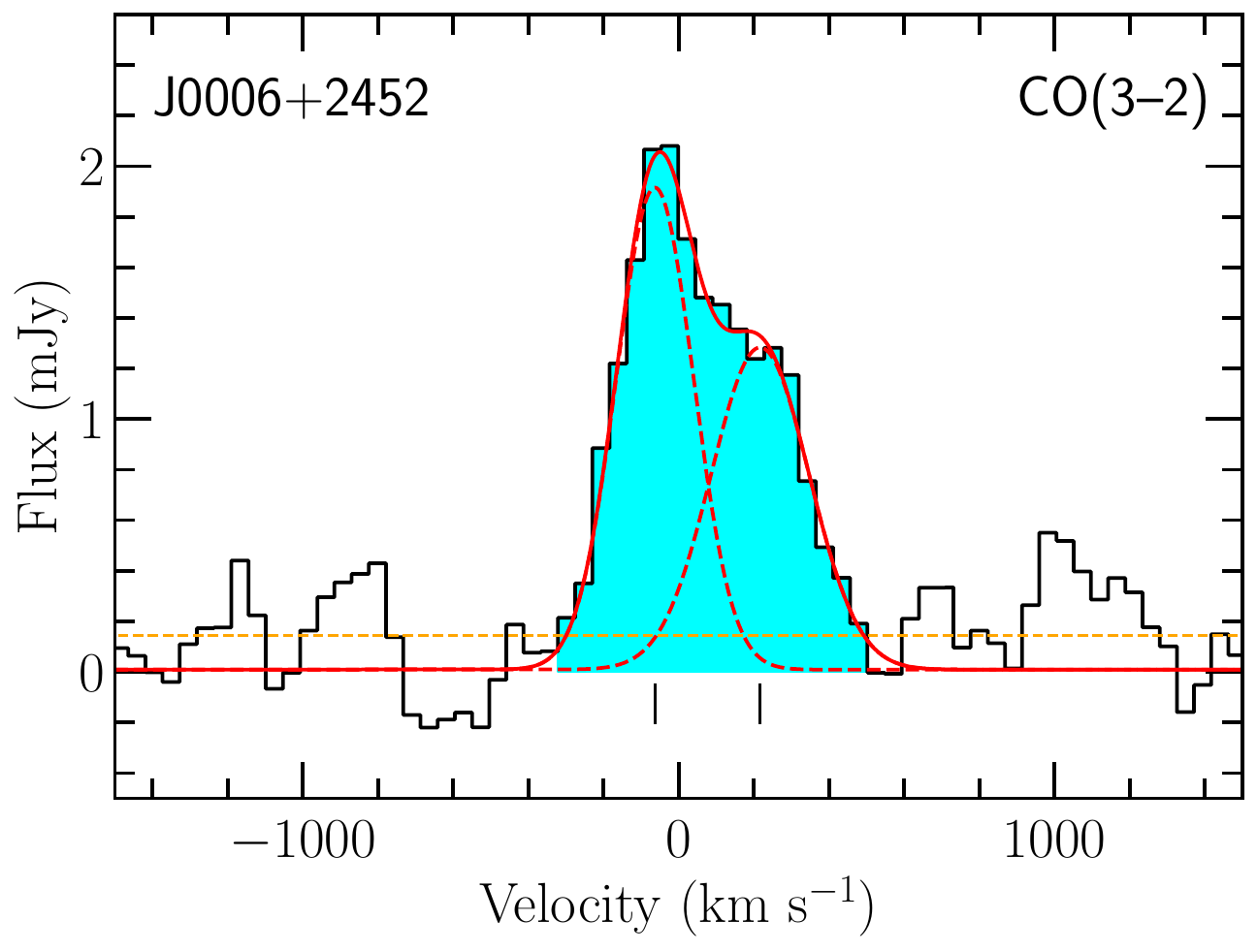}

\hspace{0.4cm}
\includegraphics[width=0.22\textwidth,clip]{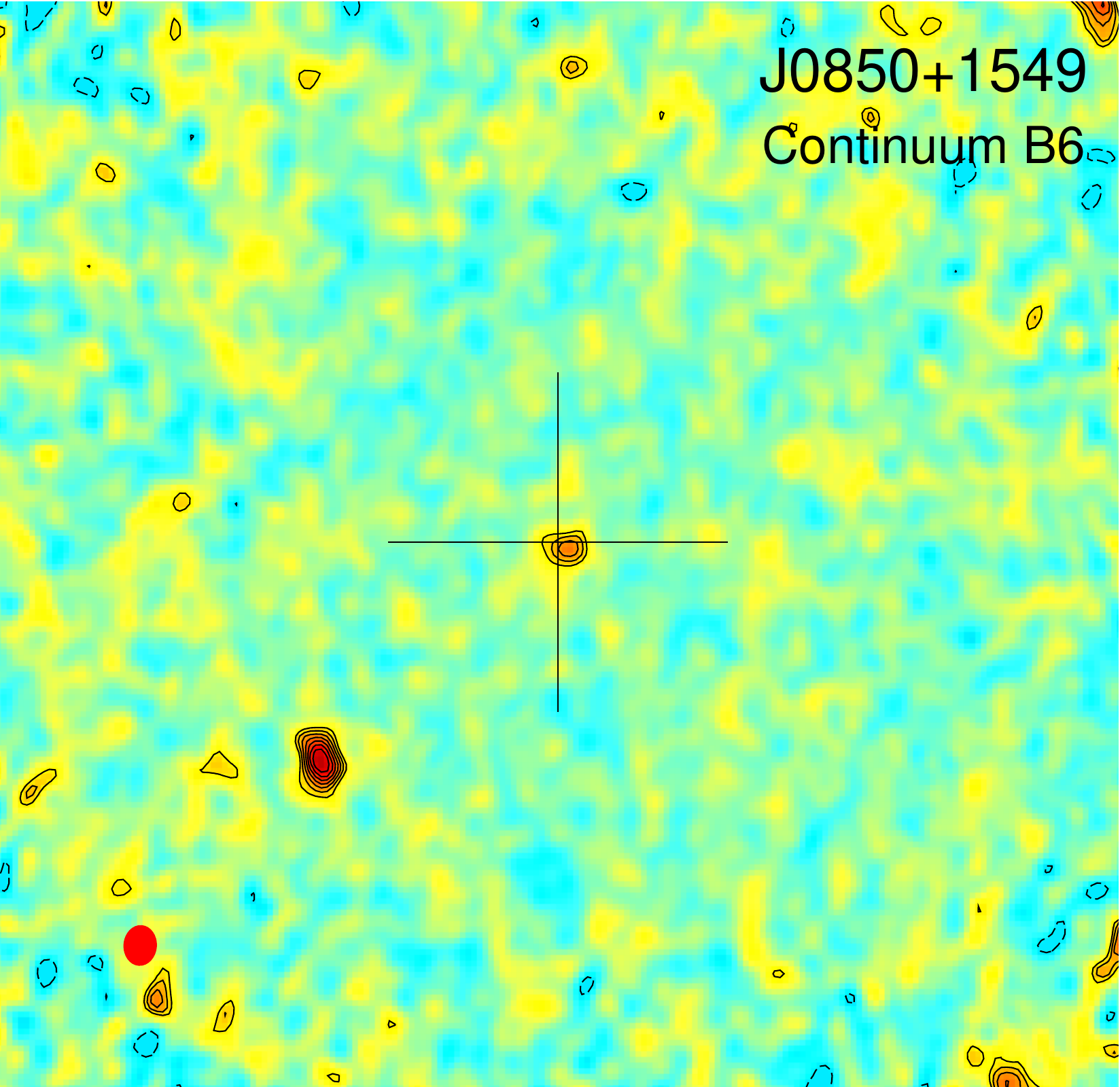}
\includegraphics[width=0.22\textwidth,clip]{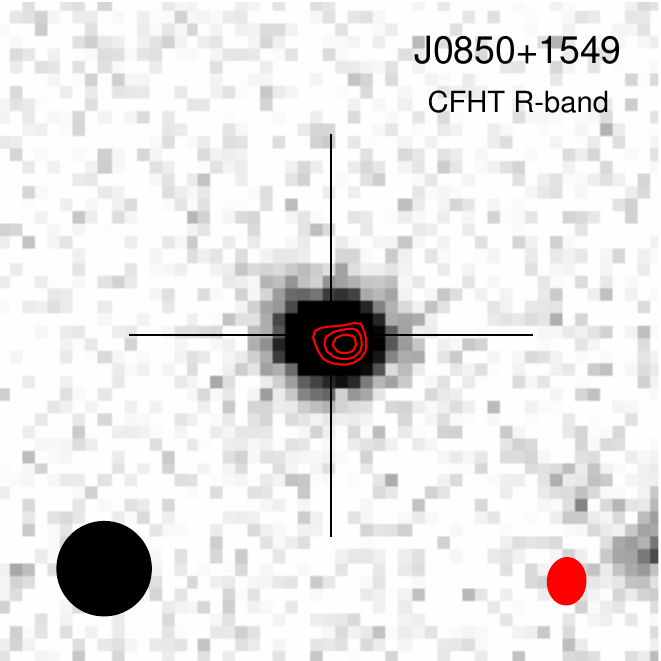}
\includegraphics[width=0.22\textwidth,clip]{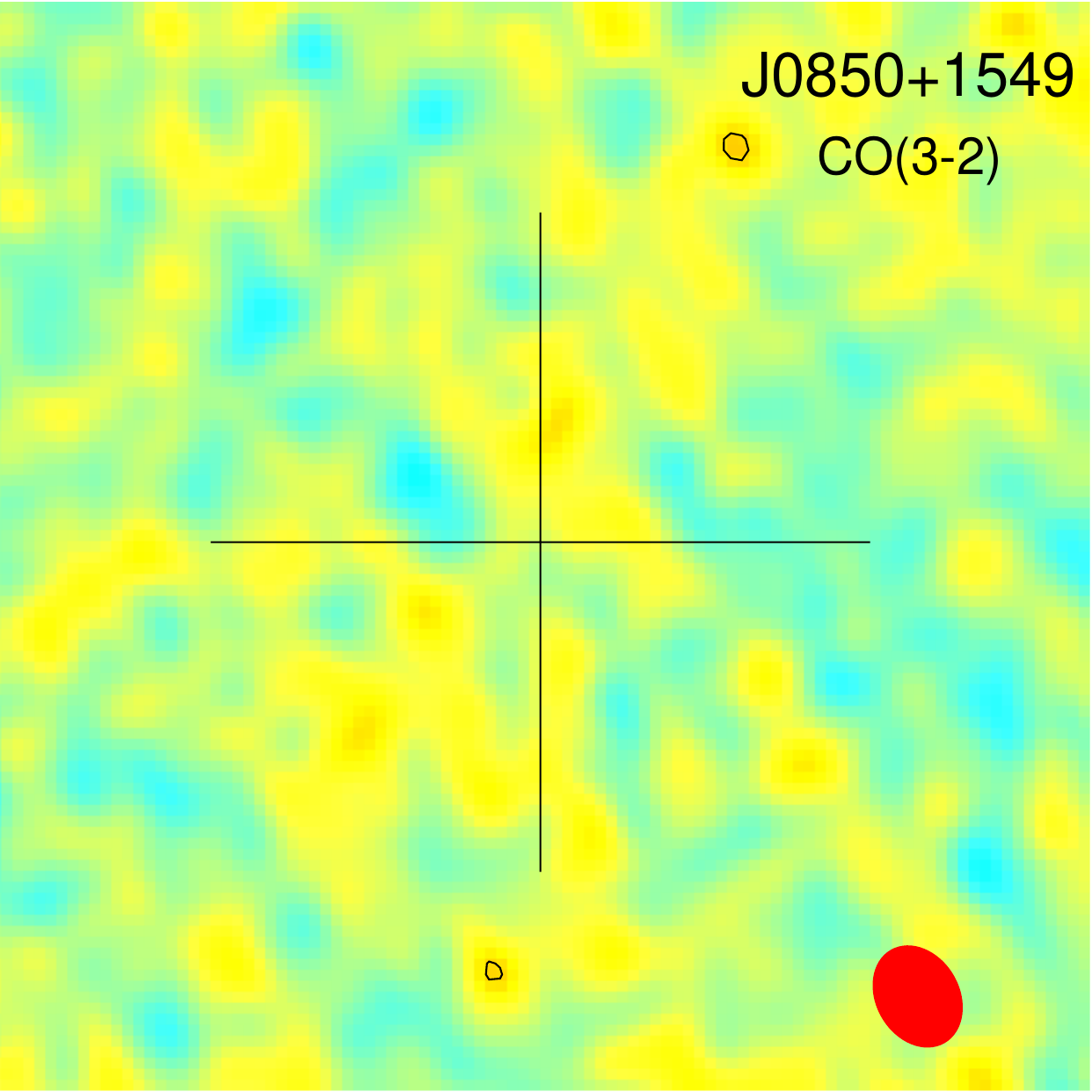}

\hspace{0.4cm}
\includegraphics[width=0.22\textwidth,clip]{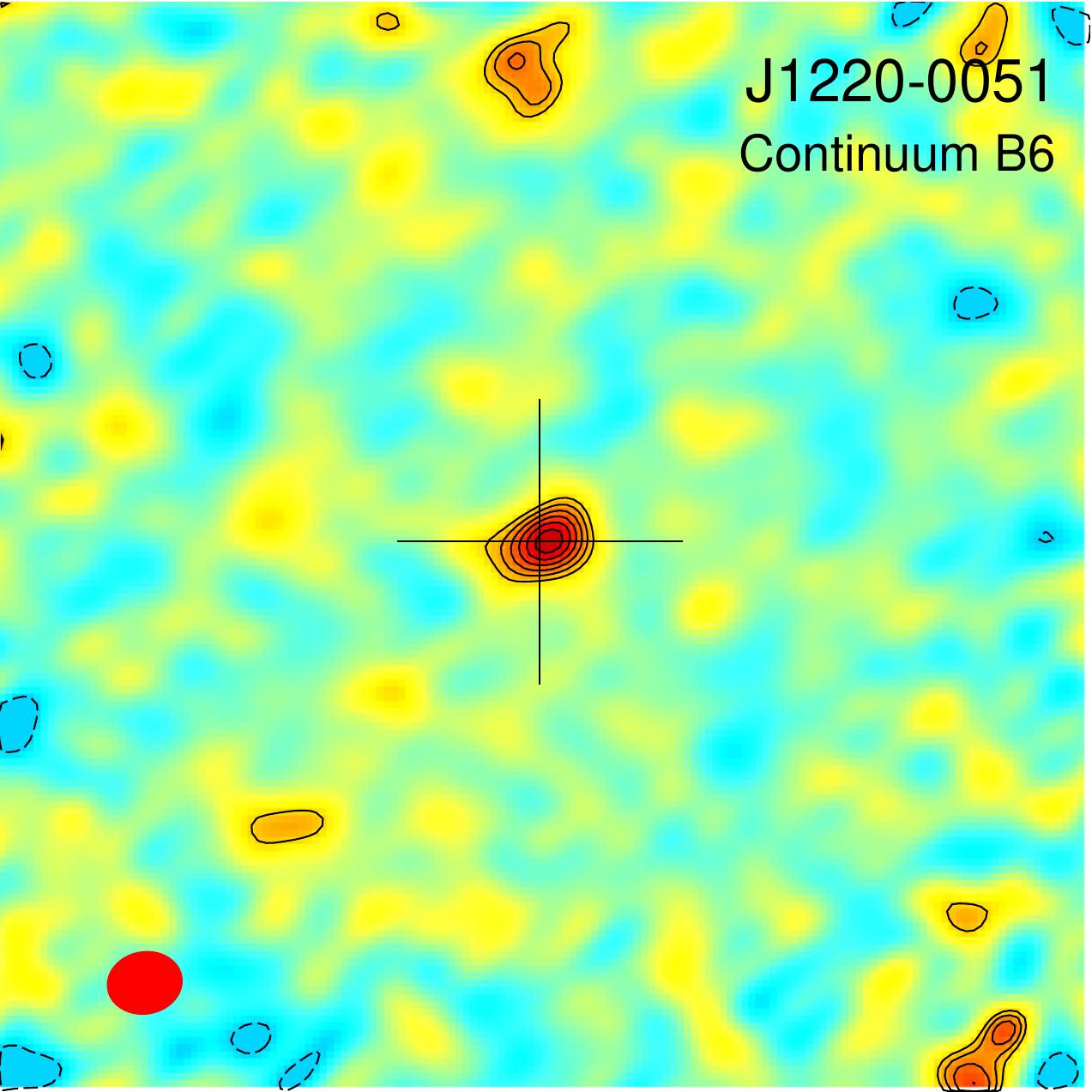}
\includegraphics[width=0.22\textwidth,clip]{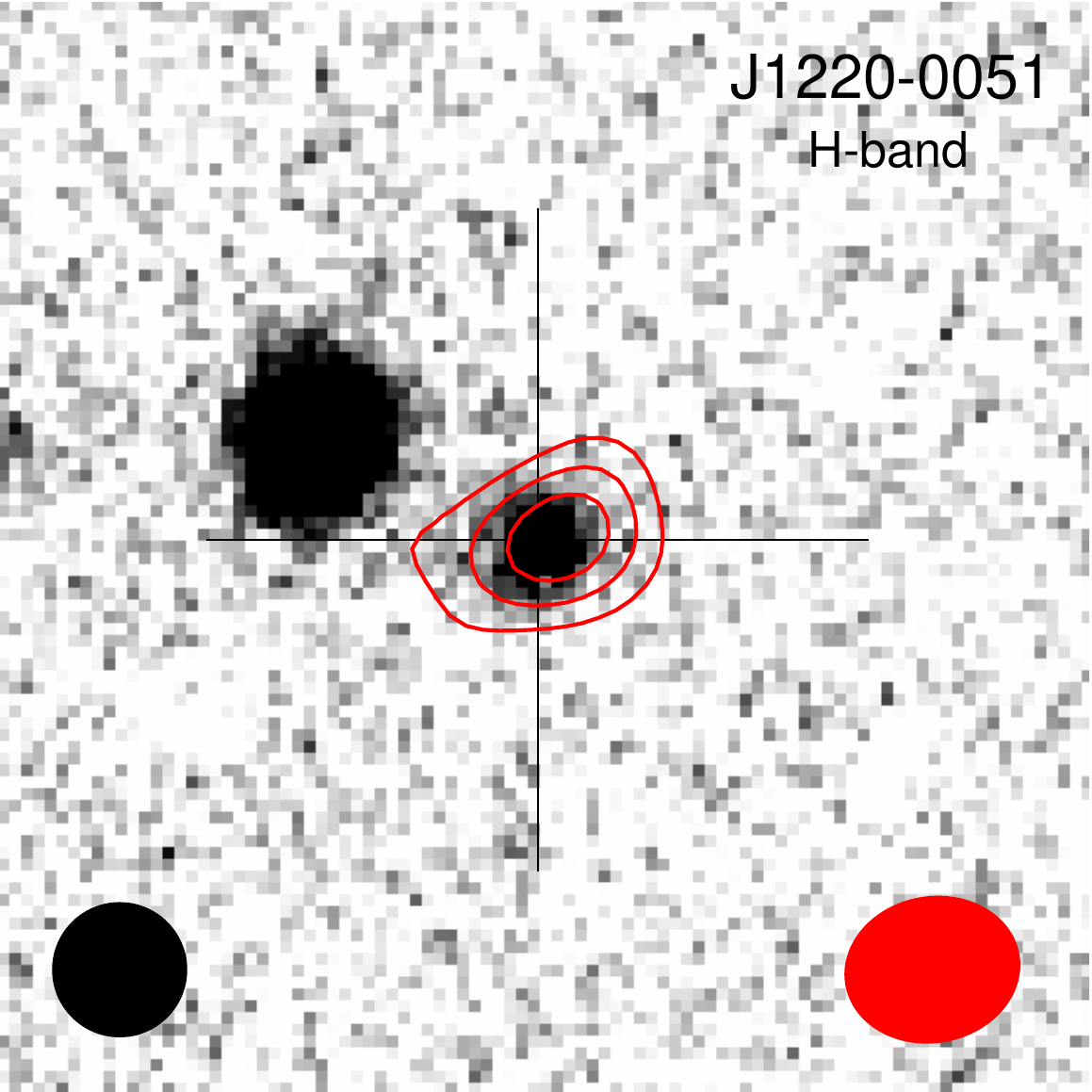}
\includegraphics[width=0.22\textwidth,clip]{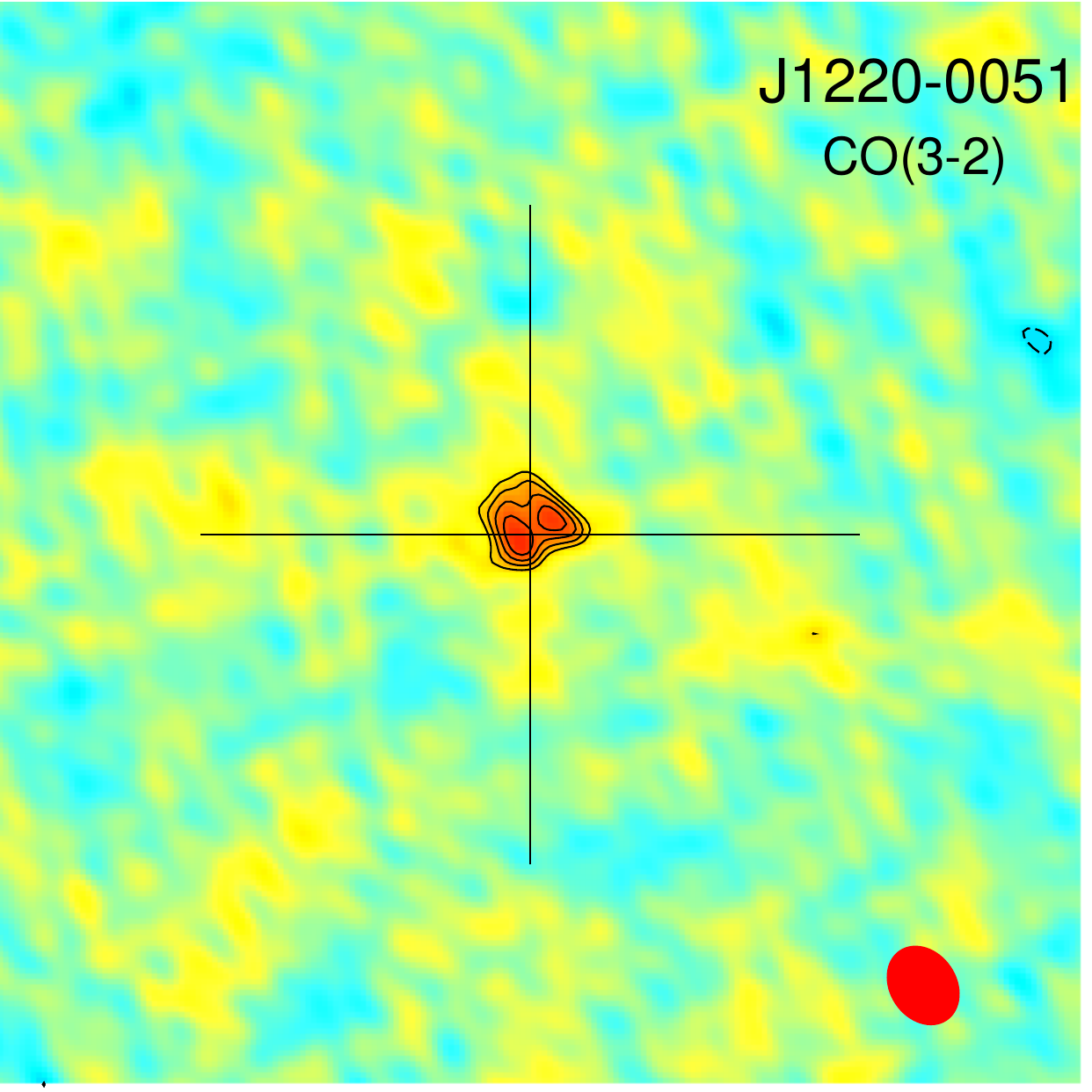}
\includegraphics[width=0.28\textwidth,clip]{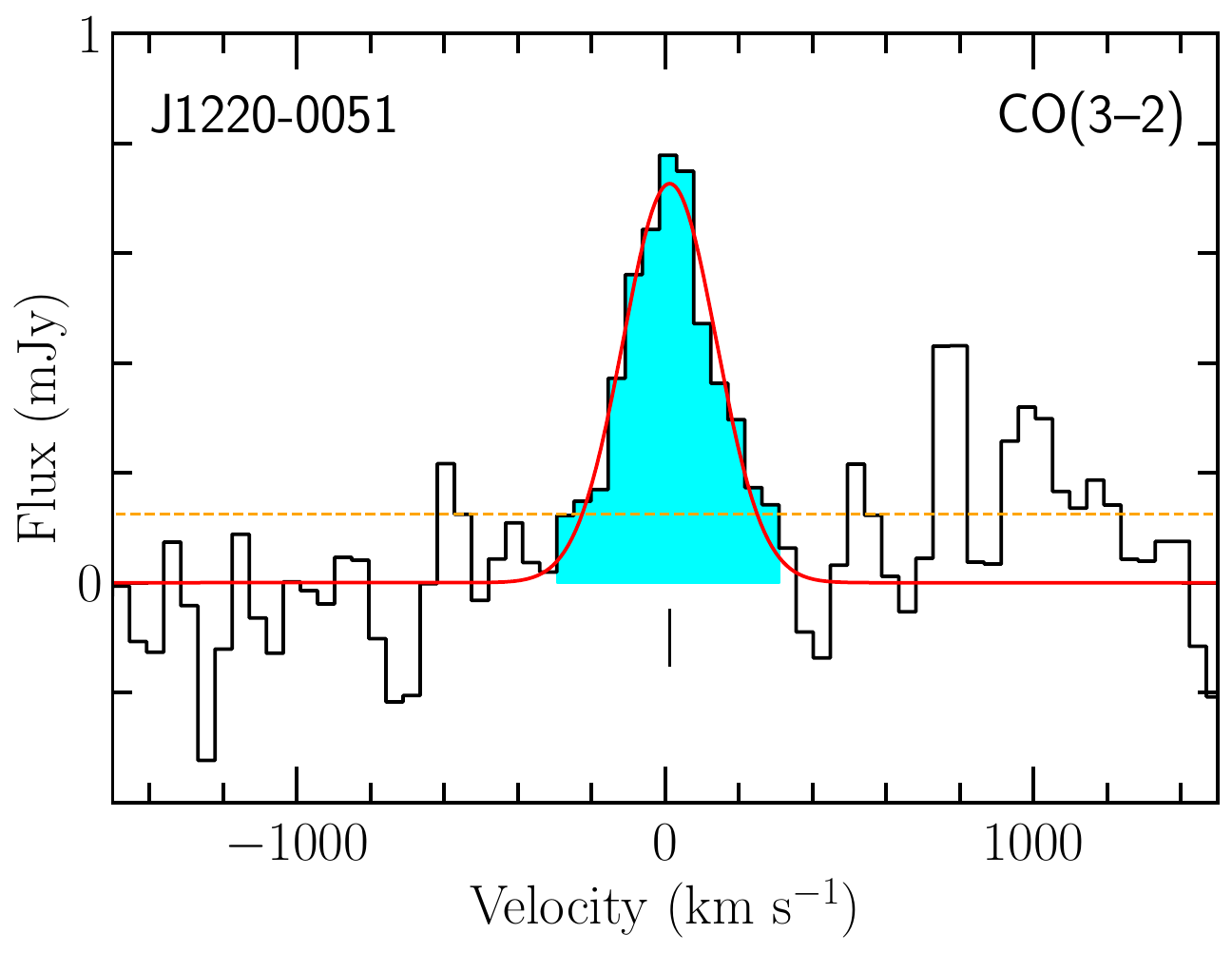}

\hspace{0.4cm}
\includegraphics[width=0.22\textwidth,clip]{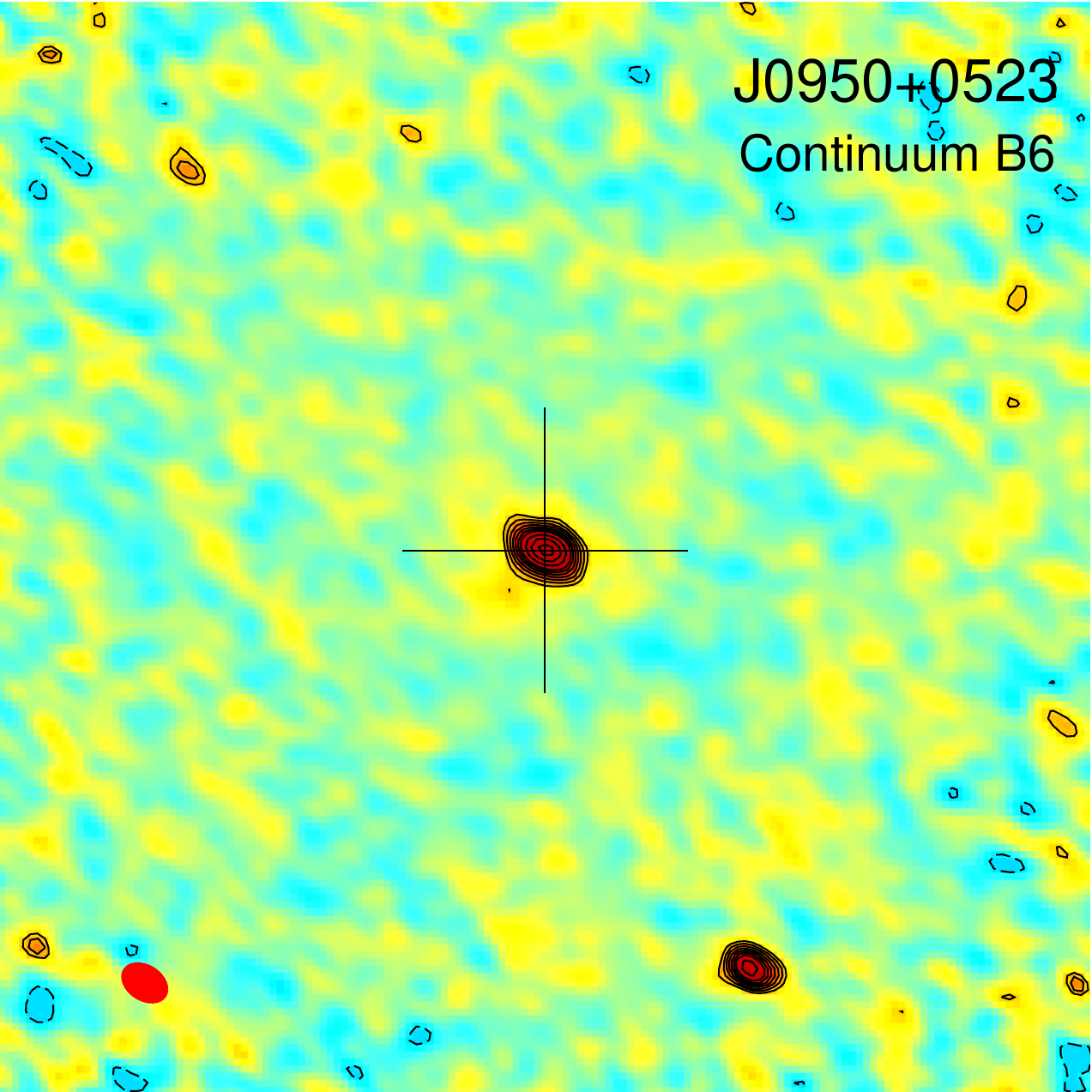}
\includegraphics[width=0.22\textwidth,clip]{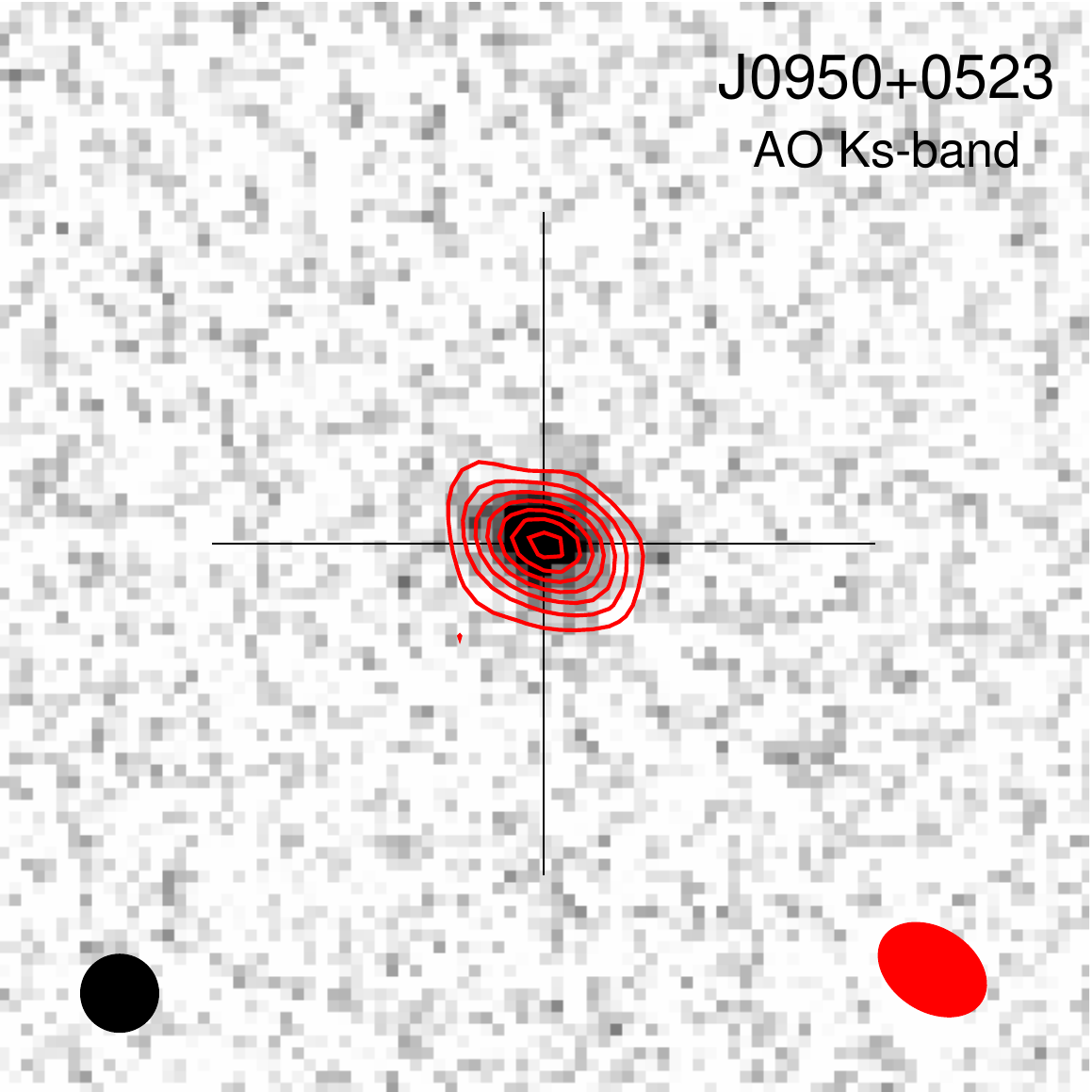}
\includegraphics[width=0.22\textwidth,clip]{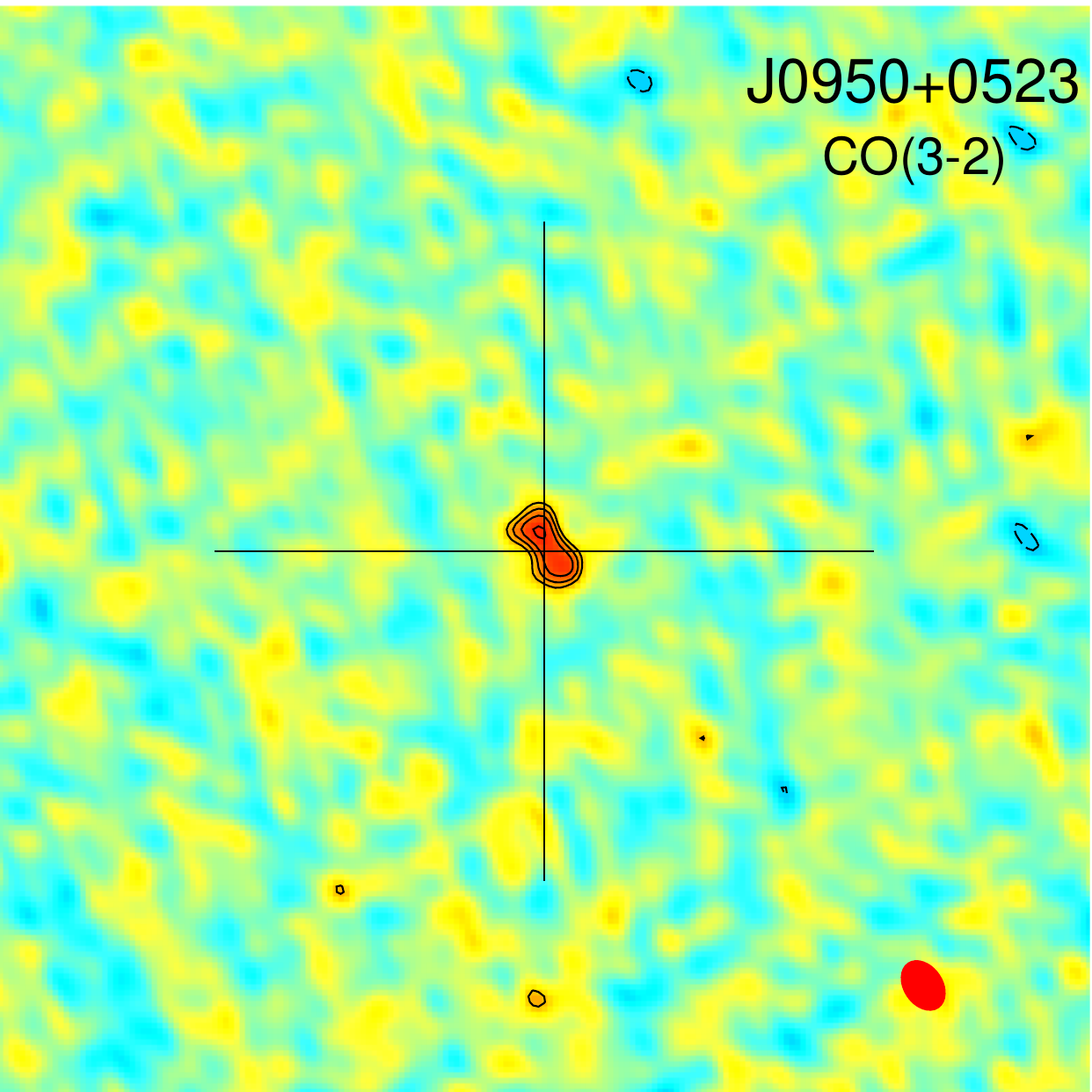}
\includegraphics[width=0.28\textwidth,clip]{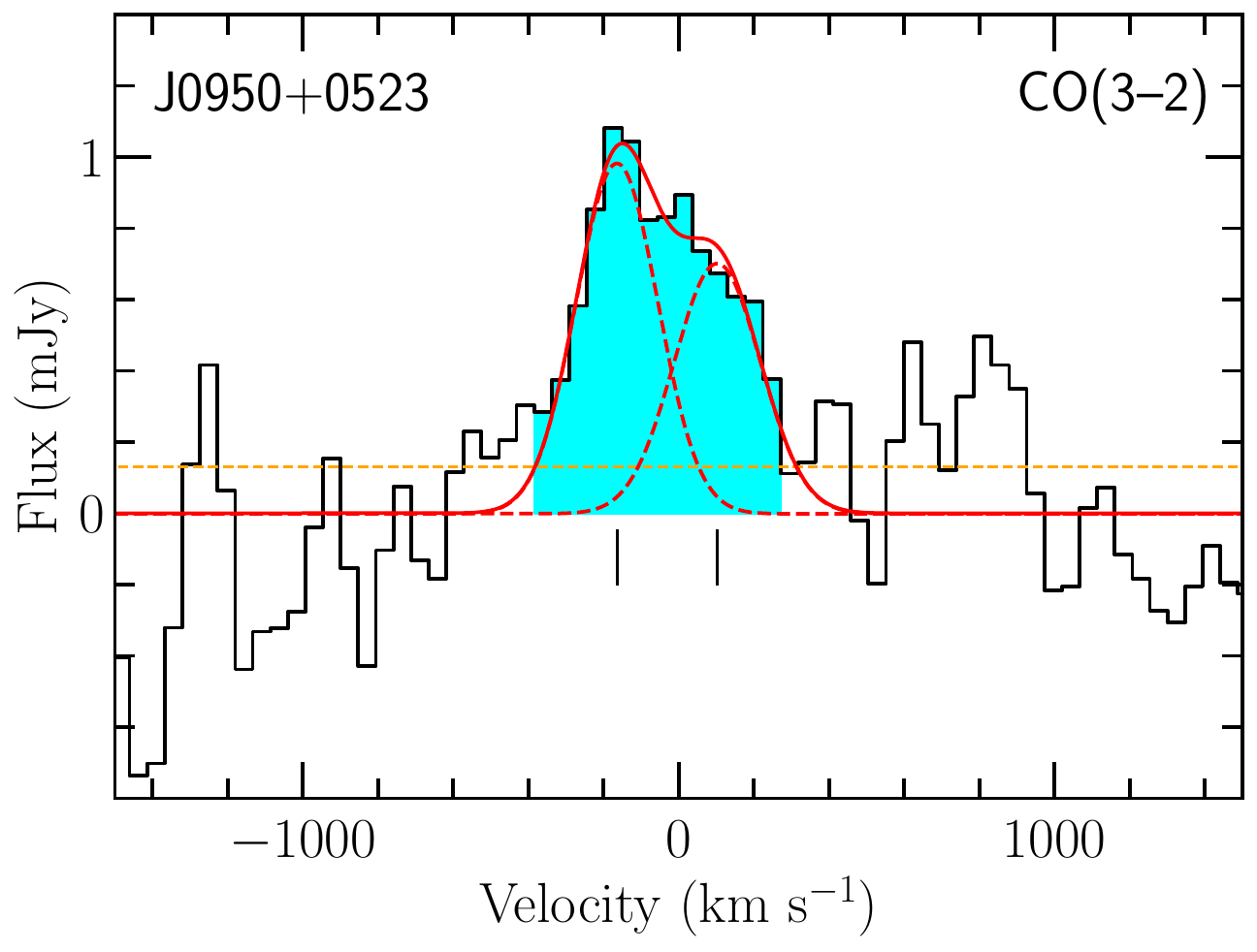}
\caption{See Fig.~\ref{fig:contB6-CO_1} for the description of panels.}
\label{fig:contB6-CO_2}
\end{figure*}
%-----------------------------------------------------------------------
\begin{figure*}[!]
\hspace{0.4cm}
\includegraphics[width=0.22\textwidth,clip]{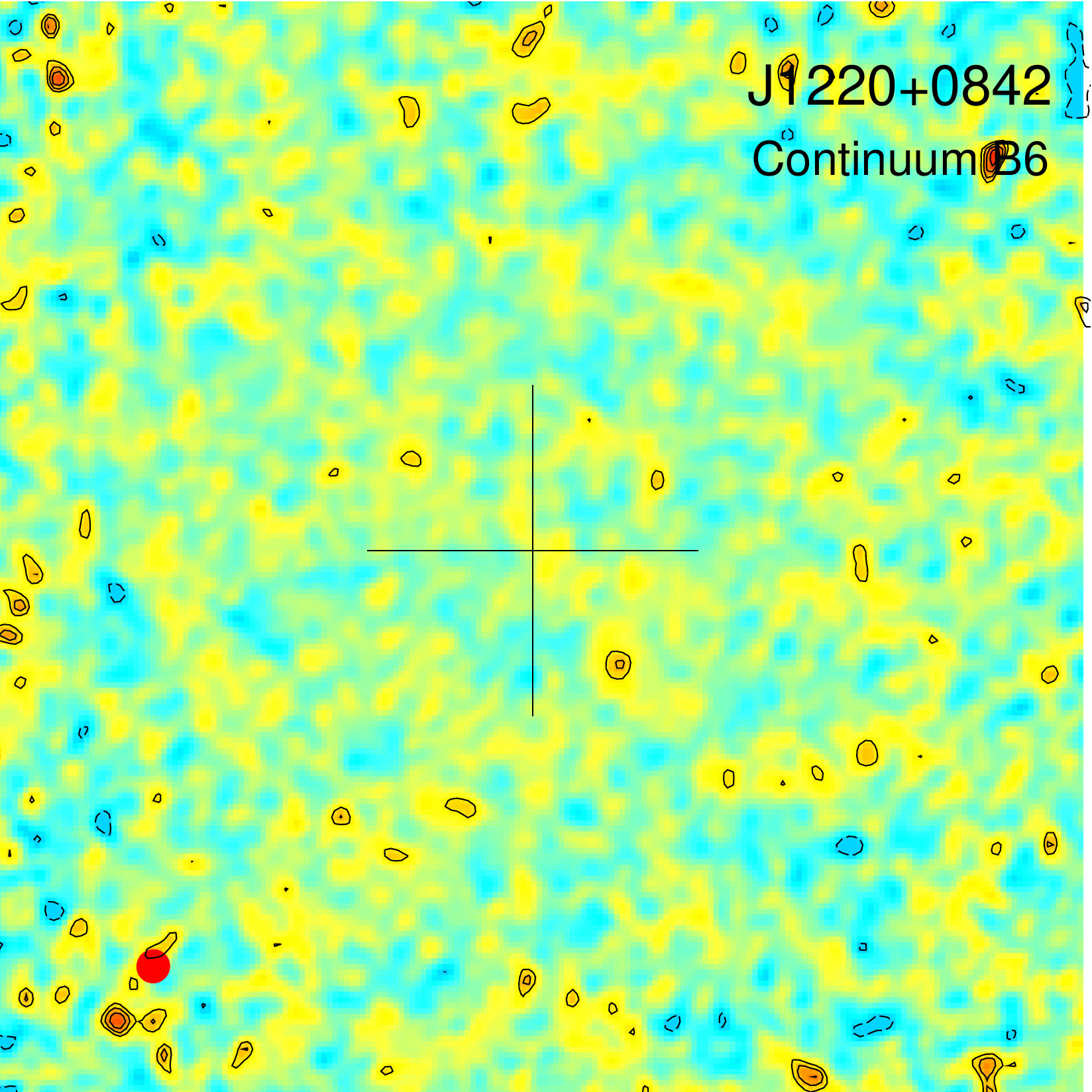}
\includegraphics[width=0.22\textwidth,clip]{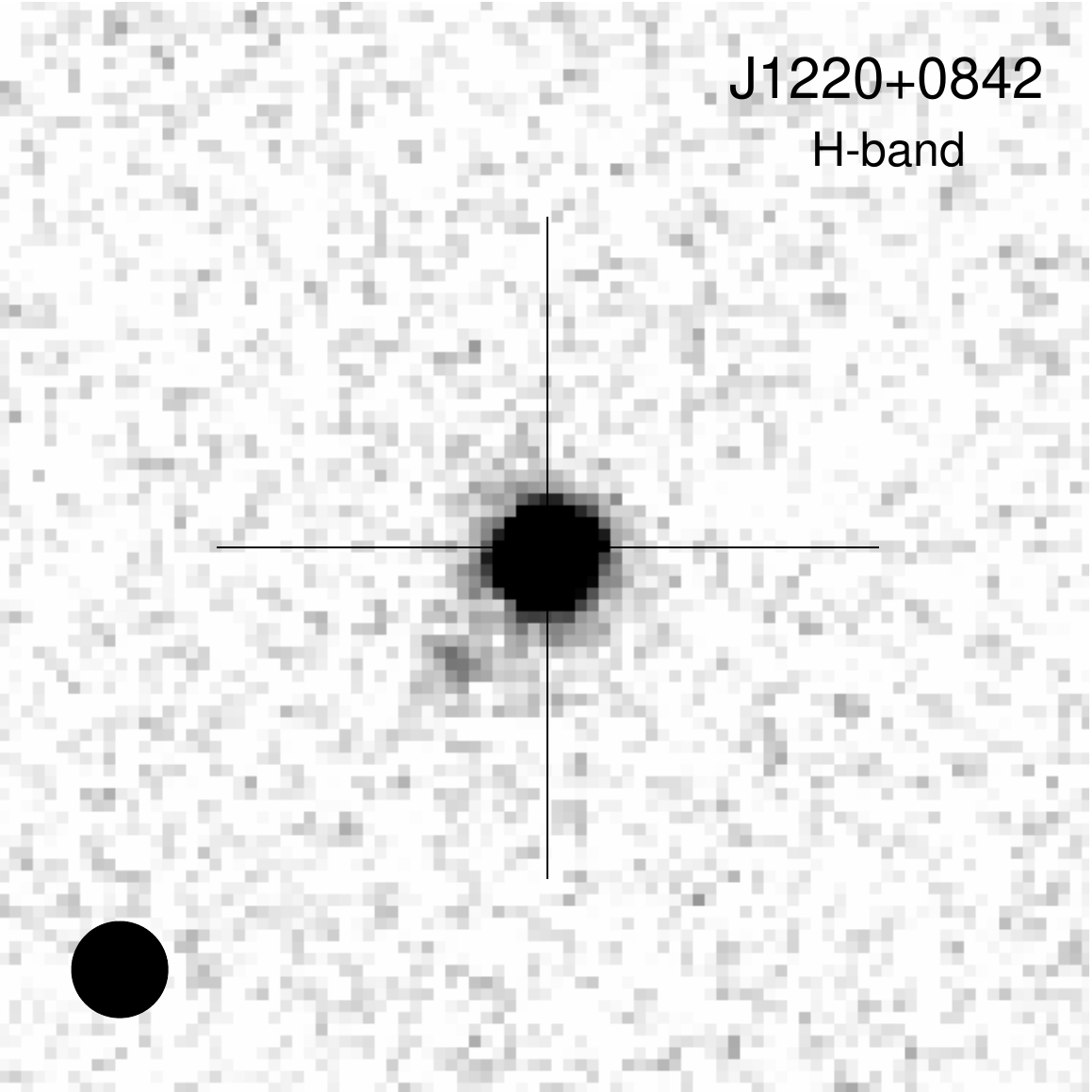}
\includegraphics[width=0.22\textwidth,clip]{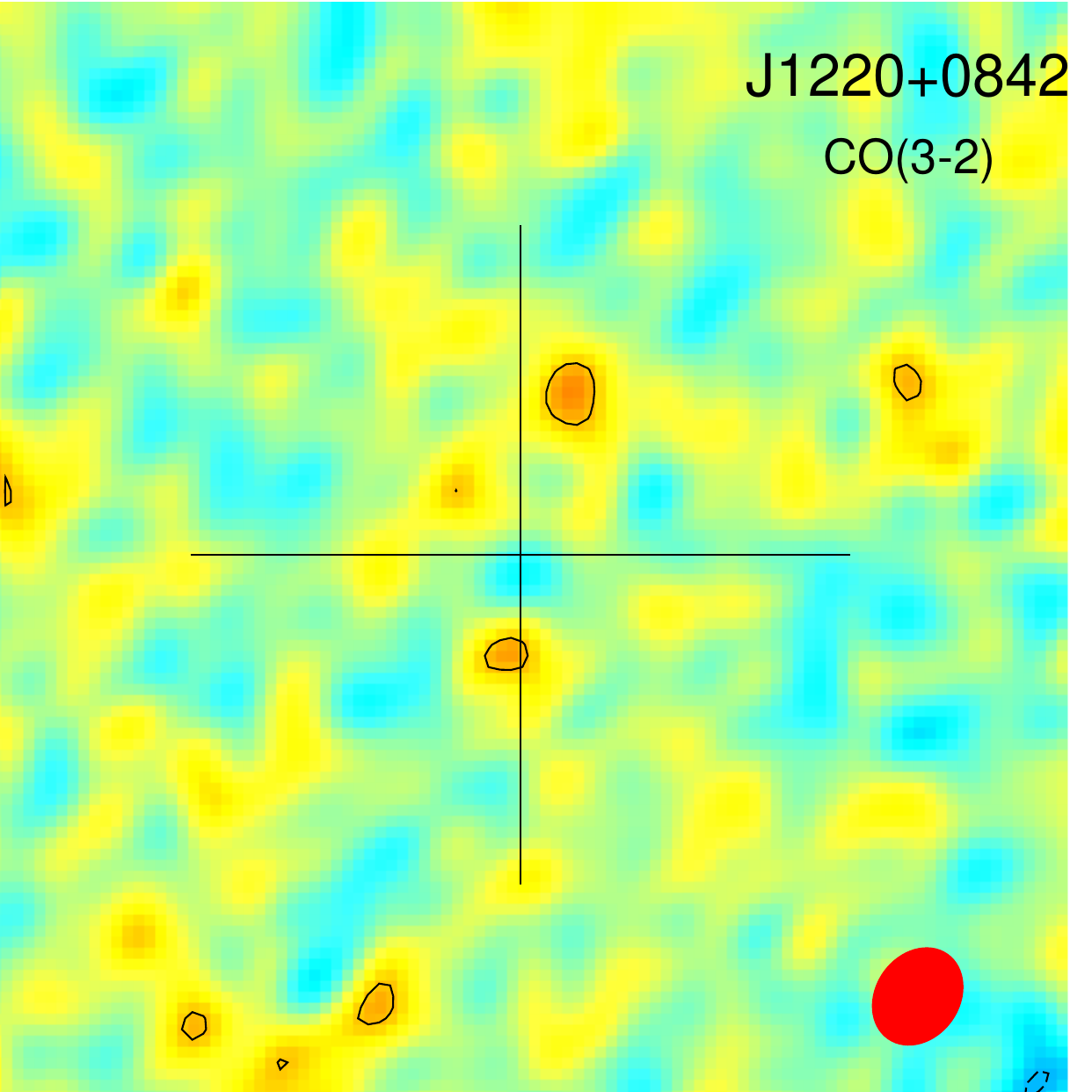}

\hspace{0.4cm}
\includegraphics[width=0.22\textwidth,clip]{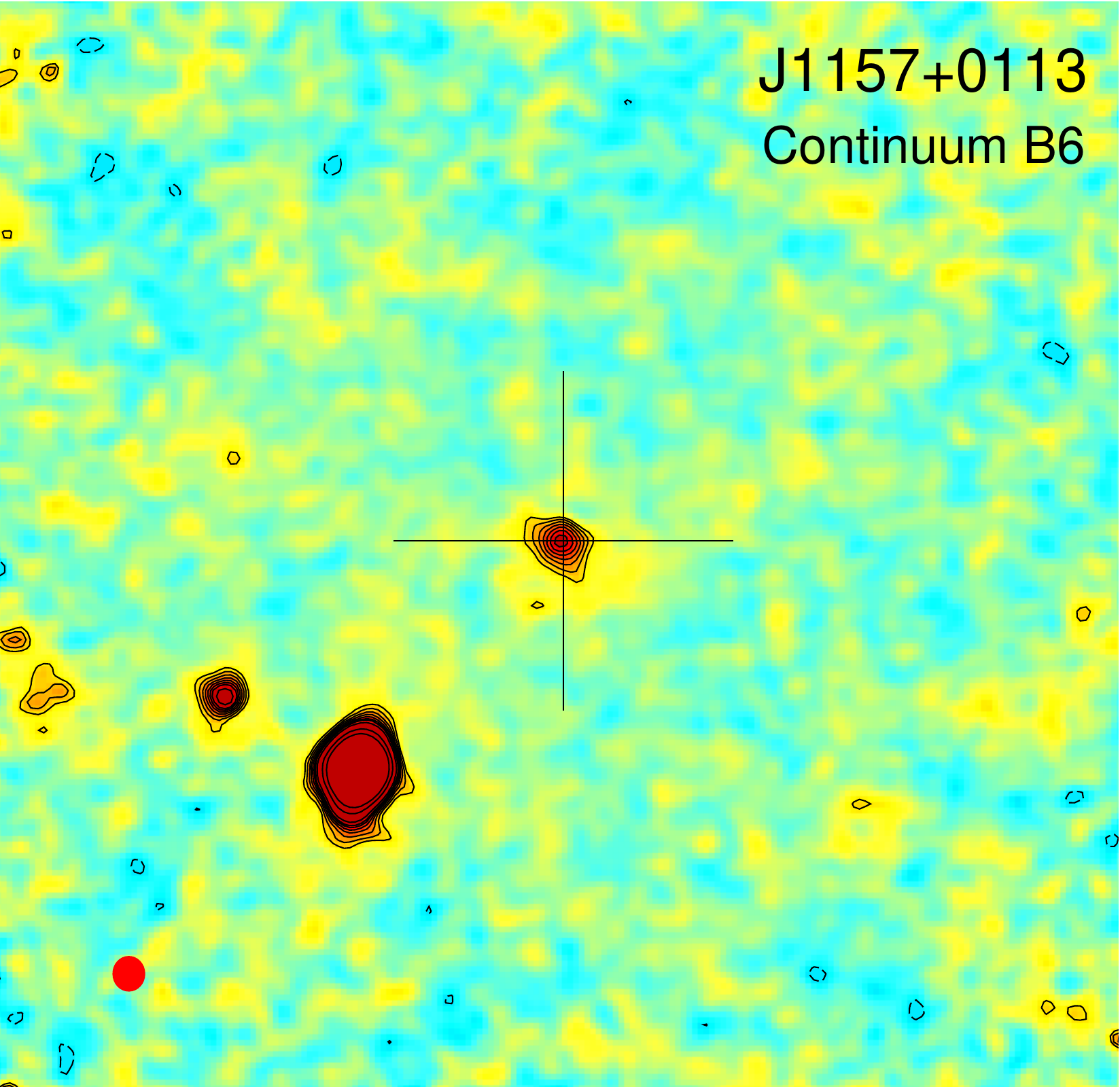}
\includegraphics[width=0.22\textwidth,clip]{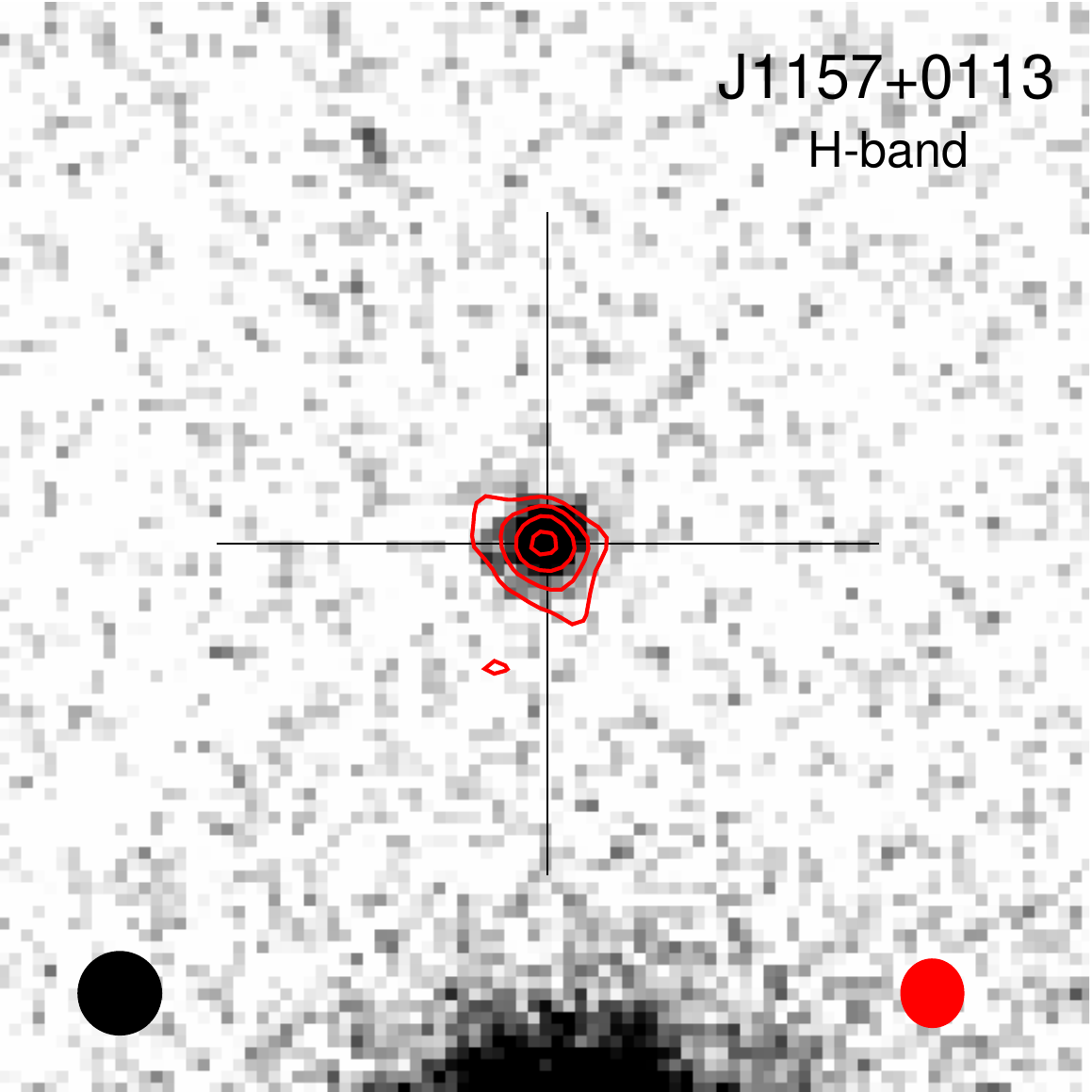}
\includegraphics[width=0.22\textwidth,clip]{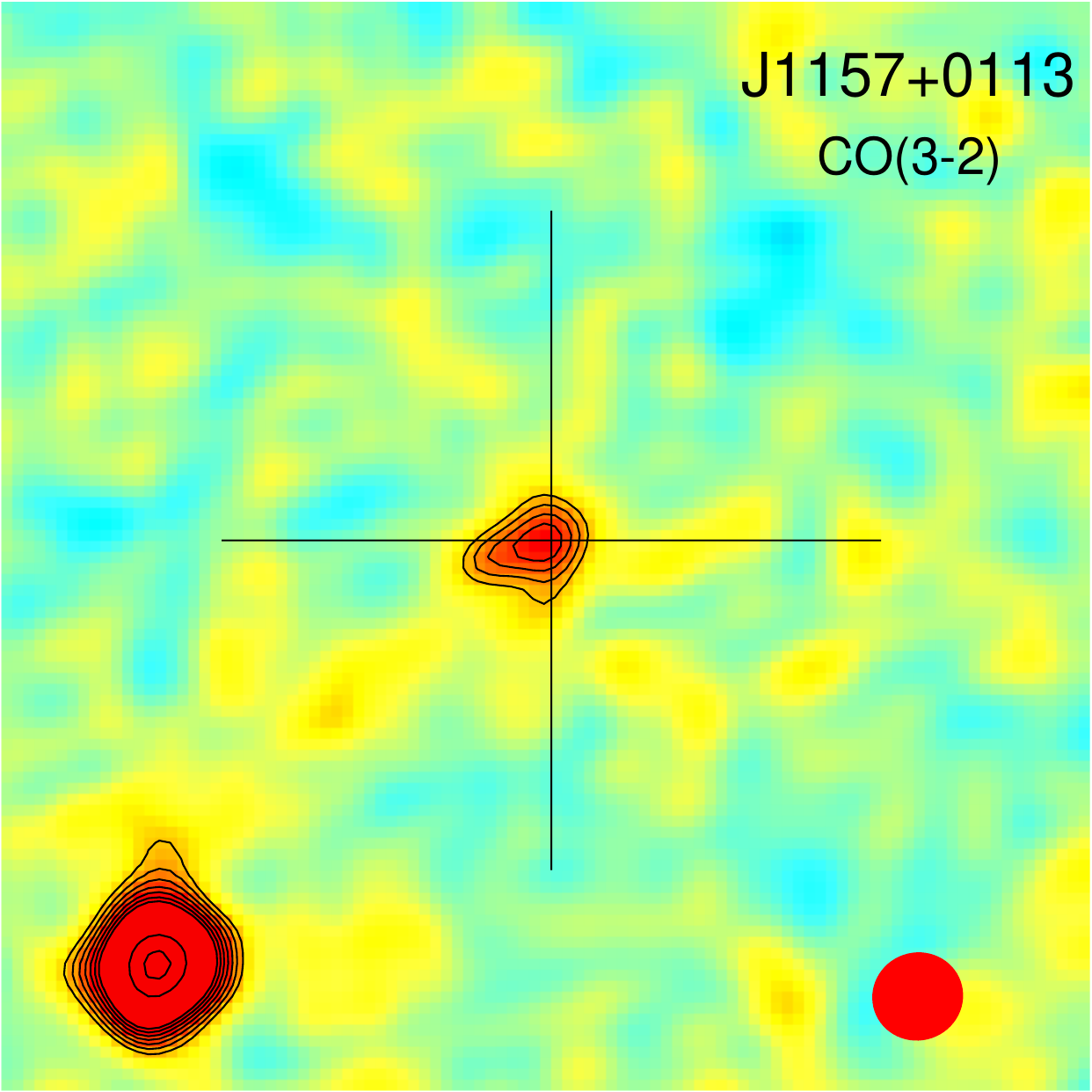}
\includegraphics[width=0.28\textwidth,clip]{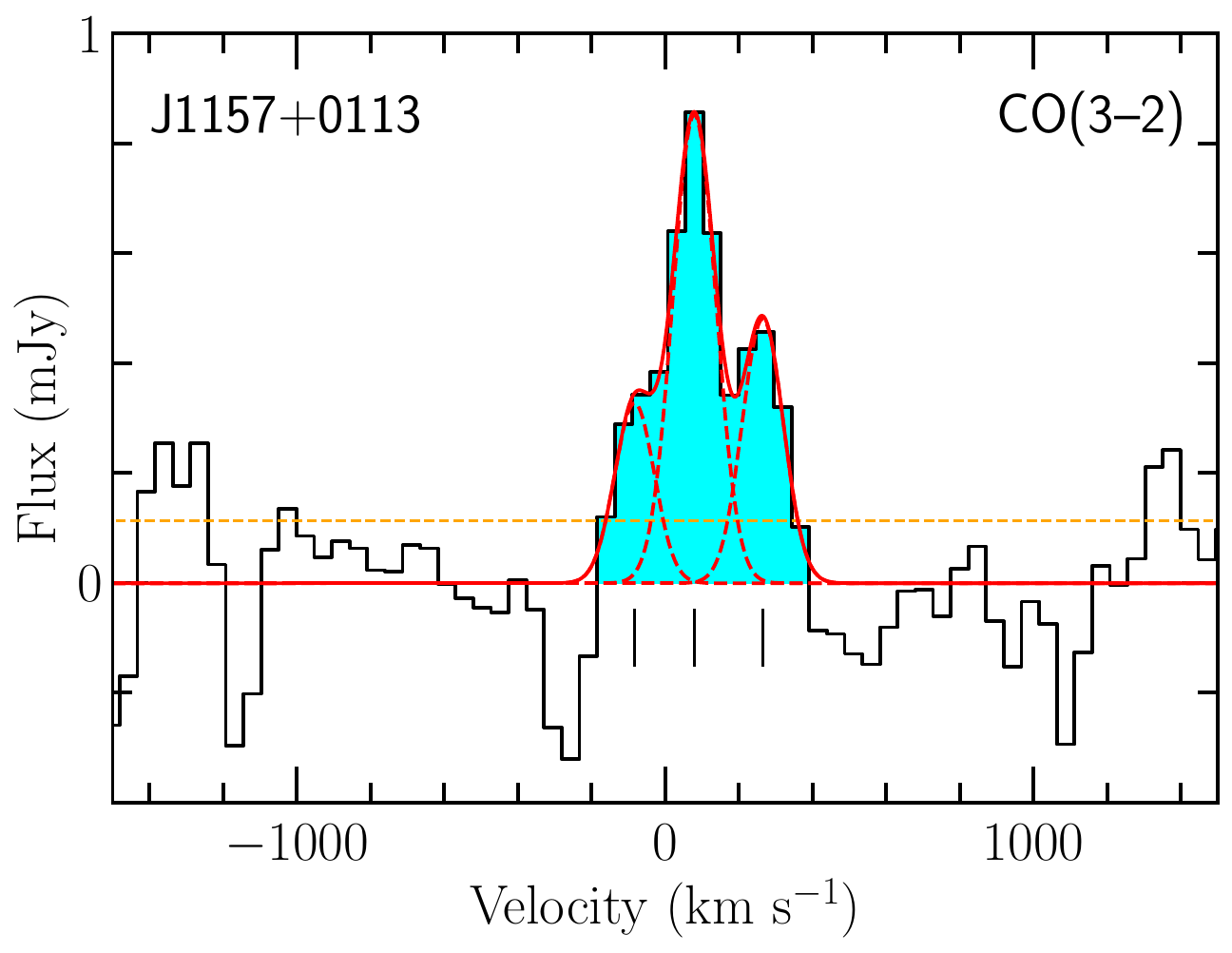}

\hspace{0.4cm}
\includegraphics[width=0.22\textwidth,clip]{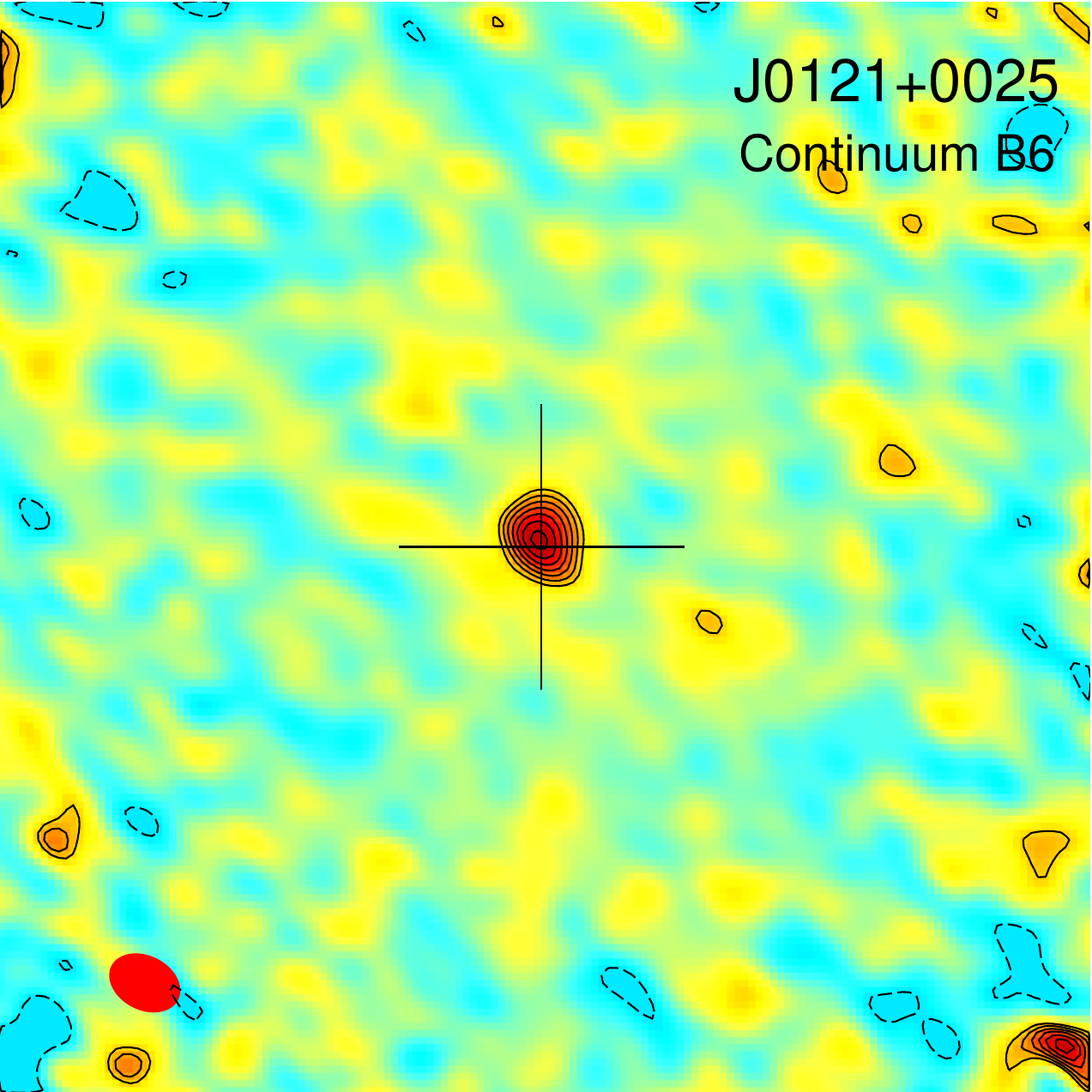}
\includegraphics[width=0.22\textwidth,clip]{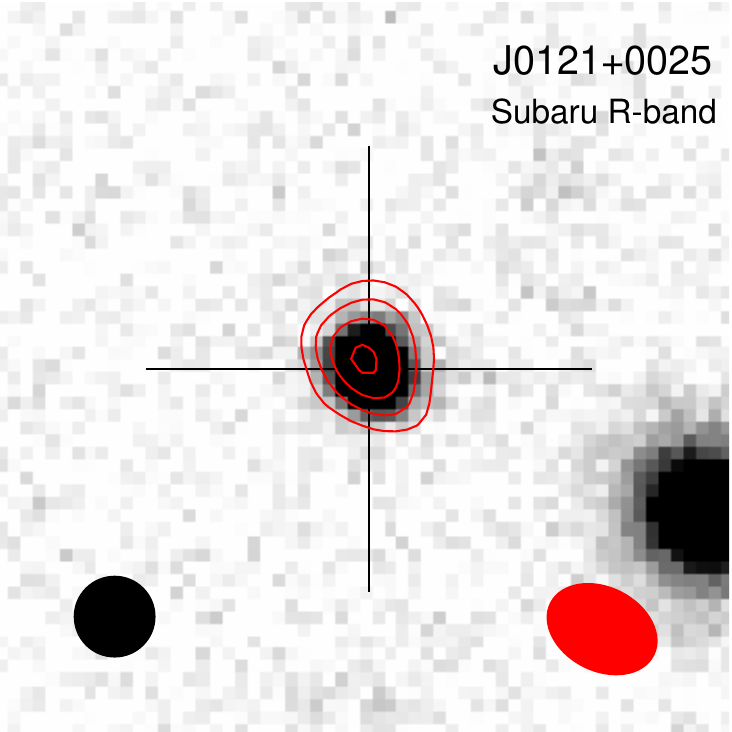}
\includegraphics[width=0.22\textwidth,clip]{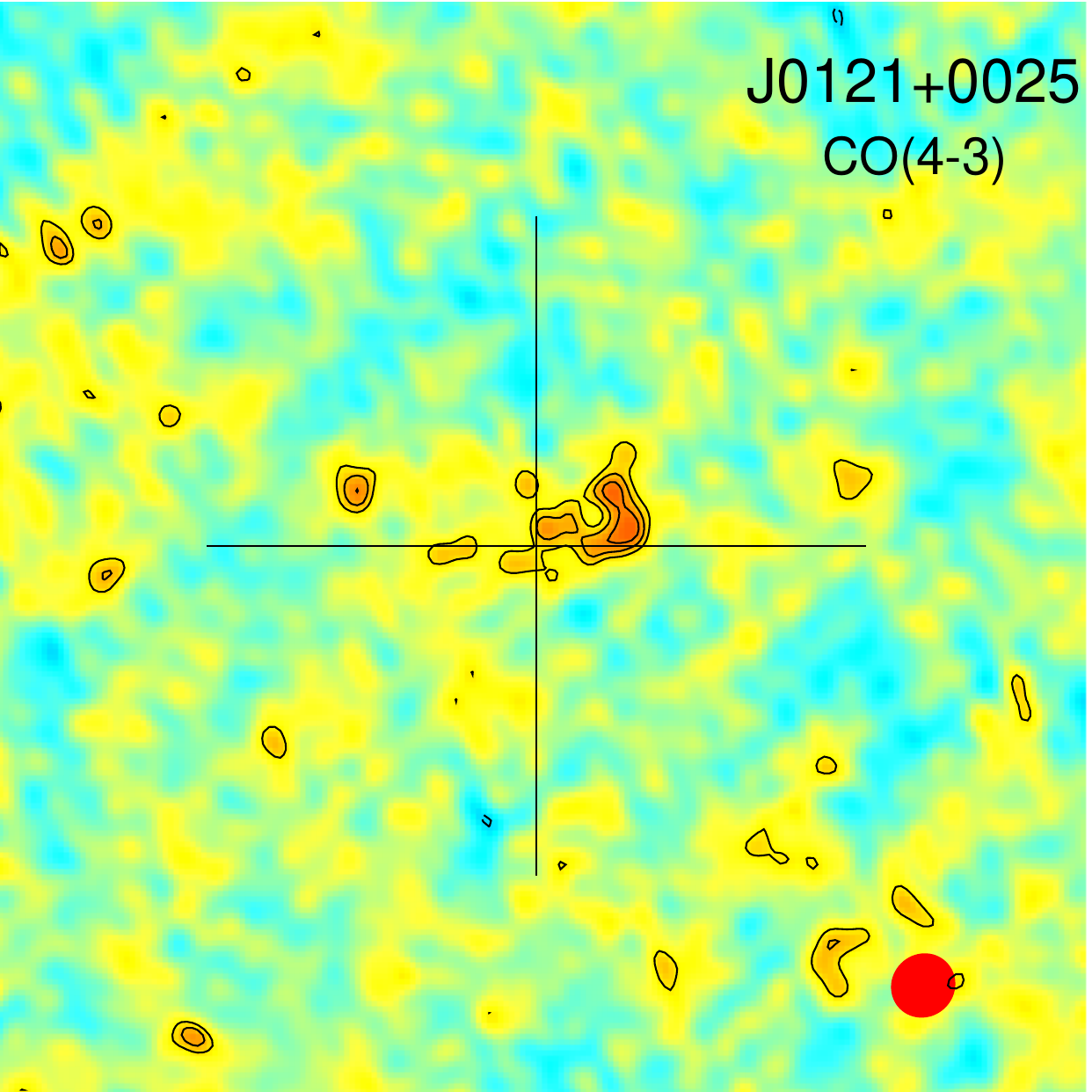}
\includegraphics[width=0.28\textwidth,clip]{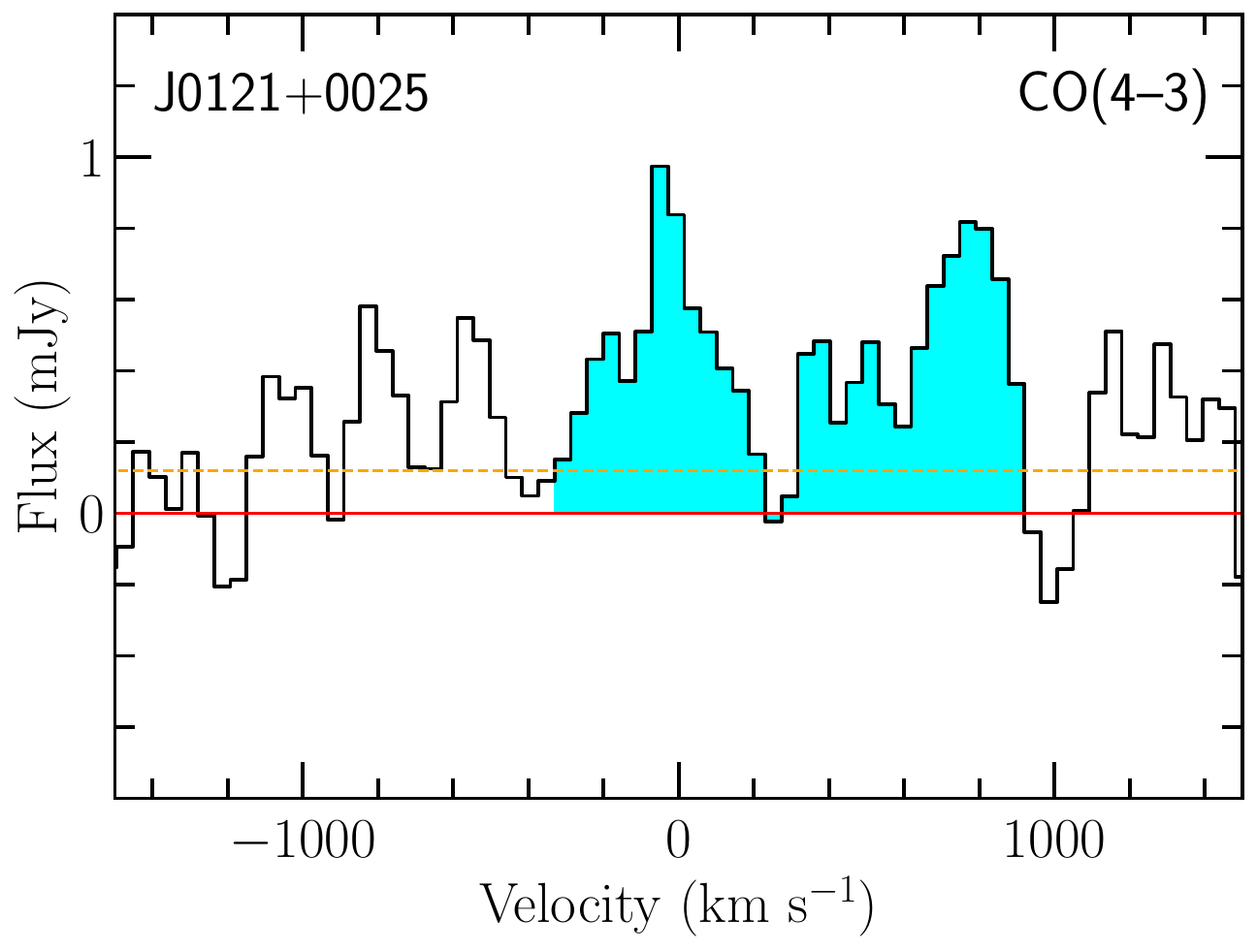}

\hspace{0.4cm}
\includegraphics[width=0.22\textwidth,clip]{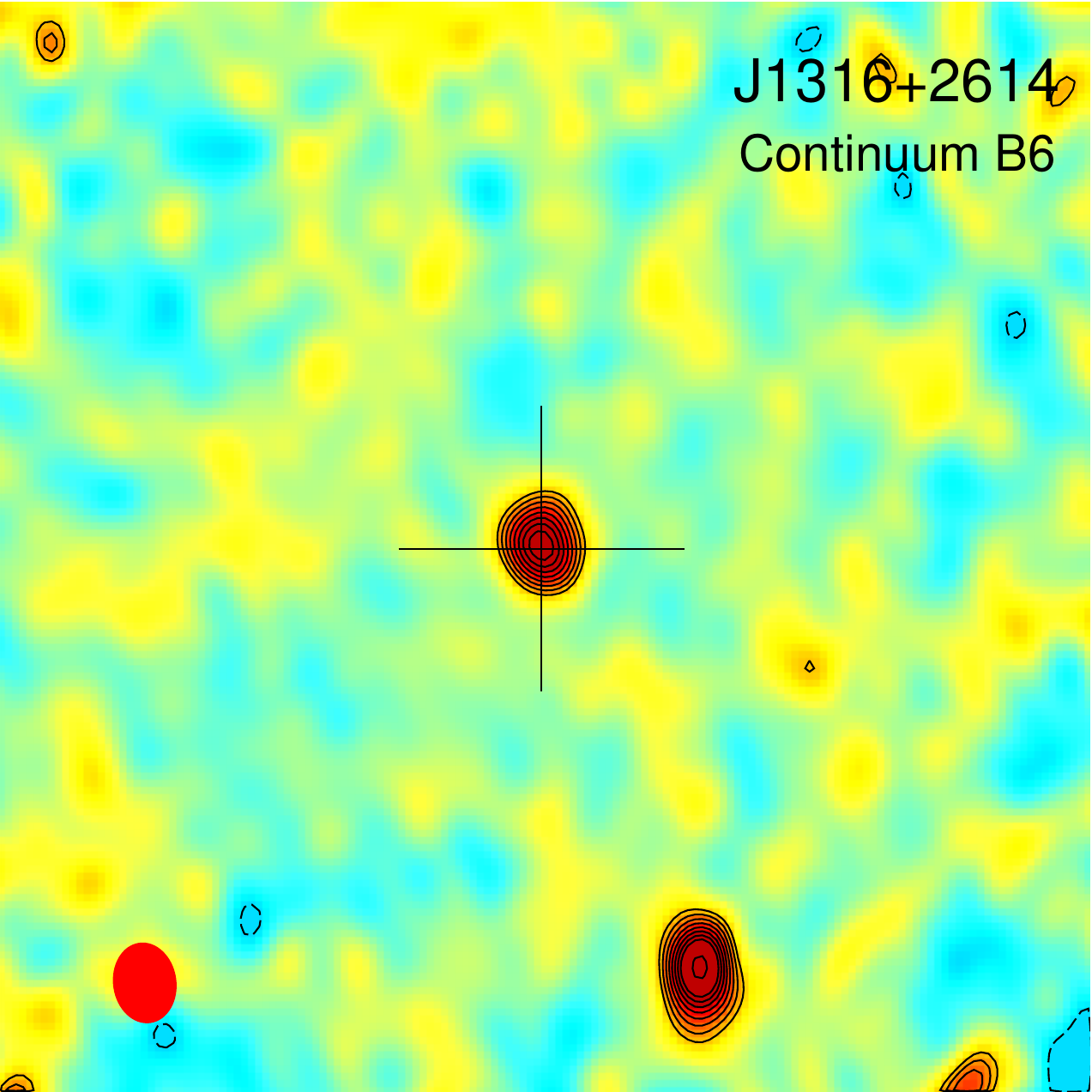}
\includegraphics[width=0.22\textwidth,clip]{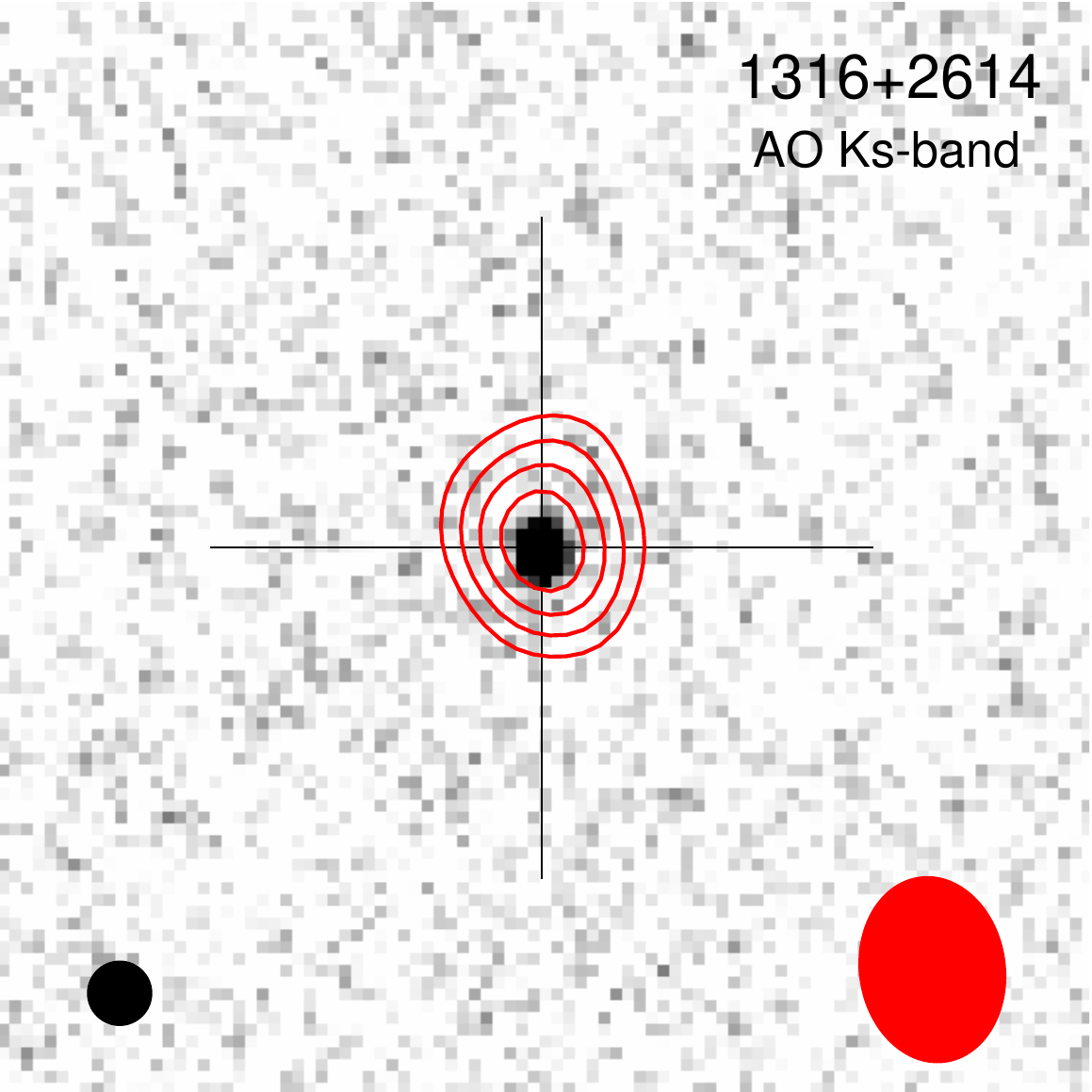}
\includegraphics[width=0.22\textwidth,clip]{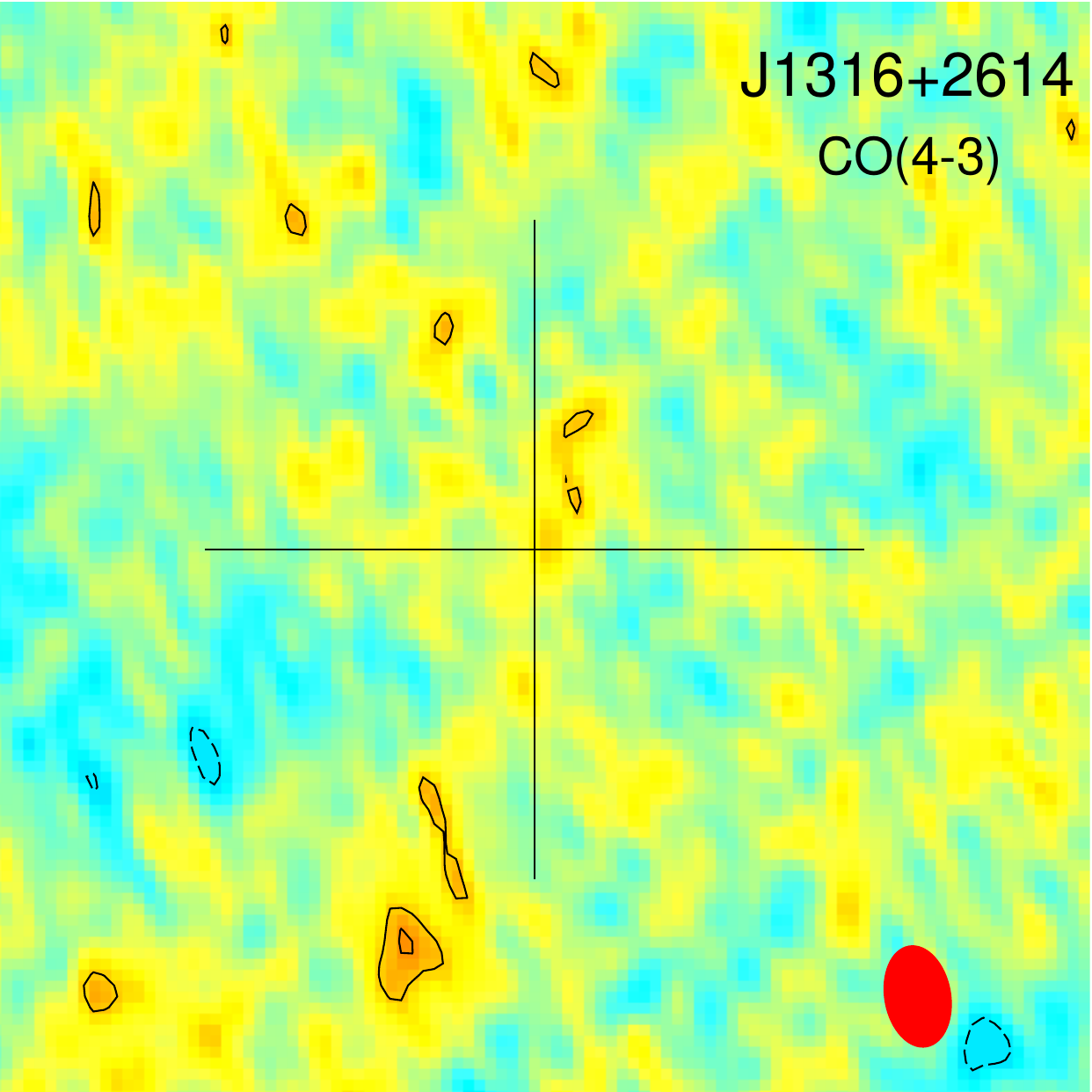}
\caption{See Fig.~\ref{fig:contB6-CO_1} for the description of panels.}
\label{fig:contB6-CO_3}
\end{figure*}

\end{appendix}
%-----------------------------------------------------------------------

\end{document}